\documentclass[twocolumn]{aastex62}
\usepackage{amsmath, amssymb, amsthm, mathtools}
\usepackage{tabularx, multirow}
\usepackage{wasysym}
\usepackage{natbib}
\usepackage{rotating}

\newcommand\SfourG{S\textsuperscript{4}G}

\received{27 August 2021}
\revised{24 March 2022}
\accepted{12 April 2022}
\submitjournal{The Astrophysical Journal}

\shorttitle{DEGS Stellar Halos}
\shortauthors{Gilhuly et al.}

\begin{document}

\title{Stellar haloes from the The Dragonfly Edge-on Galaxies Survey}


\author{Colleen Gilhuly}
\affil{Department of Astronomy \& Astrophysics, University of Toronto, Toronto, ON M5S 3H4, Canada}
\affil{Dunlap Institute for Astronomy and Astrophysics, University of Toronto, Toronto ON, M5S 3H4, Canada}
\author{Allison Merritt}
\affil{Max-Planck-Institut f{\"u}r Astronomie, D-69117 Heidelberg, Germany}
\author{Roberto Abraham}
\affil{Department of Astronomy \& Astrophysics, University of Toronto, Toronto, ON M5S 3H4, Canada}
\affil{Dunlap Institute for Astronomy and Astrophysics, University of Toronto, Toronto ON, M5S 3H4, Canada}
\author{Shany Danieli}
\thanks{NASA Hubble Fellow}
\affil{Department of Astronomy, Yale University, New Haven, CT 06511, USA}
\affil{Department of Physics, Yale University, New Haven, CT 06520, USA}
\affil{Yale Center for Astronomy and Astrophysics, Yale University, New Haven, CT 06511, USA}
\affil{Institute for Advanced Study, 1 Einstein Drive, Princeton, NJ 08540, USA}
\author{Deborah Lokhorst}
\affil{Department of Astronomy \& Astrophysics, University of Toronto, Toronto, ON M5S 3H4, Canada}
\affil{Dunlap Institute for Astronomy and Astrophysics, University of Toronto, Toronto ON, M5S 3H4, Canada}
\author{Qing Liu}
\affil{Department of Astronomy \& Astrophysics, University of Toronto, Toronto, ON M5S 3H4, Canada}
\author{Pieter van Dokkum}
\affil{Department of Astronomy, Yale University, New Haven, CT 06511, USA}
\author{Charlie Conroy}
\affil{Harvard-Smithsonian Center for Astrophysics, Cambridge, MA 02138, USA}
\author{Johnny Greco}
\affil{Center for Cosmology and Astroparticle Physics (CCAPP), The Ohio State University, Columbus, OH 43210, USA}

\begin{abstract}

We present the primary results from the Dragonfly Edge-on Galaxies Survey (DEGS), an exploration of the stellar halos of twelve nearby ($d<25$~Mpc) edge-on disc galaxies with the Dragonfly Telephoto Array. The edge-on orientation of these galaxies allows their stellar halos to be explored with minimal obscuration by or confusion with the much brighter disc light. Galaxies in the sample span a range of stellar masses from $10^{9.68}-10^{10.88}\:  \mathrm{M}_\odot$.  We confirm that the wide range of stellar halo mass fractions previously seen for Milky Way-mass galaxies is also found among less massive spiral galaxies. The scatter in stellar halo mass fraction is large but we do find a significant positive correlation between stellar halo mass fraction and total stellar mass when the former is measured beyond five half-mass radii. Reasonably good agreement is found with predictions from cosmological hydrodynamical simulations, although observed stellar halo fractions appear to be somewhat lower than expected from these simulations.

\end{abstract}


\section{Introduction} \label{introduction}

The textbook story of galaxy formation predicts that red elliptical galaxies formed in major mergers, while blue spiral galaxies have largely avoided recent mergers, preventing their thin discs with ordered rotation from being distorted or destroyed. These expectations are based on early N-body simulations \citep[e.g.][]{quinn93} where particles could only interact gravitationally. Our understanding of galaxy formation and evolution in a cosmological context has since matured. It is now well established that discs can survive mergers \citep{walker96, hopkins09, moster10, lagos18, dsouzabell18b, hammer18}; indeed, they may even be formed in major mergers \citep{springelhernquist05, robertson06}. {In-situ star formation and the accretion of stars that formed ex-situ both contribute to the growth of galaxies throughout their lifetimes. Spiral and elliptical galaxies alike grow largely by in-situ star formation at early times ($z\gtrsim2$), followed by an extended phase (since $z\lesssim3$) where accretion dominates} \citep{oser10}. More massive galaxies tend to have a higher fraction of accreted stars \citep{leeyi13} and disc-dominated galaxies have a smaller fraction of accreted stars than dispersion-dominated galaxies at fixed mass \citep{rodriguezgomez16}.

The merger history of a galaxy is imprinted onto its stellar halo, where both diffuse debris and coherent stellar streams from past accretion events and mergers linger for billions of years due to their long dynamical time scales \citep{searlezinn78, bullockjohnston05, abadi06, cooper10}. Stellar streams have been detected in many galaxies outside of the Local Group \citep{malinhadley97, shang98, forbes03, tal09, mouhcine10, martinezdelgado10, miskolczi11, atkinson13, hood18, bilek20}, implying that smooth, phase-mixed stellar halos should be common as well. However, stellar halos are notoriously difficult to observe and quantify. They can extend over 100 kpc from the center of the galaxy and have typical surface brightnesses of 28 mag/arcsec$^2$ and fainter \citep{bullockjohnston05, cooper10}, so characterizing them requires exquisite control over systematics such as the background sky level and scattered light. Failing to account for scattered light leads to contaminated measurements or even false stellar halo detections \citep{dejong08, sandin14, sandin15}.

A further complication is that while the stellar halo region is dominated by accreted stars \citep[along with dynamically heated in-situ stars in the ``inner'' stellar halo:][]{zolotov09, purcell10, tissera13}, it does not necessarily contain the majority of the total accreted stellar mass. A large fraction of the stellar mass from any given accretion event quickly settles into the center of the galaxy \citep{pillepich15}, where it becomes difficult (if not impossible) to disentangle from stars that formed in-situ. When comparing simulations and observations, it is therefore important to remember that the observed stellar halo does not directly correspond to the simulated accreted or ex-situ stellar mass \citep[though it is possible to empirically determine the relationship between the stellar halo and total accreted stellar mass; see][]{harmsen17}.

For spiral galaxies, most observational and theoretical efforts have focused on understanding the stellar halos of Milky Way mass galaxies. These efforts have led to a more complete understanding of these stellar halos. For the Milky Way itself, very detailed investigations are possible thanks to surveys such as SDSS \citep{bell08, deason11}, SEGUE \citep{kafle14, xue15}, APOGEE \citep{fernandezalvar17, mackerethbovy20}, LAMOST \citep{xu18}, HSC-SSP \citep{fukushima19}, H3 \citep{conroy19}, and Gaia \citep{gaia0, gaia1, gaia2, gaia3}. Gaia has been particularly revolutionary, supplementing stellar photometry and spectra with accurate positions, distances, and velocities. In its {early third} data release comprising {1.8} billion sources in total, parallaxes and proper motions were released for {1.5} billion sources and {updated second data release} radial velocities were released for {7.2} million sources. {The second portion of the third data release, planned for 2022, will increase the number of sources with radial velocities to approximately 30 million} \citep{gaia3}. The Gaia-Sausage-Enceladus dwarf galaxy \citep{belokurov18, haywood18, koppelman18,helmi18, fattahi19} was identified as the relic of a major progenitor of the Milky Way's halo in a direct collision 8-10 Gyr ago. Recently, \cite{naidu20} attributed 92\% of the Milky Way's stellar halo (traced by giants from the H3 Survey) to distinct structures in chemodynamical space and then determined the relative contribution of accreted and in-situ components across the Galaxy by radius and height above the midplane. Gaia data is allowing us to realize the goal of identifying individual progenitors of the stellar halo for the Milky Way.

M31's more massive stellar halo indicates {that it was likely formed by more massive accretion events than those that formed the Milky Way's stellar halo} \citep{deason15, dsouzabell18, smercina20}. Although it is not possible to obtain the same rich phase space information that Gaia has provided for the Milky Way, M31's stellar halo has been extensively mapped in resolved stars. Some noteworthy observing campaigns are the Subaru Telescope surveys \citep{tanaka10, komiyama18}, {PHAT with the Hubble Space Telescope \citep[HST,][]{dalcanton12, williams12, williams15}}, PAndAS with {the Canada-France-Hawaii Telescope} \citep[CFHT,][]{ibata14, mcconnachie18}, and the SPLASH survey with Keck/DEIMOS \citep{gilbert12, gilbert14, gilbert18}. M31 likely experienced its most recent merger approximately 2 Gyr ago, resulting in its giant stellar stream and triggering a galaxy-wide star formation burst \citep{hammer18}. M32 may be the core remnant of this stream's progenitor \citep{dsouzabell18b}.

Outside of the Local Group, resolved stars studies \citep{barker09, bailin11, greggio14, harmsen17, smercina20, cohen20, rejkuba22} reveal that Milky Way-mass spirals exhibit a range of stellar halo mass fractions. Complimentary work based on integrated light observations confirms the diversity of stellar halo masses even within a relatively small range of host galaxy stellar masses \citep{merritt16, jang20}. {Both observational approaches yield stellar halo mass fractions of $\sim0.2-6\%$.} \cite{merritt16} {also report three stellar halo non-detections and set upper limits on the order of $0.01\%$.} \cite{jang20} {revisit one of these galaxies with a stellar halo non-detection, M101, and measure a stellar halo mass fraction of $0.20^{+0.10}_{-0.08}\%$.} The diversity of stellar halos among galaxies of similar morphology and mass reflects the variety in accretion histories of these galaxies. Although the depth of SDSS imaging is insufficient to probe the stellar halo of individual galaxies, their average properties can be studied by stacking thousands of galaxies of similar mass and concentration \citep{talvandokkum11, dsouza14}. This approach reveals the underlying trends in stellar halo properties with the host galaxy's stellar mass but cannot probe the diversity of stellar halos at fixed stellar mass. 

The stellar halos of less massive spiral galaxies are little explored in comparison to Milky Way-mass halos. This is true both for observations and for theory. In the latter case, this is because predictions based on n-body simulations are limited by the mass resolution. {Simulations of cosmologically significant volumes ($\sim100\textrm{ Mpc h}^{-1}$) typically have particle masses of a few times $10^6$~M$_\odot$ or greater} \citep[e.g.][]{springel05, dubois14, vogelsberger14, khandai15, schaye15, springel18}. {At this resolution, a galaxy with a total stellar mass of $10^{10}$~M$_\odot$ with a stellar halo fraction of 1\% will have its stellar halo resolved into fewer than 100 star particles.} At {lower total stellar} masses {or} for galaxies with less active accretion histories, {resolution effects become increasingly worse} \citep[for example, see Figure~2 of][]{canas20}. 

Despite these challenges, it has been predicted that stellar halo mass fractions will decrease with host total stellar mass \citep{elias18}, matching earlier predictions of analytic models \citep{purcell07}. The outer light fraction of the stacked surface brightness profiles of observed galaxies \citep{dsouza14} agrees qualitatively with these predictions over a mass range of $10^{10.0} - 10^{11.4}\:  \mathrm{M}_\odot$, but without measurements of individual galaxies we are unable to compare the distribution of observed and simulated stellar halos. Notably, \cite{seth07} and \cite{dejong07} used {(HST)} to detect a stellar halo surrounding NGC 4244 \citep[$\mathrm{M}_* = 10^{9.12}\:  \mathrm{M}_\odot$,][]{querejeta15} with an estimated mass of $3 \times 10^6\:  \mathrm{M}_\odot$, corresponding to a stellar halo mass fraction of 0.2\%. {\cite{gilbert22} recently reported that $22\pm2$\% of their spectroscopic sample of RGB stars in M33 \citep[$\mathrm{M}_* = 10^{9.68}\:  \mathrm{M}_\odot$,][]{corbelli14} belonged to a kinematically distinct, halo-like component after PAndAS-based studies were only able to set upper limits on M33's stellar halo luminosity \citep{cockcroft13, mcmonigal16}.} With stellar halo non-detections for Milky Way-mass spiral galaxies \citep{vandokkum14, merritt16}, one could question whether or not the stellar halos of lower mass spirals are reliably detectable in integrated light. 

The Dragonfly Telephoto Array \citep{abraham14} is well-suited for exploration of the stellar halos of galaxies with Milky Way-like and lower stellar masses. The array's wide field of view coupled with state-of-the-art optical coatings designed to minimize reflections have been used to explore the faint outskirts of luminous nearby galaxies \citep{vandokkum14, merritt16, zhang18, vandokkum19, gilhuly20, miller21} but low mass targets are best investigated using samples of edge-on galaxies \citep{elias18}. We have therefore undertaken the Dragonfly Edge-on Galaxies Survey (DEGS), a complimentary observing campaign to the Dragonfly Nearby Galaxies Survey (DNGS) which focused on nearby spiral galaxies regardless of orientation \citep{merritt16}. We have produced deep $g-$ and $r-$band images for twelve edge-on spiral galaxies in the local Universe (at distances $d \lesssim 25$ Mpc), representing a wide range of stellar masses and morphologies. Two DEGS targets have been previously studied and released to the public: NGC 5907 \citep{vandokkum19} and NGC 4565 \citep{gilhuly20}. The fully reduced images for all remaining DEGS targets are publicly available\footnote{Dragonfly Data Access: \url{https://www.dragonflytelescope.org/data-access.html}} as of the publication date of this paper; please refer to Appendix~\ref{datarelease} for additional information. 

In this paper we present our work on the stellar halos of the entire DEGS sample. Section~\ref{sample_obs} describes our sample and observations. Section~\ref{analysis} pertains to our analysis procedure and Section~\ref{results} presents our resulting stellar halo mass fractions. We compare our measurements to previous work with observed and simulated stellar halos in Section~\ref{discussion} and summarize our conclusions in Section~\ref{conclusions}.

\section{Sample selection and observations} \label{sample_obs}

We constructed the DEGS sample by first defining a parent sample using objective criteria and then visually identifying edge-on spirals within that sample. The parent sample was obtained from a query to the HyperLeda database\footnote{\url{http://leda.univ-lyon1.fr/}} with four selection criteria. First, the galaxies must be visible from the telescope site (thus having declination $\delta > -20^{\circ}$). We chose to prioritize galaxies with large angular sizes in order to take advantage of Dragonfly's large field of view while simultaneously maximizing the effective resolution on physical scales (as Dragonfly has a native pixel size of $2.85''$). Therefore we required a distance modulus $m - M < 32$ (corresponding to distances $d \lesssim 25$ Mpc). To avoid most contamination by Galactic cirrus we required extinction $A_V < 0.1 \textrm{~mag}$\footnote{This roughly corresponds to $F_{100 \mu m} < 1.5$ MJy/sr. $A_V$ was favoured at this stage due to convenience. We later verified that all of our final targets meet the $F_{100 \mu m}$ criterion.}. Finally, we required $M_B < -18$ to effectively probe galaxies with luminosities (masses) similar to or larger than the Large Magellanic Cloud.

From the parent sample {of a few hundred galaxies, 40 spiral} galaxies with {axis ratios} $b/a < 0.5$ (as estimated from major and minor diameters tabulated in the NASA/IPAC Extragalactic Database\footnote{\url{http://ned.ipac.caltech.edu/}}) were reviewed visually to search for edge-on spiral galaxies. A very conservative maximum $b/a$ was chosen to avoid overlooking edge-on galaxies with large bulges or heated disks. High-inclination galaxies were rejected, as were ambiguous cases. Particular attention was given to the low luminosity end of the sample, as a very late barred spiral may appear edge-on at much milder inclinations. {Of the 15 galaxies confirmed to be edge-on, three were removed from the sample due to the presence of other galaxies located within a few optical radii in projection along the minor axis or nearby bright stars}. The final sample contains 12 galaxies, including NGC 4565 \citep[which had already been observed by Dragonfly; see][for details]{gilhuly20}. The targets and some of their basic properties are listed in Table~\ref{tab:targets}.

\begin{table*}[tbh]
\begin{center}
\begin{tabular*}{\textwidth}{c @{\extracolsep{\fill}} ccccccccccc} 
 \hline \hline
 \multirow{2}{*}{Target} & \multirow{2}{*}{$\alpha$}& \multirow{2}{*}{$\delta$} & $M_B$ & \multirow{2}{*}{$\log M_*$}\footnote{Log stellar mass from the Spitzer Survey of Stellar Structure in Galaxies \citep[\SfourG;][]{querejeta15, munozmateos15, s4g}. {Following \cite{querejeta15}, we adopt an uncertainty of 0.2 dex.}} & \multirow{2}{*}{$b/a$} & $D\footnote{{Mean redshift-independent distance from NED and standard deviation of individual distances, tabulated by \SfourG{} \citep{s4g}.}}$ & $F_{100\mu m}$\footnote{Mean 100$\mu$m emission within a $5'$ radius of target's position, accessed via \href{https://irsa.ipac.caltech.edu/applications/DUST/}{https://irsa.ipac.caltech.edu/applications/DUST/}} & $N_{obs}$ & $N_{final}$ & $3\sigma$ SB limit\footnote{$3\sigma$ surface brightness limit in mag/arcsec$^2$ {for regions of $60\times60$~arcsec$^2$} calculated using the \texttt{sbcontrast} package.}  \\ 
 & & & (mag) & & & (Mpc) & (MJy/sr) & (total) & ($g$, $r$) & ($g$, $r$) \\ \hline \hline
 \multirow{2}{*}{NGC 3044} & \multirow{2}{*}{09:53:40.88} & \multirow{2}{*}{+01:34:46.7} & \multirow{2}{*}{-19.16} & \multirow{2}{*}{10.26} & \multirow{2}{*}{0.11} & \multirow{2}{*}{$23.3\pm2.2$} & \multirow{2}{*}{1.173} & \multirow{2}{*}{2816} & 731 & $29.78 \pm 0.16$  \\ 
   & & & & & & & & & 910 & $29.13 \pm 0.16$  \\
 \multirow{2}{*}{NGC 3432} & \multirow{2}{*}{10:52:31.13} & \multirow{2}{*}{+36:37:07.6} & \multirow{2}{*}{-18.64} & \multirow{2}{*}{9.68} & \multirow{2}{*}{0.22} & \multirow{2}{*}{$12.9\pm3.1$} & \multirow{2}{*}{0.616} & \multirow{2}{*}{7838} & 1103 & $30.18 \pm 0.06$  \\
  & & & & & & & & & 1235 & $29.86 \pm 0.06$  \\
 \multirow{2}{*}{NGC 3501} & \multirow{2}{*}{11:02:47.27} & \multirow{2}{*}{+17:59:21.6} & \multirow{2}{*}{-18.13} & \multirow{2}{*}{10.16} & \multirow{2}{*}{0.13} & \multirow{2}{*}{$23.8\pm1.9$} & \multirow{2}{*}{1.179} & \multirow{2}{*}{4021} & 458 & $29.98 \pm 0.05$  \\ 
  & & & & & & & & & 620 & $29.48 \pm 0.05$  \\
 \multirow{2}{*}{NGC 3628} & \multirow{2}{*}{11:20:16.97} & \multirow{2}{*}{+13:35:22.9} & \multirow{2}{*}{-20.24} & \multirow{2}{*}{10.80} & \multirow{2}{*}{0.20} & \multirow{2}{*}{$11.3\pm2.8$} & \multirow{2}{*}{1.440} & \multirow{2}{*}{4829} & 766 & $29.90 \pm 0.04$  \\ 
  & & & & & & & & & 891 & $29.45 \pm 0.04$  \\
 \multirow{2}{*}{NGC 4010} & \multirow{2}{*}{11:58:37.89} & \multirow{2}{*}{+47:15:41.4} & \multirow{2}{*}{-18.09} & \multirow{2}{*}{9.91} & \multirow{2}{*}{0.23} & \multirow{2}{*}{$19.0\pm2.1$} & \multirow{2}{*}{1.252} & \multirow{2}{*}{3670} & 765 & $30.01 \pm 0.04$  \\ 
  & & & & & & & & & 835 & $29.54 \pm 0.04$  \\
 \multirow{2}{*}{NGC 4013} & \multirow{2}{*}{11:58:31.38} & \multirow{2}{*}{+43:56:47.7} & \multirow{2}{*}{-19.10} & \multirow{2}{*}{10.63} & \multirow{2}{*}{0.19} & \multirow{2}{*}{$18.6\pm2.5$} & \multirow{2}{*}{0.871} & \multirow{2}{*}{3773} & 588 & $29.92 \pm 0.04$  \\ 
  & & & & & & & & & 671 & $29.45 \pm 0.04$  \\
 \multirow{2}{*}{NGC 4307} & \multirow{2}{*}{12:22:05.68} & \multirow{2}{*}{+09:02:37.1} & \multirow{2}{*}{-19.02} & \multirow{2}{*}{10.37} & \multirow{2}{*}{0.23} & \multirow{2}{*}{$22.6\pm4.1$} & \multirow{2}{*}{1.103} & \multirow{2}{*}{3568} & 558 & $29.91 \pm 0.05$  \\ 
  & & & & & & & & & 669 & $29.44 \pm 0.05$  \\
 \multirow{2}{*}{NGC 4330} & \multirow{2}{*}{12:23:17.25} & \multirow{2}{*}{+11:22:04.7} & \multirow{2}{*}{-18.30} & \multirow{2}{*}{9.91} & \multirow{2}{*}{0.13} & \multirow{2}{*}{$20.4\pm0.6$} & \multirow{2}{*}{1.365} & \multirow{2}{*}{2573} & 631 & $29.90 \pm 0.10$  \\ 
  & & & & & & & & & 726 & $29.35 \pm 0.10$  \\
 \multirow{2}{*}{NGC 4437} & \multirow{2}{*}{12:32:45.59} & \multirow{2}{*}{+00:06:54.1} & \multirow{2}{*}{-18.43} & \multirow{2}{*}{10.23} & \multirow{2}{*}{0.13} & \multirow{2}{*}{$11.6\pm3.1$} & \multirow{2}{*}{1.170} & \multirow{2}{*}{3179} & 545 & $29.72 \pm 0.03$  \\ 
  & & & & & & & & & 532 & $29.23 \pm 0.03$  \\
 \multirow{2}{*}{NGC 4565} & \multirow{2}{*}{12:36:20.78} & \multirow{2}{*}{+25:59:15.6} & \multirow{2}{*}{-20.35} & \multirow{2}{*}{10.88} & \multirow{2}{*}{0.12} & \multirow{2}{*}{$11.7\pm4.8$} & \multirow{2}{*}{0.743} & \multirow{2}{*}{26168} & 4448 & $30.31 \pm 0.07$  \\
  & & & & & & & & & 4416 & $29.81 \pm 0.07$  \\
 \multirow{2}{*}{NGC 4634} & \multirow{2}{*}{12:42:40.96} & \multirow{2}{*}{+14:17:45.0} & \multirow{2}{*}{-18.50} & \multirow{2}{*}{10.12\footnote{The \SfourG{} stellar mass for NGC~4634 was unusually small relative to the total stellar mass measured by integrating the Dragonfly stellar mass surface density profiles. This may be related to Galactic cirrus contamination. We therefore adopt the mean difference between Dragonfly and \SfourG{} stellar masses for the other DEGS targets to empirically correct the Dragonfly stellar mass for NGC~4634.}} & \multirow{2}{*}{0.27} & \multirow{2}{*}{$20.2\pm1.3$} & \multirow{2}{*}{1.380} & \multirow{2}{*}{2951} & 636 & $29.95 \pm 0.06$ \\ 
  & & & & & & & & & 782 & $29.41 \pm 0.06$   \\
 \multirow{2}{*}{NGC 5907} & \multirow{2}{*}{15:15:53.77} & \multirow{2}{*}{+56:19:43.6} & \multirow{2}{*}{-20.11} & \multirow{2}{*}{10.87} & \multirow{2}{*}{0.11} & \multirow{2}{*}{$16.6\pm1.9$} & \multirow{2}{*}{0.496} & \multirow{2}{*}{2920} & 607 & $29.97 \pm 0.04$  \\ 
  & & & & & & & & & 767 & $29.45 \pm 0.04$  \\
 \hline \hline
\end{tabular*}
\end{center}
\caption{Table of basic parameters of all observed DEGS targets. The total number of observed frames, $N_{obs}$, is approximately evenly split between $g$ and $r$ filters. The exact number of frames that pass all checks in the pipeline and contribute to the final stacked images, $N_{final}$, is listed for each band. }
\label{tab:targets}
\end{table*}

Observations for the 11 galaxies hitherto unobserved by Dragonfly in the DEGS sample were carried out with the Dragonfly Telephoto Array between February 2018 and May 2019. At this time, Dragonfly was configured with 48 Canon 400mm $f$/2.8L IS II USM telephoto lenses divided equally between two independent mounts. There was an even split of $g$ and $r$ filters on each mount. A nine-point dither pattern with displacements of $25'$ in right ascension and declination was used to regularize over detector- or lens-dependent spatial variation and any variation in the sky background due to effects originating in the Earth's atmosphere (such as airglow).

Based on previous photometry obtained when Dragonfly had fewer lenses in the array  \citep{vandokkum14, merritt16, zhang18}, 5 hour integrations are needed to reach the smooth accreted component of stellar halos (around $30-32$ mag/arcsec$^2$). 
Anticipating roughly $40-60$\% of the individual exposures would be of sufficient quality to be retained by our pipeline \citep{danieli20}, our goal was to collect 10 hours of raw exposure time (under largely clear conditions) for each target. This is equivalent to 2880 total 600-second exposures by individual lens-camera systems in Dragonfly, approximately evenly divided between $g$ and $r$. The number of raw exposures and the final number of exposures in each stacked image is listed in Table~\ref{tab:targets}. 

\subsection{Data reduction} \label{datareduction}

The individual exposures, or frames, collected for each target were processed, assessed, and combined using the Dragonfly reduction pipeline. The interested reader is referred to \cite{danieli20}, but in brief, the pipeline creates master calibration frames, calibrates and registers frames, stacks sky-subtracted frames to produce a deeper source mask, and then repeats the sky modelling and subtraction process using the deeper source mask before creating the final stacks for each band. Frames can be rejected for failing to meet one of several quality checks. Most checks are relatively simple: Does the frame contain non-zero data? Are a reasonable number of stars detected? Are these stars round or elongated? In addition to these simple checks early in the pipeline, we later accept or reject frames based on their photometric zeropoints. Zeropoints that deviate significantly from the mean indicate an enhanced wide angle point-spread function (PSF), likely due to unfavourable high atmospheric conditions. 

\subsubsection{Correcting poorly flattened frames}

 {Although sky flats were attempted at twilight and dawn every night, some nights did not have a master flat available. This could happen if an insufficient number of flat exposures were obtained (e.g. due to poor weather or the domes being closed) or if they failed to pass all quality checks. Frames were flattened with the master flat from the closest available night when there was no master flat available for that night.} In some cases, individual frames were poorly flattened due to a poor match with the nearest available master flat. Depending on the number of observing nights affected for a given galaxy, this poor flattening could result in bowl-like residual light about the center of the final stacked image. An optional module, \texttt{check\_frame\_flatness}, was added to the \cite{danieli20} pipeline to review flattened frames for this problem and to find a more appropriate master flat if possible.  

{The new optional module was called after dark subtraction, flat fielding, plate solving, and several quality checks for a given night's data. For each frame, we calculated the median pixel value of each column and each row to produce horizontal and vertical profiles of flux across the frame. Both profiles were fitted with a second order polynomial $f(x)=ax^2+bx+c$. This function was not necessarily a good fit to the bowl-like residuals we wished to identify, but we found it could successfully separate poorly flattened frames from well flattened frames. The $x^2$ coefficients $a$ and the $x$ coordinate of the parabola vertex $x_0 = -0.5 b/a$ from each fit were stored. Once every frame taken during the night was fitted, the standard deviation of $a$, $\sigma_{a}$, was calculated for the ensemble of all horizontal fits and all vertical fits.}

{Next, each frame was assigned a quality flag according to several possible conditions:}
\begin{itemize}
    \item {$|a_h| > 3\sigma_{a_h}$ AND $|a_v| > 3\sigma_{a_v}$: Indicated a severe mismatch between master flat and raw frame. The frame was assigned a ``bad'' quality flag.}
    \item {$|a_h| > 3\sigma_{a_h}$ XOR $|a_v| > 3\sigma_{a_v}$: Indicated a strong flux gradient due to shutter malfunction. The frame was assigned an ``unsalvageable'' quality flag because this problem could not be fixed by reflattening.}
    \item {$a_h < 0$ AND $a_v < 0$: Could indicate a slight central excess due to mild mismatch between master flat and raw frame. If $x_{0,h}$ and $x_{0,v}$ were both within 25\% of the image centre along the given axis, the frame was assigned a ``bad'' quality flag. If this second condition was not met, the frame was assigned a ``borderline'' quality flag.} 
    \item {Frames meeting none of the above conditions were assigned a ``good'' quality flag.} 
\end{itemize}

{After assigning a quality flag to each individual frame, the frames were then grouped according to their camera ID. Since all exposures were the same length, the same master flat was used to calibrate all frames taken by a given camera on a given night. If the master flat was a poor match for a significant proportion of the frames for that camera and night, it was likely a poor match for all of those frames even if they had passed the checks described above. If $\geq30\%$ of the frames for a given camera were flagged as bad, all borderline cases were upgraded to bad. Alternatively, if $\geq70\%$ of frames for a given camera were marked as bad or borderline, all good and borderline frames were upgraded to bad.}

{At this stage, all frames for a given camera with good or borderline quality flags were set aside for the next stage of the pipeline. Frames marked as unsalvageable were rejected from any further processing. The frames with a bad quality flag were reprocessed through the initial stages of the pipeline as a group. The key difference in this rerun was that the function finding matching master calibration frames was passed an additional argument indicating that the first matching master flat should be ignored and the second should be used instead. The reprocessed frames were then assigned a new quality flag according to the conditions described above. Any reprocessed frames found to be acceptable were set aside while frames still marked with a bad quality flag were reprocessed again, this time skipping over the first two matching master flats. This process was done iteratively until all frames passed the flatness quality checks or until reflattening attempts had exhausted all master flats within three months of the observation date. Any frames that could not be reflattened adequately were rejected from further stages of the pipeline.}

This technique was very successful and significantly improved the flatness of the most strongly affected final stacked images. {We were also able to improve small numbers of poorly flattened frames contributing to final stacked images that did not seem to be affected by this problem. The tests and thresholds used above were developed by exploring the raw and calibrated frames for one of our galaxies with the worst bowl-like residuals in the final images. If this module is incorporated into the Dragonfly data reduction pipeline, it should be tested on a broader set of images first.}

\begin{figure*}[p]
\begin{centering}
\includegraphics[width=0.9\textwidth]{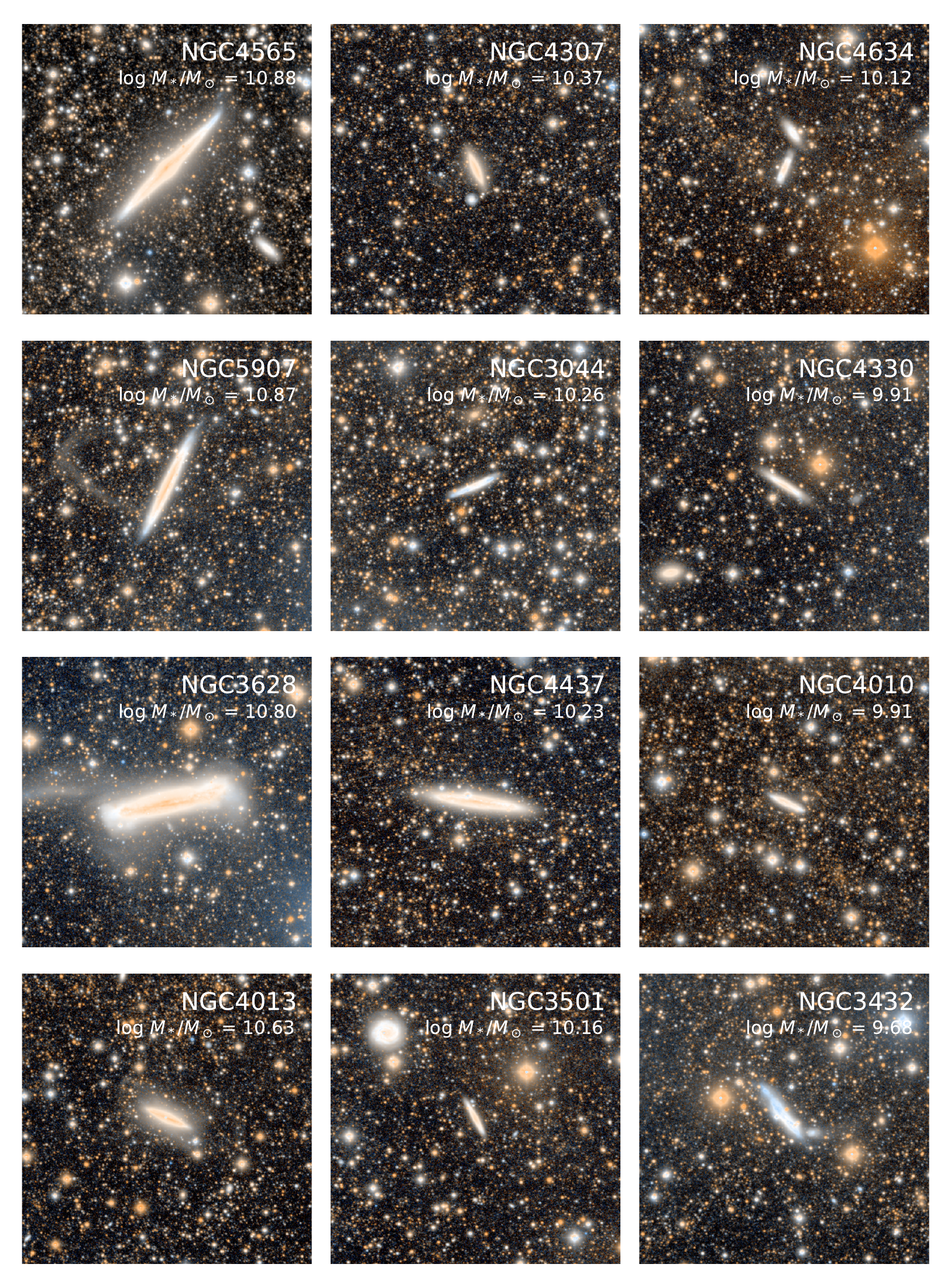}
\caption{$0.25^\circ \times 0.25^\circ$ cutouts of false colour images of each DEGS target, constructed using Dragonfly $g-$ and $r-$band images. {We used the} \cite{lupton04} {method as implemented in} \texttt{astropy.visualization.make\_lupton\_rgb()}{, with $Q=7$ and} \texttt{stretch}$=10$. {The three image colour channels are as follows: $R=r$, $G=0.5(r+g)$, $B=1.2g$.} The galaxies are shown in order of decreasing stellar mass.}
\label{fig:colour_cutouts}
\end{centering}
\end{figure*}

\subsubsection{Fully reduced images}

The final stacked images are $3.99^\circ \times 2.95^\circ$ in angular size. Please refer to Appendix~\ref{datarelease} for details on accessing these images. Central cutouts in false colour for all DEGS targets are shown in Figure~\ref{fig:colour_cutouts}. These images illustrate the diversity of discs in our sample. It can be seen that most of these galaxies do not possess prominent bulges. Variety is evident in the amount of light in their outskirts and this light's flattening in projection. Each galaxy is unique. Even for three similarly massive galaxies in our sample that each host a prominent stellar stream or tidal tail, the shape of their outskirts varies from boxy (NGC~4013) to asymmetric about the midplane and lobed (NGC~3628) to comparatively quiet (NGC~5907). The tidal disruption of UGC~05983 by NGC~3432 is striking as well. The full false colour images for each field are shown in composite figures in Appendix~\ref{fig_appendix}. 

For each field, we use the \texttt{sbcontrast}\footnote{This package is bundled with the \texttt{mrf} package. The source code is available at  \url{https://github.com/AstroJacobLi/mrf/blob/master/mrf/sbcontrast.py}. {A description of the algorithm is available in Appendix~B of} \url{https://arxiv.org/pdf/1910.12867v1.pdf}.} package to estimate the {surface brightness detection limit} as a way to quantify the depth of the images. {We calculated the $3\sigma$ depth for regions of $60\times60$~arcsec$^2$ on each image after subtracting compact sources. Our source subtraction method is described in Section}~\ref{analysis}. {The mask that we use during our analysis was used while assessing image depth. The central galaxies themselves are unmasked; we found that masking the central galaxies did not significantly change the depth measurements.} Typical $3\sigma$ depths are $\sim$29.9 mag/arcsec$^2$ for $g-$band images and $\sim$29.4 mag/arcsec$^2$ for $r-$band images. These depths are reported for each individual image in the last column of Table~\ref{tab:targets}.

\section{Analysis} \label{analysis}

After having obtained stacked images from the image reduction pipeline, some additional steps were required to prepare these images for analysis. Wide angle PSF characterization was a critical first step. Scattered light is the perennial bane of low surface brightness photometry, and edge-on galaxies are especially susceptible to scattered light contamination along the minor axis where the true light distribution may drop off very steeply \citep{dejong08, sandin14, sandin15}. The \texttt{elderflower} package\footnote{\url{https://github.com/NGC4676/elderflower}} \citep{liu21} was used to fit a three-component power law PSF {with a Moffat core} within a $0.42^\circ \times 0.42^\circ$ central cutout of each $g-$ and $r-$band image. The PSF model was extrapolated to a radius of $0.56^\circ$ or 2000~arcsec. \texttt{elderflower} simultaneously {models} the light distribution of many bright stars and the background, operating on a pixel-by-pixel basis across the entire input field.  While characterizing the PSF, any large galaxies within the cutout were masked, as were any regions with bright or high contrast/structured cirrus.

This holistic 2D approach {required adopting a parametric model for the PSF, unlike non-parametric methods based on} azimuthally averaged profiles {or stacks} of bright point sources and planets, {but in exchange it enabled} local PSF characterization even in crowded fields.  {It is advantageous to measure the PSF on the science images themselves because the PSF is not constant over time. For example, \cite{liu21} showed that the Dragonfly wide angle PSF is measureably impacted by the cleanliness of the lenses by comparing the PSFs of the M44 field observed on two clear nights immediately before and after the lenses were cleaned. The pre-cleaning PSF was much shallower than the post-cleaning PSF beyond $3'$, with fitted power law indices of 1.35 and 1.9 respectively}\footnote{{The NMS staff were instructed to delay regular lens cleaning specifically for this test.}}. {This means that a single standard, deep Dragonfly PSF would not necessarily provide a good match to a given image at these scales and larger, regardless of whether it was determined in a parametric or a non-parametric way. Furthermore, the data for individual targets in our survey typically spanned two weeks to two months, with some fields that were revisited in following years. The PSFs of the final stacked images were therefore complicated combinations of the PSFs from many individual nights and analytically deriving a combined PSF from hypothetical nightly PSFs would be non-trivial. With these considerations in mind, we believed that the best approach to obtain a PSF faithful to our data was by measuring it on the data themselves.} 

\begin{figure*}[p]
\begin{centering}
\includegraphics[width=\textwidth]{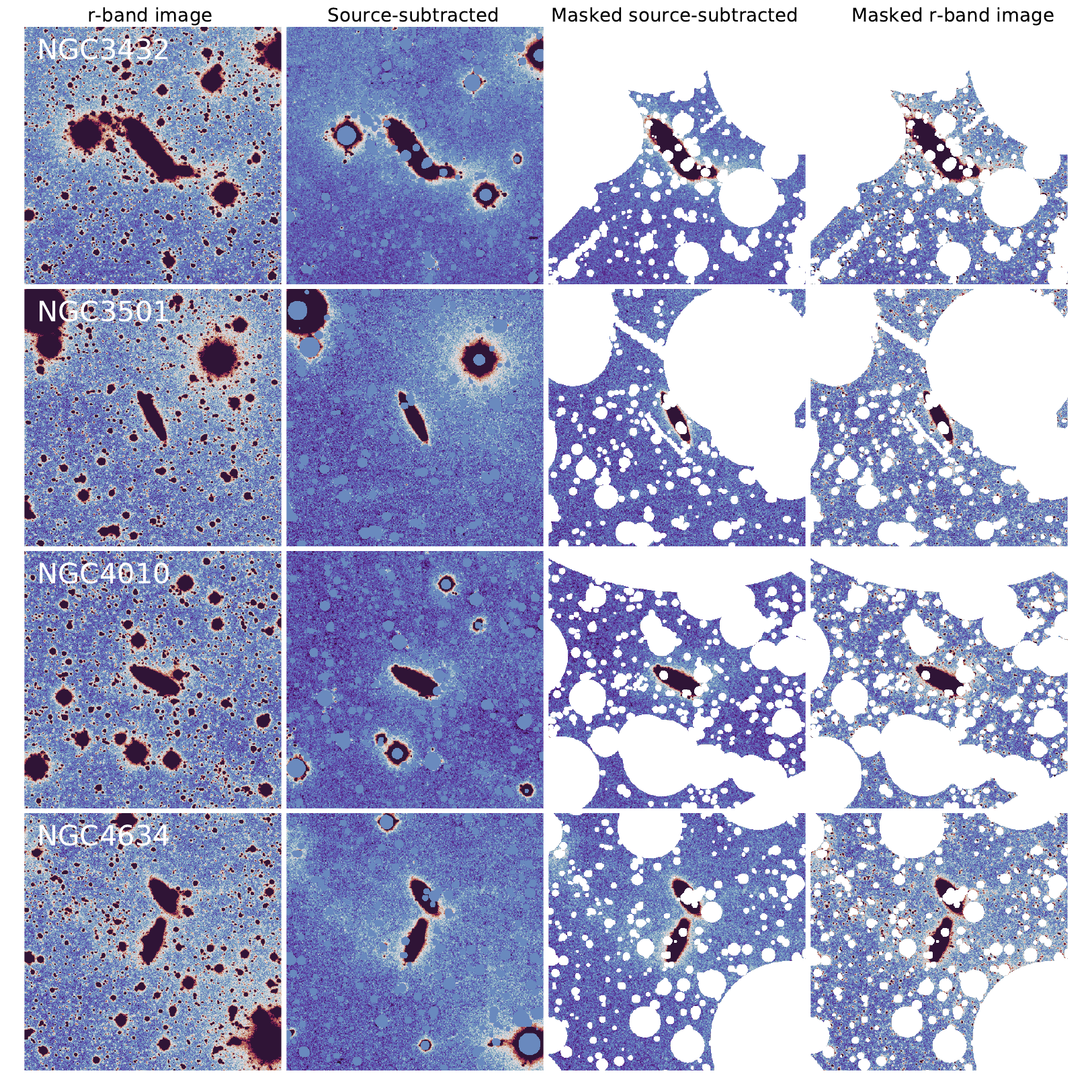}
\caption{Central cutouts showing the $r-$band Dragonfly images before {(first column)} and after {(second column)} subtracting compact sources with \texttt{mrf} for four example fields. {The third column shows the source-subtracted $r-$band images with} \texttt{mrf} residuals and the stellar aureoles of saturated stars {masked. For comparison, the same mask is applied to the original $r-$band images in the fourth column.} All images are log-scaled.}
\label{fig:mrf}
\end{centering}
\end{figure*}

\begin{figure*}[p]
\begin{centering}
\includegraphics[width=\textwidth]{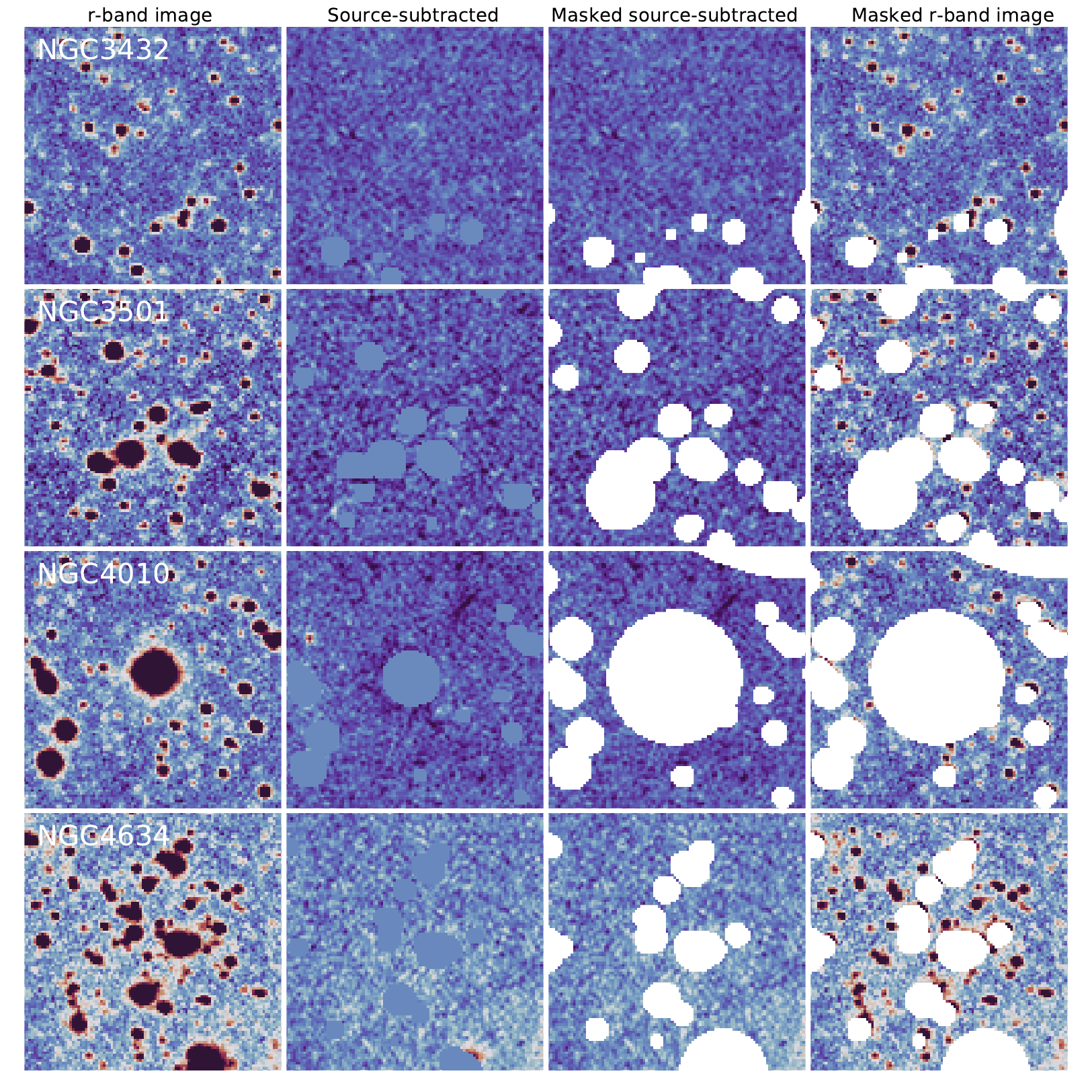}
\caption{{Same as Figure~}\ref{fig:mrf} {but for smaller cutouts highlighting the capabilities of} \texttt{mrf} {when subtracting faint compact sources. The background level in the fourth row is higher as it lies along the prominent cirrus filament seen in the fourth row of Figure}~\ref{fig:mrf}.}
\label{fig:mrf2}
\end{centering}
\end{figure*}

The final images used in our analysis were source-subtracted images generated by the \texttt{mrf} package \citep{vandokkum20}. This package takes advantage of secondary high resolution data sets to model closely-spaced compact sources in primary low resolution data sets (i.e. our Dragonfly images). We used Dark Energy Camera Legacy Survey (DECaLS) imaging as our high resolution data set for its wide coverage and good point source depth. \texttt{mrf} {employed a non-parametric kernel within $30''$.} The wide angle PSF models fitted by \texttt{elderflower} {were used to model scattered light beyond $30''$ for compact sources with $m<16.5$~mag}. Examples of the results of \texttt{mrf} are shown in Figure~\ref{fig:mrf}~{and}~\ref{fig:mrf2}. {The flux calibration for saturated stars depended on the flux in an annulus with inner and outer radii of 5 and 8 pixels, respectively. If the star was saturated or in the nonlinear regime within part or all of this flux annulus, the flux of that star was underestimated, leading to bright residuals surrounding the brightest saturated stars. These residuals were reflective of the flux calibration and did not indicate that the} \texttt{elderflower} {wide angle PSF was underestimated.} 

Source-subtraction residuals, the aureoles of saturated stars, and any sources not deblended or subtracted properly were masked, along with image artifacts from satellite trails\footnote{The Dragonfly pipeline is designed to exclude bright pixels in individual frames that deviate from a median coadd of all frames, in order to exclude cosmic rays or trails left by aircraft or satellites. The masks for these erroneous bright sources are generated by expanding the SExtractor segmentation image. Upon reducing the DEGS data, it became clear that this approach would benefit from further refinement. Two dramatic examples of the current limitations of this algorithm are NGC 4307 (with prominent residual streaks on either side of a masked asteroid trail) and NGC 4437 (which is littered with many faint satellite trails).} and areas with obvious cirrus. {Although some star masking was required after running} \texttt{mrf}, {this technique still offers an advantage over simply masking all stars and compact sources. Unsaturated compact sources were very well subtracted (see Figure~}\ref{fig:mrf2}), {greatly reducing the amount of the image that must be masked and therefore increasing the area over which low surface brightness emission may be measured.} With the PSF models, source-subtracted images, and masks in place, we were {then} able to extract surface brightness profiles from {our} data {and proceed with our analysis}.

\subsection{Profiles} \label{profiles}

{We now describe our procedure for extracting surface brightness profiles, correcting these profiles for scattered disc light, and calculating stellar mass surface density profiles. The sources of error we considered are noted throughout. We display purely random error separately from combined random + systematic error when plotting azimuthally averaged profiles (Figure}~\ref{fig:SB}, { Figure}~\ref{fig:surfacedensity}, {and in Appendix}~\ref{fig_appendix}{) but we always tabulate and display the combined random + systematic error on our final measurements of stellar halo mass fractions.  These errors are combined in quadrature (occasionally weighted by the number of contributing unmasked pixels); please refer to Appendix}~\ref{errcalcs} {for full error calculations and additional remarks on neglected sources of error. Jupyter notebooks containing our analysis for each galaxy are available at } \url{https://github.com/cgilhuly/papers/tree/master/2021/DEGS_halos}.

 We used the \texttt{isophote} package from Photutils\footnote{\url{https://photutils.readthedocs.io/en/stable/isophote.html}} to {fit elliptical isophotes to our images and} extract azimuthally averaged profiles. Elliptical isophotes were fitted to the {masked} $r-$band source-subtracted image and then imposed on the {masked} $g-$band source-subtracted image. {The position angle and ellipticity of the isophotes were allowed to vary.} The semi-major axis length of the isophotes was set to increase non-linearly in order to counteract decreasing signal-to-noise (S/N) in the outer profiles. Despite this, we found the isophote fitting algorithm was unable to converge towards the galaxy outskirts and a fixed position angle and ellipticity were adopted within {a regime where the light distribution remains highly flattened and therefore more reflective of the disc than the stellar halo}. This isophote geometry did not fairly trace the transition from the edge-on disc to disc outskirts and stellar halo. {We therefore determined the outer ellipticity from isophotal fits to masked, binned coadds of the $g-$ and $r-$band source subtracted images. Four slightly different coadded and binned images were used (either binning the coadd by a factor of four or five, and either applying the mask before or after binning).} The {mean of the four resulting outer ellipticities was adopted as the outer ellipticity and their standard deviation was adopted} as an estimate of the uncertainty. These values are tabulated in Table~\ref{tab:ellip}, along with the radius where the outer ellipticity was fitted. {We note that the estimated uncertainty of the outer ellipticity is not included in any further error calculations and is only tabulated to document our procedure. The original isophotes fitted to the masked $r-$band image were then modified with a smooth decrease in ellipticity over a fixed radial range, ending with the outer ellipticity determined from the coadded and binned isophotal fits. The isophote position angles were left unchanged.}

 There were two galaxies for which we instead selected the best outer ellipticity by visually inspecting the binned images in an extreme stretch. The first of these was NGC 3432. This galaxy's high asymmetry and interacting dwarf companion were not handled well by the \texttt{isophote} package. We assigned an outer ellipticity of 0.45, in contrast to the value measured as described above ($0.72 \pm 0.02$). The second galaxy requiring intervention was NGC 4330. This was a more minor case; heavy masking of two bright stars north of the galaxy midplane likely led to a premature stop in attempts to fit the isophote geometry. In this case we adopted an outer ellipticity of 0.50, a relatively modest adjustment from the previously measured value of $0.63 \pm 0.05$.

 \begin{table}[tb]
\begin{center}
\begin{tabular}{cccc c} 
 \hline \hline
 Target & $\epsilon$ & \multicolumn{2}{c}{$r_\epsilon$} & $r_{bkgd}$ \\ \hline \hline
 NGC 3044 & $0.36 \pm 0.08$ & $269''$ & 27.3 kpc & $250''$ \\
 NGC 3432 & $0.72 \pm 0.02$ & $374''$ & 21.2 kpc & $437.5''$ \\
 NGC 3501 & $0.45 \pm 0.06$ & $173''$ & 19.7 kpc & $250''$ \\
 NGC 3628 & $0.13 \pm 0.05$ & $678''$ & 32.1 kpc & $1250''$ \\
 NGC 4010 & $0.46 \pm 0.02$ & $197''$ & 17.5 kpc & $250''$ \\
 NGC 4013 & $0.25 \pm 0.03$ & $367''$ & 32.1 kpc & $625''$ \\
 NGC 4307 & $0.13 \pm 0.11$ & $196''$ & 20.8 kpc & $312.5''$ \\
 NGC 4330 & $0.63 \pm 0.05$ & $262''$ & 24.8 kpc & $375''$ \\
 NGC 4437 & $0.63 \pm 0.01$ & $482''$ & 22.6 kpc & $625''$ \\
 NGC 4565 & $0.47 \pm 0.02$ & $728''$ & 41.2 kpc & $750''$ \\
 NGC 4634 & $0.36 \pm 0.07$ & $162''$ & 15.8 kpc & $375''$ \\
 NGC 5907 & $0.53 \pm 0.01$ & $587''$ & 40.5 kpc & $750''$ \\
 \hline \hline
\end{tabular}
\end{center}
\caption{Outer ellipticities ($\epsilon$), {the semi-major axis length at which $\epsilon$ is measured ($r_\epsilon$, in angular and physical units), and the inner radius of the background annulus ($r_{bkgd}$)} for each DEGS target.}
\label{tab:ellip}
\end{table}

 \begin{table*}[tb]
\begin{center}
\begin{tabular}{cl} 
 \hline \hline
 Target & Imfit model components  \\ \hline \hline
 NGC 3044 & \texttt{BrokenExponentialDisk3D} \\
 NGC 3432 & \texttt{BrokenExponentialDisk3D} + \texttt{Exponential} + \texttt{Exponential}\footnote{The second \texttt{Exponential} component for NGC 3432 has an offset of $\sim10$~pixels from the previous two components. It was included because of this galaxy's asymmetry.} \\
 NGC 3501 & \texttt{BrokenExponentialDisk3D} \\
 NGC 3628 & \texttt{BrokenExponentialDisk3D} + \texttt{BrokenExponential} + \texttt{Exponential}\footnote{The \texttt{BrokenExponential} component can be thought of as the thick disc of NGC 3628. The \texttt{Exponential} component fits one half of the X-shaped bulge. Attempts were made to add a second \texttt{Exponential} for the other half but either the intensity would tend to zero or the position angle would snap back to that of the disc.}\\
 NGC 4010 & \texttt{BrokenExponentialDisk3D} \\
 NGC 4013 & \texttt{BrokenExponentialDisk3D} + \texttt{Exponential} + \texttt{PointSource}\footnote{The \texttt{PointSource} component of the NGC~4013 model was included for a moderately bright star near the center of the galaxy.} \\
 NGC 4307 & \texttt{BrokenExponentialDisk3D} + \texttt{Exponential} \\
 NGC 4330 & \texttt{BrokenExponentialDisk3D} + \texttt{Exponential} \\
 NGC 4437 & \texttt{BrokenExponentialDisk3D}  \\
 NGC 4565 & \texttt{BrokenExponentialDisk3D} + \texttt{Sersic} \\
 NGC 4634 & \texttt{BrokenExponential} + \texttt{Exponential} \\
 NGC 5907 & \texttt{BrokenExponentialDisk3D} + \texttt{Sersic} \\
 \hline \hline
\end{tabular}
\end{center}
\caption{A summary of the Imfit model components used to fit the bright regions of each galaxy. The $g-$ and $r-$band images for each galaxy were fit separately, but with the same set of model components and with very similar initial conditions. The majority of our sample are well-described by a \texttt{BrokenExponentialDisk3D} model, either alone or with one additional component. This complicated function fits the disc inclination and vertical scaleheight in addition to the position angle, inner/outer disc scalelengths, and the break radius (as well as parameters describing the break softening and the shape of the vertical light distribution). The individual model parameters are omitted for brevity. It is expected that the exact model is not important for our PSF-correcting method as long as the model is fairly reasonable \protect\citep{szomoru10, szomoru12}.}
\label{tab:models}
\end{table*}

Background subtraction must be done very carefully when studying stellar halos. Undersubtracting or oversubtracting by a small amount has a dramatic impact on the resulting profile. To address the possibility that the background may vary across the frame (e.g. due to differing amounts of cirrus or small deviations from flatness), we divided the {masked} source-subtracted images into quadrants along the major and minor axes of the target galaxy and measured the background separately in each quadrant. {Extended tidal structures were also masked during this procedure.} We defined a background annulus with an inner radius slightly larger than the extent of faint emission plausibly associated with the galaxy {upon visual inspection} and an outer radius four times larger than the inner radius. {The inner radius of the background annulus for each galaxy is tabulated in Table}~\ref{tab:ellip}. The annulus had the same inner and outer radii for each quadrant. The annulus was divided into approximately 20 radial bins and the median flux was measured within each bin (and within each quadrant). The local slope of the median flux profile was determined by fitting a line centered at each position with a sliding window seven bins wide. The goal was to identify the flattest local slope within the quadrant and thus identify the most suitable background value. This approach sometimes did not work well if a symmetric local maximum or minimum occurs in the median flux profile. To combat this, we defined a slope ``score'' as the product of the best fit slope and the fit's residuals and, for some galaxies, identified the most suitable background value where this score was lowest. In other cases, the median flux profile rose almost monotonically with radius due to encroaching cirrus and background measurements at progressively larger radii were expected to be less representative of the background at the location of the galaxy of interest. In these cases, the best background was identified at the innermost radial bin where the slope or slope score met an acceptable threshold. 

{We defined the measurement error of the best background value as the standard error of the median within the selected radial bin. Due to the subjective nature of this approach (both in the specific metric chosen and the threshold used, if applicable), a reasonable systematic error of the background measurement should generally exceed the random error. We assumed that at least one of the radial bins provided a reasonable estimate of the local background at the position of the galaxy, and that greater variation among the radial bins correlated with decreased certainty in the identification of the best bin. The standard deviation of the median fluxes in each radial bin was therefore adopted as an estimate of the systematic error of the background measurement. A possible alternative metric could have been [half of] the difference between the highest and lowest background measurements within the annulus quadrant. }

With individual background measurements for each quadrant, surface brightness profiles were re-extracted within each quadrant using the previously fitted isophotes. The result was four independent surface brightness profiles tracing similar subsections of the galaxy, modulo any asymmetries. \texttt{photutils.isophote.ellipse} {provided accompanying radial profiles of flux error, based on the root mean square deviation of flux along each isophote. The random and systematic errors in each quadrant's background measurement were combined in quadrature with the flux error profiles.} For each quadrant profile, the maximum reliable radius was determined based on the relative number of unmasked pixels as a function of radius and where the profile was robust (to within $\sim 0.2$ mag/arcsec$^2$) when adopting a mask with expanded coverage of cirrus and bright stars. Finally, the four quadrant profiles were recombined into one profile {with its corresponding errors}. At each radius, the quadrant {flux and error} profiles were {averaged with relative weights} according to their number of unmasked pixels.

In order to avoid overestimating the contributions of the stellar halo, light scattered from the bright disc must be considered and corrected \citep{dejong08, sandin14, sandin15}. We adopted the same procedure used in \cite{gilhuly20}; briefly, we fitted 2D models and then generated model images with and without PSF convolution to empirically remove the PSF influence on the galaxy's light distribution. This is very similar to the approach developed by \cite{szomoru10, szomoru12} for high redshift compact galaxies. Correcting for PSF effects is very important in this regime as the galaxies of interest are not large compared to the PSF and their effective radii are strongly impacted by PSF effects. Many others have translated this technique to local or lower-redshift galaxies to remove the influence of scattered light from the outskirts of galaxies \citep[to name a few]{talvandokkum11, trujillofliri16, borlaff17, peters17, martinezlombilla19, wang19}; this is another regime in which PSF effects become important. 

We fitted the $r-$ and $g-$band images individually with Imfit \citep{erwin15}. {The PSF produced by} \texttt{elderflower} {was convolved with the image models during fitting.} {We began} with a single exponential model and increas{ed} the model complexity in response to structure in the residual images. {A model was found to be satisfactory when attempts to introduce new components or increase the component complexity led to decreases of $\lesssim10\%$ in the reduced chi-square and the structure of the residuals remained qualitatively unchanged. These criteria were somewhat arbitrary but justifiable as the method we used was insensitive to the exact model parameterization and therefore more complex or more rigorously selected models were unnecessary} \citep[see our summary of the tests of][below]{szomoru10, szomoru12}. Typically only one or two components were used per galaxy, plus a flat sky, but some of these components themselves were fairly complicated (e.g. an inclined broken exponential disc). The model components used for each galaxy are listed in Table~\ref{tab:models}. We generated model images with and without PSF convolution and extracted surface brightness profiles along the same elliptical isophotes as the observed profiles. We subtracted the difference between the model {flux} profiles with and without PSF convolution from the observed {flux} profiles to approximately correct for PSF effects {such that for a given band $\Lambda$,}

\begin{equation}
    f_{\Lambda,\textrm{corr}} = f_{\Lambda,\textrm{obs}} - (f_{\Lambda,\textrm{model\_PSF}} - f_{\Lambda,\textrm{model\_noPSF}}).
\end{equation}

\begin{figure*}[tbp]
\begin{centering}
\includegraphics[width=\textwidth]{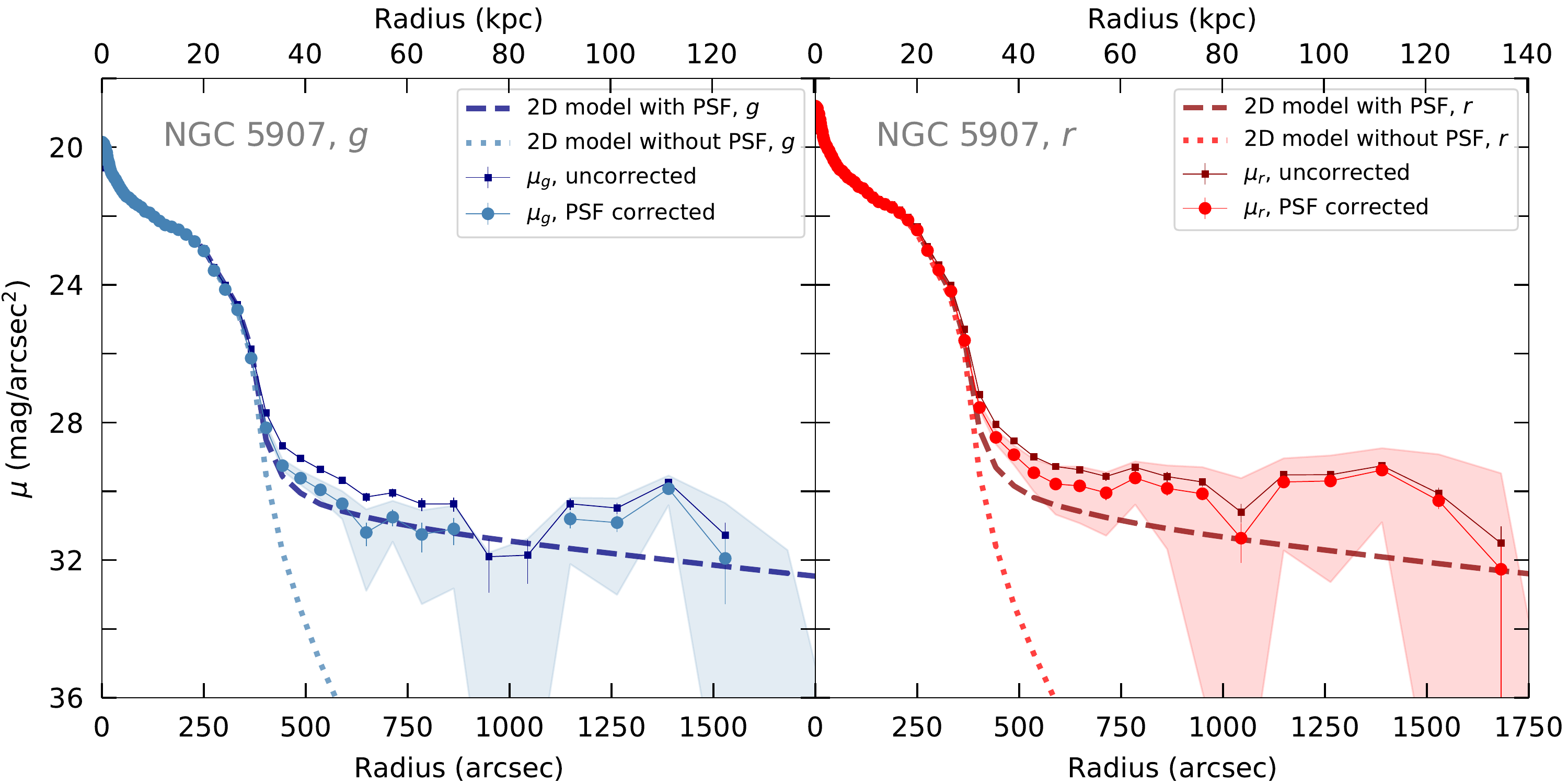}
\caption{{Corrected $g-$ and $r-$band surface brightness profiles for NGC~5907, shown in the left and right panels, respectively, as large round points. For comparison, the uncorrected $g-$ and $r-$band surface brightness profiles are shown with small dark square points. The error bars on individual points depict only random error while the shaded error envelopes include random and systematic errors. The latter are shown only for the corrected profiles. The model profiles used to correct the surface brightness profiles are shown with thick lines matching the colour of the observed profiles. The profiles extracted from model images generated with PSF convolution are shown with dashed lines with a dark colour matching the uncorrected observed profiles, and those extracted from model images generated without PSF convolution are shown with dotted lines with a brighter colour matching the corrected observed profiles. }}
\label{fig:profile_corr}
\end{centering}
\end{figure*}

{Figure}~\ref{fig:profile_corr} {shows the $g-$ and $r-$band surface brightness profiles before and after applying this correction for one of our targets. The model image profiles with and without PSF convolution are included for reference.}

\cite{szomoru10} demonstrated that changing the structural parameters of the 2D models leads to smaller differences in the final corrected profiles, and argued that this technique has a weak dependence on the specific model used. Their tests focused on a $z=1.91$ compact galaxy, which they fitted with a single S\'ersic function. Their suite of models included  an array of models with fixed S\'ersic index $n$ as well as a best-fit model where $n$ was allowed to vary. They found 0.3\% error in the integrated magnitude of their corrected surface brightness profile based on the range of possible values when adopting the various fixed-$n$ models instead of the best-fit model. The effective radius was more sensitive to model choice but still varied by less than 20\% from the best fit results (see their Figure~2). \cite{szomoru12} further tested this technique with mock observations of a variety of two-component synthetic galaxies and showed they were able to recover the original surface brightness profiles very well (see their Figure~3). They again reported that integrated quantities were better recovered from the corrected profiles than sizes. 

Given the high redshift and compact nature of the galaxies in these works, the relative impact of the PSF on observed galaxy properties is expected to be larger than for our galaxies. We also note that we are largely concerned with integrated quantities, which have been demonstrated to be robustly recovered and insensitive to specific model choice. We therefore assume that any error introduced by this method is small compared to random error in the flux measurements in the surface brightness profiles (particularly in the outskirts, where this correction is most relevant) and the systematic uncertainty in the background. 

Our corrected surface brightness profiles are shown in Figure~\ref{fig:SB}. The error bars on individual points reflect random error in the flux measurements extracted using \texttt{photutils.isophote.ellipse} as well as random error in the background measurements. Shaded envelopes are used to display the combination of the aforementioned random errors with the systematic uncertainty in background measurements (due to spatial variation in the background and the uncertainty in the task of identifying the most appropriate or representative background value). This is done to draw attention to the relative importance of the latter source of uncertainty. We maintain the same visual convention for all profiles derived from the surface brightness profiles (i.e. colour and stellar mass surface density profiles, which will be described next). 

All of our surface brightness profiles are ``Z-shaped'' to some degree, with a transition from the inner disc to a more steeply declining region that subsequently turns over to the gradually decreasing outskirts (which we expect to be dominated by the stellar halo). Some of these transitions resemble truncations and anti-truncations. However, this common feature is best understood as the transition from the disc-dominated to the stellar halo-dominated regime, or from high ellipticity discy isophotes to rounder ones. The truncation-like downturns all correspond to decreases in isophote ellipticity. We have confirmed this by comparing our final surface brightness profiles with those extracted from isophotes as fitted (i.e. without decreasing the outer isophote ellipticity). Major axis profiles would be needed to determine whether or not the discs of our target galaxies are truncated.

\begin{figure*}[tbp]
\includegraphics[width=0.9\textwidth]{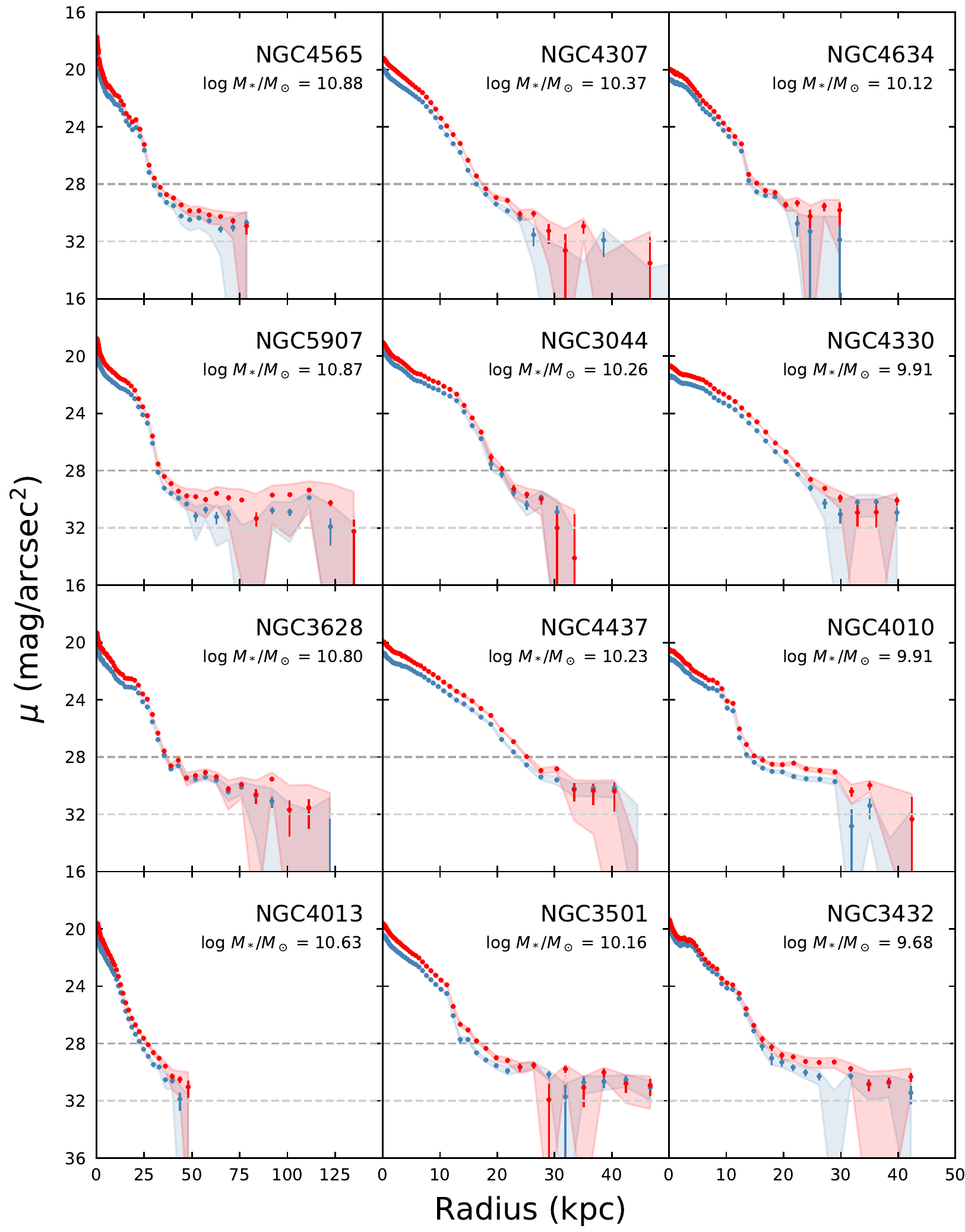}
\caption{Surface brightness profiles for all DEGS galaxies. The galaxies are shown in order of decreasing stellar mass. The red points depict $r-$band profiles and the blue points depict $g-$band profiles. Each column shares a common radial scale. Horizontal dashed lines at surface brightnesses of 28~mag/arcsec$^2$ and 32~mag/arcsec$^2$ are included to highlight the depth of these profiles.}
\label{fig:SB}
\end{figure*}

After correcting the surface brightness profiles as described above, we produced $g-r$ colour profiles and their associated uncertainties. We then calculated stellar mass surface density ($\Sigma_*$) profiles using the $(g-r, g)$ colour-mass-to-light ratio (CMLR) relation from \cite{roediger15}. The random and systematic uncertainties associated with this CMLR are 0.15 dex and 0.3 dex, respectively. These were added in quadrature with {the propagated uncertainty of the $g-$ and $r-$band profiles}. In the outer profile when colour errors and point-to-point variation are high, the colour profile was replaced with the average $g-r$ of the last five reliable points. We defined the uncertainty in the fixed outer colour as the sum in quadrature of the standard deviation of these points and their mean uncertainty. In a few cases where the outer $g-$band profile was missing points or became brighter than the $r-$band profile, $\mu_g$ was replaced with ($g-r$ + $\mu_r$) in the $\Sigma_*$ calculation. This is exactly equivalent to $\mu_g$ except where the observed colour has been replaced with a fixed value. {Our final $\Sigma_*$ profiles are shown in Figure}~\ref{fig:surfacedensity}.

These stellar mass surface density profiles and the preceding surface brightness profiles were not corrected for self-extinction. Dust poses a major challenge to optical photometry of the midplane of edge-on galaxies. Our galaxies are no exception; indeed, several DEGS targets have prominent dust lanes. Self-extinction is not expected to pose a problem in the disc outskirts and stellar halo region, our regions of greatest interest, but one cannot take the inner $\Sigma_*$ profiles at face value. This makes it complicated to determine the total stellar mass of the galaxies from Dragonfly imaging alone. To avoid this issue, we instead adopted the total stellar masses from the Spitzer Survey of Stellar Structure in Galaxies (\SfourG) catalog \citep{sheth10, querejeta15, munozmateos15, s4g}. These stellar masses are based on infrared photometry and therefore trace old stellar populations with greatly reduced dust interference. We assumed an uncertainty of {0.2 dex} for these stellar masses, following \cite{querejeta15}.

\begin{figure*}[tbp]
\includegraphics[width=0.9\textwidth]{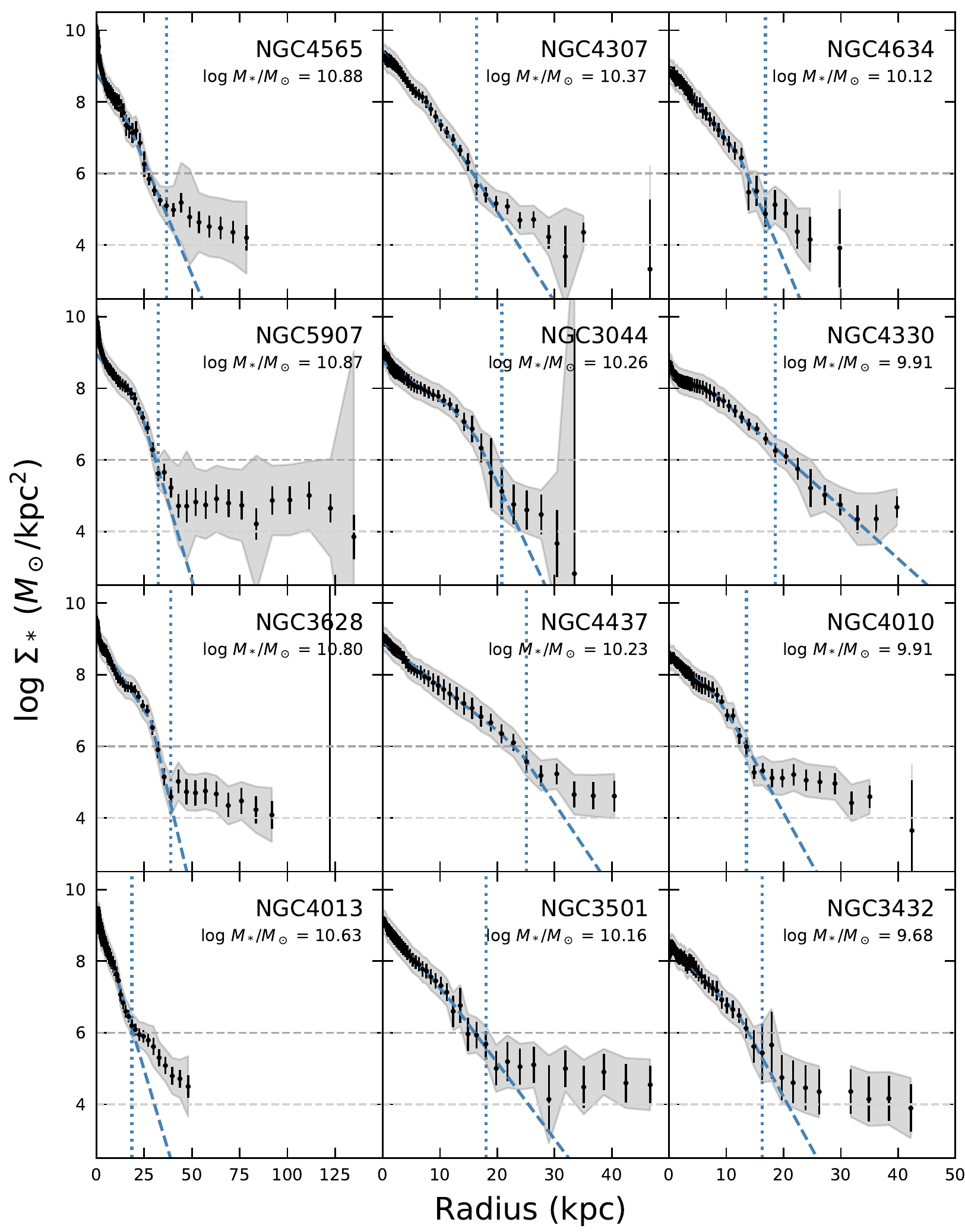}
\caption{Stellar mass surface density profiles for all DEGS galaxies. Broken exponential fits to the inner profile, representing the disc contribution, are shown in dashed blue. The vertical blue dotted line marks the outer edge of the disc fitting region. The galaxies are shown in order of decreasing stellar mass. Horizontal dashed lines at stellar mass surface densities of $10^6\:  \mathrm{M}_\odot/\mathrm{kpc}^2$ and $10^4\:  \mathrm{M}_\odot/\mathrm{kpc}^2$ are included for reference.}
\label{fig:surfacedensity}
\end{figure*}

\subsection{Separating discs from stellar halos} \label{halos}

{Separating the in-situ and accreted stellar populations in observed galaxies poses a great challenge. When only broadband integrated light observations are available, the total accreted stellar mass cannot be recovered directly. Simulations show that most accreted star particles are centrally concentrated and coincide with the bulge and disc \citep{pillepich15}. From an observer's standpoint, this mass is obscured by or highly confused with the in-situ stellar component. Rather than attempting to recover the total accreted stellar mass, our goal in this work is to recover the accreted stellar mass in the target galaxies' outskirts where the in-situ population diminishes and the accreted population becomes dominant. In this section, we describe how we defined the stellar halo-dominated region and how we estimated (and removed) the estimated contribution from the disc in this region.}

{For edge-on disc galaxies, the in-situ population drops off most quickly along the minor axis \citep[e.g.][]{monachesi19}. Considering only the starlight along the minor axis of the galaxy should therefore enable a relatively clean separation of the in-situ disc and the accreted stellar halo. However, we also wish to leverage our wide field broadband images to capture the entire stellar halo-dominated region. We considered the possibility that our use of azimuthally averaged profiles weakened our ability to separate the disc and halo components in a galaxy's outskirts. We will show in Section~\ref{minorprofs} that the resulting stellar halo masses and mass fractions are comparable with values measured along the minor axis only.}

There are many possible empirical definitions of the stellar halo-dominated region and none can definitively separate in-situ-dominated regions from accreted or ex-situ-dominated regions for all galaxies.  After studying the stellar mass surface density profiles of 17 Milky Way-mass galaxies from the FIRE-2 suite of simulations, \cite{sanderson18} found no threshold or inflection point that consistently indicated a transition to an accretion-dominated regime. This was directly attributed to the natural variety among accretion histories and therefore the amount and distribution of accreted stars. The half-mass radius of accreted stellar mass varied from $5-50$ kpc within their sample (see their Figure~2), and the accreted component began to dominate over the in-situ component at distances of $10-50$ kpc from the galaxies' centers (see their Figure~3). If there is no simple rule to cleanly separate the stellar halo-dominated region for disc-dominated galaxies of similar mass, this separation can only become less clear when considering a larger mass range.

With the above caveats in mind, we saw no a priori justification to adopt one definition of the stellar halo-dominated region over another. Instead, we explored four tractable definitions for the approximate onset of the stellar halo-dominated region: 5 times the disc scalelength, $R_d$ \citep{sanderson18}, a fixed distance of 20 kpc \citep{merritt20}, the radius where the stellar mass surface density reaches $10^6 \:  \mathrm{M}_\odot/\textrm{kpc}^{2}$ \citep{cooper13}, and 5 times the half-mass radius, $R_{1/2}$ \citep{pillepich14, merritt16}. We adopted the inner scalelength of a broken exponential model fitted to the stellar mass surface density profile as the characteristic disc scalelength. The fitting procedure is described below. The half-mass radii were determined from our stellar mass surface density profiles. Since we did not correct these profiles for self-extinction, our half-mass radii may be overestimated. This would result in underestimated stellar halo masses and mass fractions for this particular definition of the stellar halo-dominated region. 

Others have favoured 30 kpc over 20 kpc as a fixed distance cutoff to approximately separate galaxies from their stellar halos. We view this as an arbitrary choice due to the variation among galaxies. \cite{tissera13} found the transition from the inner halo (formed in place or populated by heated disc stars) to the outer halo (where accreted material dominates) occurred around $15-30$ kpc. \cite{sanderson18} found a very wide range of $10-50$ kpc for the in-situ to accretion-dominated transition radius. Observationally, \cite{dsouza14} found transition radii between inner and outer fitted components over an even wider range, varying with galaxy stellar mass and concentration. As the choice is ultimately arbitrary, we preferred the smaller distance because we subtracted the extrapolated disc contribution instead of simply attributing all stellar mass beyond a given radius to the stellar halo. 

For each galaxy and each definition of the stellar halo-dominated region, we integrated the observed stellar mass surface density profile starting at the inner radial limit. In order to account for contributions from the disc extending into these regions, we fitted either a single exponential or broken exponential model to the stellar mass surface density profile between 1 kpc and the visual extent of the disc. We excluded the inner 1 kpc to avoid introducing a bulge model. The majority of our galaxies do not possess large bulges (see Figure~\ref{fig:colour_cutouts}); furthermore, this region of the profile is the least reliable due to self-extinction. For NGC 4565 and NGC 5907, the two most massive galaxies in the DEGS sample, the inner 5 kpc was excluded instead. 

The disc model was extrapolated and integrated over the same radial range as the observed stellar mass surface density profile to obtain the estimated mass contributed by the disc in the supposed stellar halo-dominated region, for each definition of this region. We subtracted the estimated disc mass contribution from the previous integral, leaving the total mass found in the stellar halo-dominated region in excess of the expected disc mass. We designated all of this excess mass as the stellar halo. 

To estimate the uncertainty in the estimated disc mass contribution, we used the \texttt{emcee} package \citep{emcee} to explore the posterior probability density function of the fitted model parameters. We initialized 32 ``walkers'' at random states slightly offset from the maximum likelihood fit and then generated large chains of 30,000 steps per walker. A burn-in period of 1000 steps was discarded before thinning each chain by approximately the mean autocorrelation time in order to obtain a collection of independent samples (where each sample was a possible set of model parameters). For each definition of the stellar halo-dominated region, we calculated the extrapolated disc contribution for all samples and took the 16th and 84th percentiles of the resulting distribution of masses as the $1\sigma$ limits of the disc mass contribution.

\section{Results}\label{results}

\begin{sidewaystable*}[p]
\begin{center}
\vspace{2in}
\small
\begin{tabular}{ccccccccccccc} 
 \hline \hline
 \multirow{2}{*}{Target} & \multicolumn{3}{c}{$r>5R_d$} & \multicolumn{3}{c}{$r>20\mathrm{ kpc}$} & \multicolumn{3}{c}{$\Sigma_* < 10^6\:  \mathrm{M}_\odot/\mathrm{kpc}^2$} & \multicolumn{3}{c}{$r>5R_{1/2}$} \\
  & $\log f_{outer}$ & $\log f_{halo}$ & $\log \mathrm{M}_{*,halo}$ & $\log f_{outer}$ & $\log f_{halo}$ & $\log \mathrm{M}_{*,halo}$ & $\log f_{outer}$ & $\log f_{halo}$ & $\log \mathrm{M}_{*,halo}$ & $\log f_{outer}$ & $\log f_{halo}$ & $\log \mathrm{M}_{*,halo}$\\ \hline \hline
NGC 3044 & $-2.4\pm0.4$ & $-2.7^{+0.4}_{-1.7}$ & $7.6^{+0.4}_{-1.7}$ & $-2.4\pm0.4$ & $-2.7^{+0.4}_{-1.7}$ & $7.6^{+0.4}_{-1.7}$ & $-2.2\pm0.5$ & $-2.9^{+0.6}_{-1.5}$ & $7.3^{+0.5}_{-1.5}$ & $-3.9\pm2.1$ & $-4.0^{+2.1}_{-3.4}$ & $6.3^{+2.1}_{-3.4}$ \\
NGC 3432 & $-1.5\pm0.4$ & $-1.7^{+0.5}_{-1.0}$ & $8.0^{+0.4}_{-1.0}$ & $-2.1\pm0.3$ & $-2.1^{+0.4}_{-1.7}$ & $7.6^{+0.3}_{-1.7}$ & $-1.5\pm0.4$ & $-1.7^{+0.5}_{-1.0}$ & $8.0^{+0.4}_{-1.0}$ & $-2.3\pm0.4$ & $-2.3^{+0.5}_{-1.9}$ & $7.4^{+0.4}_{-1.9}$ \\
NGC 3501 & $-1.2\pm0.4$ & $-1.7^{+0.4}_{-0.5}$ & $8.4^{+0.3}_{-0.5}$ & $-1.9\pm0.3$ & $-1.9^{+0.4}_{-1.0}$ & $8.3^{+0.4}_{-1.0}$ & $-1.5\pm0.3$ & $-2.1^{+0.3}_{-0.5}$ & $8.0^{+0.2}_{-0.5}$ & $-1.9\pm0.3$ & $-1.9^{+0.4}_{-1.0}$ & $8.3^{+0.4}_{-1.0}$ \\
NGC 3628 & $-1.5\pm0.3$ & $-2.0^{+0.3}_{-0.5}$ & $8.8^{+0.2}_{-0.4}$ & $-1.0\pm0.2$ & $-1.6^{+0.3}_{-0.3}$ & $9.2^{+0.2}_{-0.2}$ & $-1.7\pm0.3$ & $-2.0^{+0.3}_{-0.7}$ & $8.8^{+0.2}_{-0.6}$ & $-1.9\pm0.3$ & $-2.0^{+0.3}_{-1.1}$ & $8.8^{+0.2}_{-1.1}$ \\
NGC 4010 & $-1.7\pm0.3$ & $-1.8^{+0.3}_{-1.1}$ & $8.1^{+0.2}_{-1.1}$ & $-1.9\pm0.3$ & $-1.9^{+0.4}_{-1.6}$ & $8.0^{+0.3}_{-1.5}$ & $-1.5\pm0.3$ & $-1.9^{+0.3}_{-0.8}$ & $8.0^{+0.2}_{-0.8}$ & $-2.0\pm0.3$ & $-2.0^{+0.4}_{-1.7}$ & $7.9^{+0.4}_{-1.6}$ \\
NGC 4013 & $-1.2\pm0.2$ & $-1.6^{+0.3}_{-0.4}$ & $9.0^{+0.2}_{-0.3}$ & $-1.5\pm0.3$ & $-1.6^{+0.3}_{-0.6}$ & $9.0^{+0.3}_{-0.5}$ & $-1.6\pm0.3$ & $-1.6^{+0.4}_{-0.7}$ & $9.0^{+0.3}_{-0.6}$ & $-1.4\pm0.2$ & $-1.6^{+0.3}_{-0.5}$ & $9.1^{+0.2}_{-0.5}$ \\
NGC 4307 & $-1.3\pm0.3$ & $-2.1^{+0.3}_{-0.4}$ & $8.3^{+0.2}_{-0.4}$ & $-2.5\pm0.3$ & $-2.5^{+0.4}_{-0.9}$ & $7.9^{+0.4}_{-0.9}$ & $-1.9\pm0.3$ & $-3.0^{+0.3}_{-0.7}$ & $7.4^{+0.3}_{-0.6}$ & $-2.3\pm0.3$ & $-2.5^{+0.4}_{-0.8}$ & $7.9^{+0.3}_{-0.8}$ \\
NGC 4330 & $-2.4\pm0.4$ & $-2.5^{+0.6}_{-1.0}$ & $7.4^{+0.5}_{-1.0}$ & $-1.5\pm0.3$ & $-2.6^{+0.5}_{-0.8}$ & $7.3^{+0.4}_{-0.8}$ & $-1.6\pm0.3$ & $-2.6^{+0.5}_{-0.8}$ & $7.3^{+0.4}_{-0.8}$ & $-2.6\pm0.6$ & $-2.6^{+0.7}_{-1.1}$ & $7.3^{+0.7}_{-1.1}$ \\
NGC 4437 & $-1.3\pm0.3$ & $-2.2^{+0.3}_{-0.5}$ & $8.1^{+0.2}_{-0.4}$ & $-1.5\pm0.3$ & $-2.3^{+0.3}_{-0.5}$ & $7.9^{+0.2}_{-0.5}$ & $-2.0\pm0.3$ & $-2.3^{+0.3}_{-1.0}$ & $7.9^{+0.2}_{-0.9}$ & $-2.7\pm0.4$ & $-2.7^{+0.4}_{-1.9}$ & $7.5^{+0.4}_{-1.9}$ \\
NGC 4565 & $-2.2\pm0.4$ & $-2.3^{+0.4}_{-0.5}$ & $8.5^{+0.3}_{-0.4}$ & $-1.6\pm0.3$ & $-2.4^{+0.3}_{-0.3}$ & $8.5^{+0.2}_{-0.3}$ & $-1.9\pm0.3$ & $-2.6^{+0.3}_{-0.4}$ & $8.3^{+0.2}_{-0.3}$ & $-1.4\pm0.3$ & $-2.5^{+0.3}_{-0.3}$ & $8.4^{+0.2}_{-0.2}$ \\
NGC 4634 & $-1.7\pm0.3$ & $-2.4^{+0.4}_{-1.3}$ & $7.7^{+0.3}_{-1.2}$ & $-2.8\pm0.4$ & $-2.8^{+0.4}_{-2.2}$ & $7.3^{+0.4}_{-2.2}$ & $-2.0\pm0.3$ & $-2.4^{+0.4}_{-1.3}$ & $7.7^{+0.3}_{-1.2}$ & $-2.6\pm0.4$ & $-2.6^{+0.4}_{-1.9}$ & $7.5^{+0.3}_{-1.9}$ \\
NGC 5907 & $-1.7\pm0.4$ & $-1.7^{+0.5}_{-0.9}$ & $9.1^{+0.5}_{-0.9}$ & $-1.1\pm0.3$ & $-1.7^{+0.3}_{-0.3}$ & $9.1^{+0.2}_{-0.2}$ & $-1.6\pm0.4$ & $-1.7^{+0.4}_{-0.7}$ & $9.2^{+0.3}_{-0.6}$ & $-1.5\pm0.3$ & $-1.7^{+0.3}_{-0.5}$ & $9.2^{+0.3}_{-0.5}$ \\
\hline 
NGC1042 & $-1.5\pm0.2$ & -- & -- & $-2.5\pm0.2$ & -- & -- & $-1.8\pm0.2$ & -- & -- & $-4.4\pm0.6$ & -- & -- \\
NGC1084 & $-1.0\pm0.2$ & $-1.3^{+0.2}_{-0.2}$ & $9.5^{+0.1}_{-0.1}$ & $-1.5\pm0.2$ & $-1.5^{+0.2}_{-0.2}$ & $9.2^{+0.1}_{-0.1}$ & $-1.7\pm0.2$ & $-1.7^{+0.2}_{-0.2}$ & $9.1^{+0.1}_{-0.1}$ & $-1.4\pm0.2$ & $-1.4^{+0.2}_{-0.2}$ & $9.3^{+0.1}_{-0.1}$ \\
NGC2903 & $-1.3\pm0.2$ & $-2.1^{+0.2}_{-0.3}$ & $8.6^{+0.1}_{-0.2}$& $-1.8\pm0.2$ & $-2.1^{+0.2}_{-0.5}$ & $8.6^{+0.1}_{-0.5}$ & $-2.1\pm0.2$ & $-2.2^{+0.2}_{-0.7}$ & $8.4^{+0.1}_{-0.7}$ & $-1.7\pm0.2$ & $-2.0^{+0.2}_{-0.5}$ & $8.6^{+0.1}_{-0.4}$ \\
NGC3351 & $-1.4\pm0.2$ & -- & -- & $-2.9\pm0.2$ & -- & -- & $-2.1\pm0.2$ & -- & -- & $-1.1\pm0.2$ & -- & -- \\
NGC3368 & $-1.3\pm0.2$ & $-2.3^{+0.2}_{-0.2}$ & $8.5^{+0.1}_{-0.1}$ & $-1.7\pm0.2$ & $-2.2^{+0.2}_{-0.2}$ & $8.5^{+0.1}_{-0.1}$ & $-2.0\pm0.2$ & $-2.2^{+0.2}_{-0.3}$ & $8.5^{+0.1}_{-0.2}$ & $-1.0\pm0.2$ & $-2.2^{+0.2}_{-0.2}$ & $8.6^{+0.1}_{-0.1}$\\
NGC4220 & $-0.9\pm0.2$ & $-1.4^{+0.2}_{-0.2}$ & $9.2^{+0.1}_{-0.1}$ & $-1.5\pm0.2$ & $-1.6^{+0.2}_{-0.2}$ & $9.0^{+0.1}_{-0.1}$ & $-1.6\pm0.2$ & $-1.6^{+0.2}_{-0.2}$ & $9.0^{+0.1}_{-0.1}$ & $-1.1\pm0.2$ & $-1.3^{+0.2}_{-0.2}$ & $9.2^{+0.1}_{-0.1}$ \\
NGC4258 & $-1.0\pm0.2$ & $-1.8^{+0.2}_{-0.2}$ & $9.1^{+0.1}_{-0.1}$ & $-1.0\pm0.2$ & $-1.8^{+0.2}_{-0.2}$ & $9.1^{+0.1}_{-0.1}$ & $-1.7\pm0.2$ & $-1.9^{+0.2}_{-0.5}$ & $9.1^{+0.1}_{-0.5}$ & $-1.0\pm0.2$ & $-1.8^{+0.2}_{-0.3}$ & $9.1^{+0.1}_{-0.2}$ \\
M101 & $-1.2\pm0.2$ & $-2.0^{+0.2}_{-0.2}$ & $8.7^{+0.1}_{-0.1}$ & $-1.1\pm0.2$ & $-1.9^{+0.2}_{-0.2}$ & $8.8^{+0.1}_{-0.1}$ & $-1.9\pm0.2$ & -- & -- & $-1.1\pm0.2$ & $-1.9^{+0.2}_{-0.2}$ & $8.7^{+0.1}_{-0.1}$ \\
 \hline \hline
\end{tabular}
\end{center}
\caption{Stellar halo mass fractions for DEGS and DNGS targets for four different definitions of the stellar halo-dominated region. Missing lower error limits indicate a lower bound on $f_{halo}$ consistent with zero (undefined in log space). Missing entries for $f_{halo}$ indicate stellar halo non-detections. We note that the measurements for M101 are erroneously high; see Section~\ref{comparison} for an explanation and an alternative measurement of $f_{halo}$. }
\label{tab:results}
\end{sidewaystable*}

The stellar halo mass fractions ($f_{halo}$) are presented in Table~\ref{tab:results}. The upper and lower error bounds include all previously discussed sources of uncertainty: random measurement error in the surface brightness profiles, random and systematic uncertainty in the background measurement, random and systematic uncertainty in the CMLR relation used from \cite{roediger15}, the uncertainty in the \SfourG{} stellar masses \citep{querejeta15, s4g}, and the empirically determined random uncertainty in the stellar mass contributed by the disc in the proposed stellar halo-dominated regions.

To facilitate comparison with past and future work on simulated (or observed) galaxies which share one or more of our stellar halo-dominated region definitions but which attributes \emph{all} stellar mass within the region to the stellar halo, we also tabulate outer stellar mass fractions ($f_{outer}$) where no attempt has been made to model and subtract a disc component from the outskirts. The uncertainties on these values therefore include all the sources listed above except for the one related to the estimated disc contribution. 

In addition to the DEGS targets, we apply our procedure to stellar mass surface density profiles for Dragonfly Nearby Galaxies Survey (DNGS) targets previously studied by \cite{merritt16}. They adopted a slightly different approach to modeling the profiles (among other minor differences), fitting a S\'ersic bulge and exponential disc to the extent of the spiral arms and extrapolating a power law fit to the stellar halo to a uniform limiting stellar mass surface density of $10^4\:  \mathrm{M}_\odot/\textrm{kpc}^2$. For our common definition of the stellar halo-dominated region ($r>5R_{1/2}$), there is generally good agreement between the stellar halo mass fractions of \cite{merritt16} and the re-measured values in this work (see Table~\ref{tab:DNGS_fracs}).  This demonstrates that our decision to not model the inner profile nor extrapolate the stellar mass surface density profiles has had a minimal impact for these galaxies. Our repeated analysis of DNGS sample supports the conclusion that NGC 1042 and NGC 3551 are consistent with no stellar halo.  We will discuss the discrepancy for M101 in Section~\ref{discussion}; this is explained by some disc mass being erroneously attributed to the stellar halo, and a stellar halo-less M101 is likely consistent with our analysis.

\begin{table}[tbh]
\begin{center}
\begin{tabular}{ccc} 
 \hline \hline
 \multirow{2}{*}{Target} & log $f_{halo} $ & log $f_{halo} $\\
  & \citep{merritt16} & (this work)\\ \hline \hline
 NGC 1042 & ($-4.0^{+0.3}_{--}$) & (-4.4) \\ 
 NGC 1084 & $-1.3^{+0.2}_{-0.3}$ & $-1.4^{+0.2}_{-0.2}$ \\ 
 NGC 2903 & $-2.0^{+0.2}_{-0.5}$ & $-2.0^{+0.2}_{-0.5}$ \\ 
 NGC 3351 & ($-3.7^{+2.1}_{--}$) & (-1.1) \\ 
 NGC 3368 & $-2.4^{+1.0}_{--}$ & $-2.2^{+0.2}_{-0.2}$ \\ 
 NGC 4220 & $-1.9^{+0.3}_{-0.7}$ & $-1.3^{+0.2}_{-0.2}$ \\ 
 NGC 4258 & $-2.5^{+0.8}_{--}$ & $-1.8^{+0.2}_{-0.3}$ \\ 
 M101 & ($-3.4^{+0.5}_{--}$) & $-1.9^{+0.2}_{-0.2}$ \\ 
 \hline \hline
\end{tabular}
\end{center}
\caption{Stellar halo mass fractions for DNGS targets, where the stellar halo mass is measured beyond 5 half-mass radii. Numbers in brackets are upper limits; in this work, we adopt $f_{outer}$ as a generous upper limit in the case of non-detections.}
\label{tab:DNGS_fracs}
\end{table}

\subsection{Relation between $f_{halo}$ and total stellar mass}

\begin{figure*}[tb]
\includegraphics[width=0.9\textwidth]{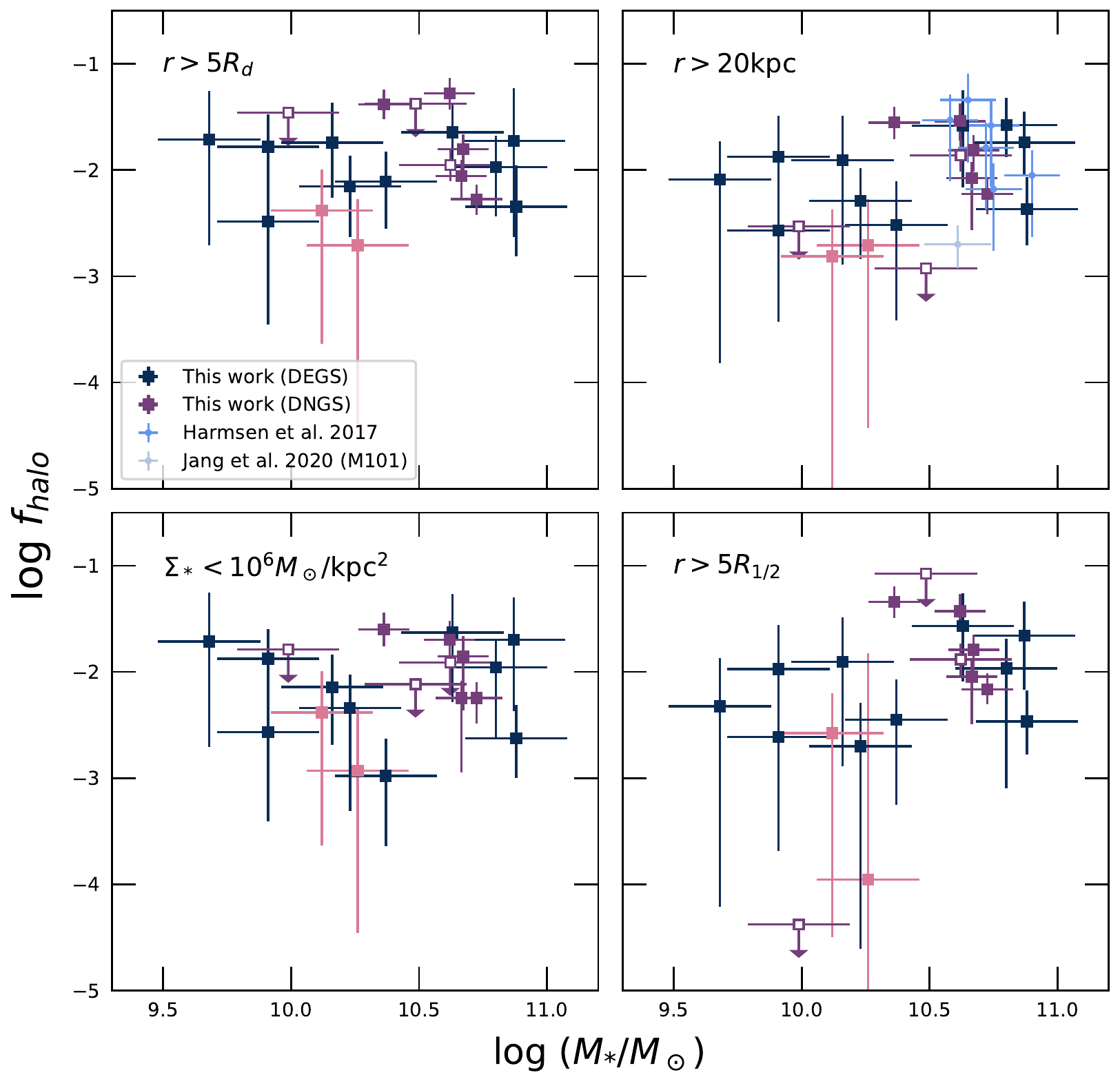}
\caption{Stellar halo mass fraction versus total stellar mass for four different definitions of the stellar halo-dominated region. Square symbols indicate galaxies from this work. Open squares indicate galaxies consistent with no stellar halo; for these galaxies, the outer stellar mass fraction is plotted as an upper limit. The pink squares indicate two DEGS fields that we consider less reliable due to cirrus contamination and other complicating factors (NGC 3044 and NGC 4634). In the top right panel, results from \protect\cite{harmsen17} and \protect\cite{jang20} are included. They are shown in the panel most closely matching their definition of the stellar halo-dominated region.}
\label{fig:halofrac}
\end{figure*}

Figure~\ref{fig:halofrac} shows the relation between stellar halo mass fraction ($f_{halo}$) and host galaxy stellar mass for all four definitions of the stellar halo-dominated region that we consider in this work. Included alongside our measurements are a recent re-measurement of M101's stellar halo mass fraction combining {HST} and Dragonfly data \citep{jang20} and the galaxies from GHOSTS \citep{harmsen17}. The GHOSTS stellar halo masses measured between $10-40$ kpc (along the minor axis) are most closely analogous to ours, as we do not attempt to account for accreted stellar mass outside the proposed stellar halo-dominated regions. In their analysis, Harmsen~et~al. empirically extrapolate their stellar halo mass measurements by applying their analysis technique to simulated stellar halos from  \cite{bullockjohnston05}. They find that their observations and analysis procedure recover $32\pm10$\% of the total mass of the simulated stellar halos, and accordingly increase their stellar halo masses measured within $10-40$ kpc by a factor of $\frac{1}{0.32} \sim 3.1$ (equivalent to an increase of 0.49 dex) to obtain estimates of the total accreted stellar mass. For simplicity, we compare our measurements with their uncorrected stellar halo masses rather than attempting to correct our measurements in a similar way. 

In general, the four panels of Figure~\ref{fig:halofrac} all tell a similar story: the galaxies in our sample vary in the relative proportion of mass found in their stellar halos, spanning a range of values similar to the stellar halos of Milky Way-mass galaxies. This diversity in stellar halo mass fraction appears to be stronger than any correlation with the host galaxies' stellar masses. Visually, there is a weak positive correlation between stellar halo mass fraction and total stellar mass. To formalize this statement, we calculate the Pearson correlation coefficient $r$ and its $p$-value for each definition of the stellar halo-dominated region, including all unique galaxies represented in Figure~\ref{fig:halofrac}\footnote{As the GHOSTS stellar halo mass fractions are determined for one stellar halo definition, it may be ill-advised to include these measurements alongside our measurements for all four definitions of the stellar halo-dominated region instead of only the closest matching definition. However, this would then make it difficult to fairly compare the relative correlation strength and significance, as one data set would contain an additional 5 data points grouped at high masses. To confirm that the inclusion of the GHOSTS galaxies does not introduce any biases or qualitatively change the result of this test, we also calculate $r$ and its $p$-value for DEGS and DNGS galaxies alone. We confirm that the relative strength and significance of the correlations remains unchanged, though all are offset to weaker strengths and significances.}. We exclude the measurement of the stellar halo of M101 produced by \cite{jang20} and the measurement of the stellar halo of NGC 4565 produced by \cite{harmsen17} because these galaxies are already represented in the DNGS and DEGS samples, respectively. For simplicity, we do not include upper limits for any non-detections. This leaves 25 unique galaxies, with 2-3 of these excluded as non-detections depending on the stellar halo definition in question. 

The strongest and most significant correlation with total stellar mass is found for $f_{halo}$ {measured beyond 20 kpc ($r=0.42$,  $p=0.05$). The stellar halo masses measured beyond 5 half-mass radii yield a similarly strong correlation ($r=0.40, p=0.06$).} The other two sets of stellar halo mass fractions have weaker correlation strengths {($r\sim0.2$) and much lower significance ($p\sim 0.4$)}. We also calculate the Spearman rank correlation coefficient, which assesses the degree to which the data are monotonic but otherwise requires no particular relationship. The correlations between $f_{halo}$ and total stellar mass for the various definitions of the stellar halo-dominated region follow the same order in terms of correlation strength and significance.

\begin{figure}[tb]
\includegraphics[width=0.45\textwidth]{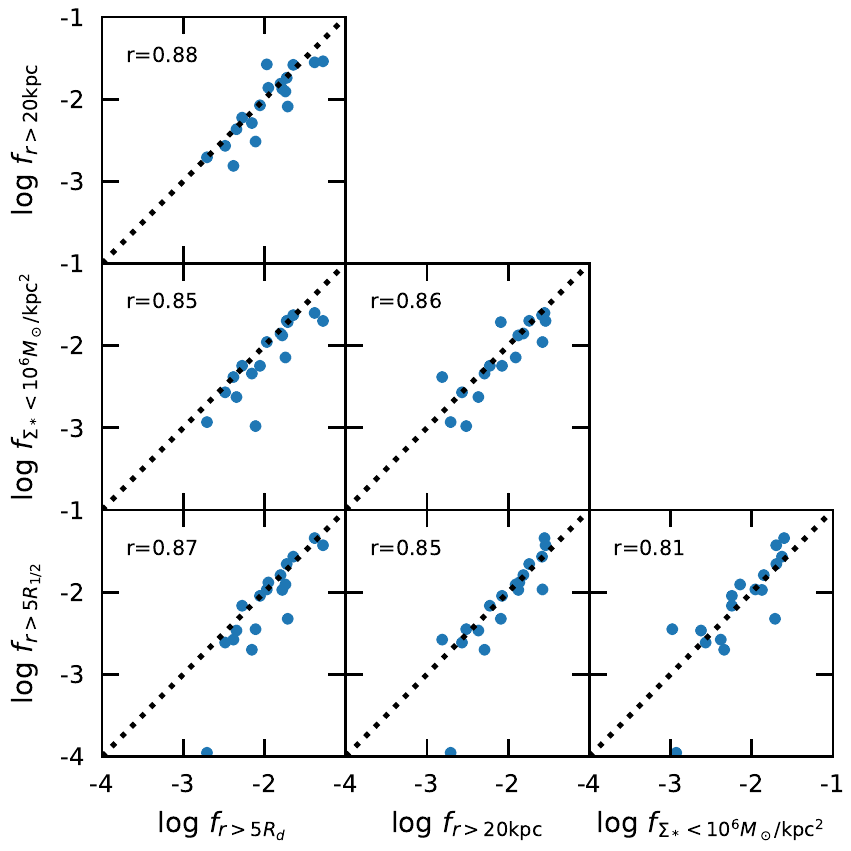}
\caption{Comparison of all combinations of stellar halo mass fractions for DEGS and DNGS targets. Pearson $r$ correlation coefficients are shown in the upper left corner of each panel. The black dashed lines indicate a 1:1 relationship.}
\label{fig:cornerplot}
\end{figure}

Although some definitions of the stellar halo-dominated region yield stellar halo mass fractions that correlate more strongly and significantly with total stellar mass, the difference alone is not necessarily enough to justify recommending one definition over the others for future observational work. Stronger correlations may reflect the influence of a relationship between the radius defining the approximate galaxy-halo boundary and total stellar mass rather than improved galaxy-halo separation. For instance, more massive galaxies tend to have larger central mass concentrations and therefore smaller half-mass radii, thereby decreasing the radius of the galaxy-halo boundary ($r > 5 R_{1/2}$) and increasing the amount of mass attributed to the stellar halo. 

Figure~\ref{fig:cornerplot} shows that the resulting stellar halo mass fractions are strongly correlated with one another, making them largely interchangeable. The definitions of the stellar halo-dominated region that we consider are somewhat arbitrary rules of thumb, and individual galaxies do not necessarily adhere to any given definition \citep[Section 5]{sanderson18}. 

\subsection{{Minor axis profiles}} \label{minorprofs}

\begin{figure*}[tbp]
\includegraphics[width=0.9\textwidth]{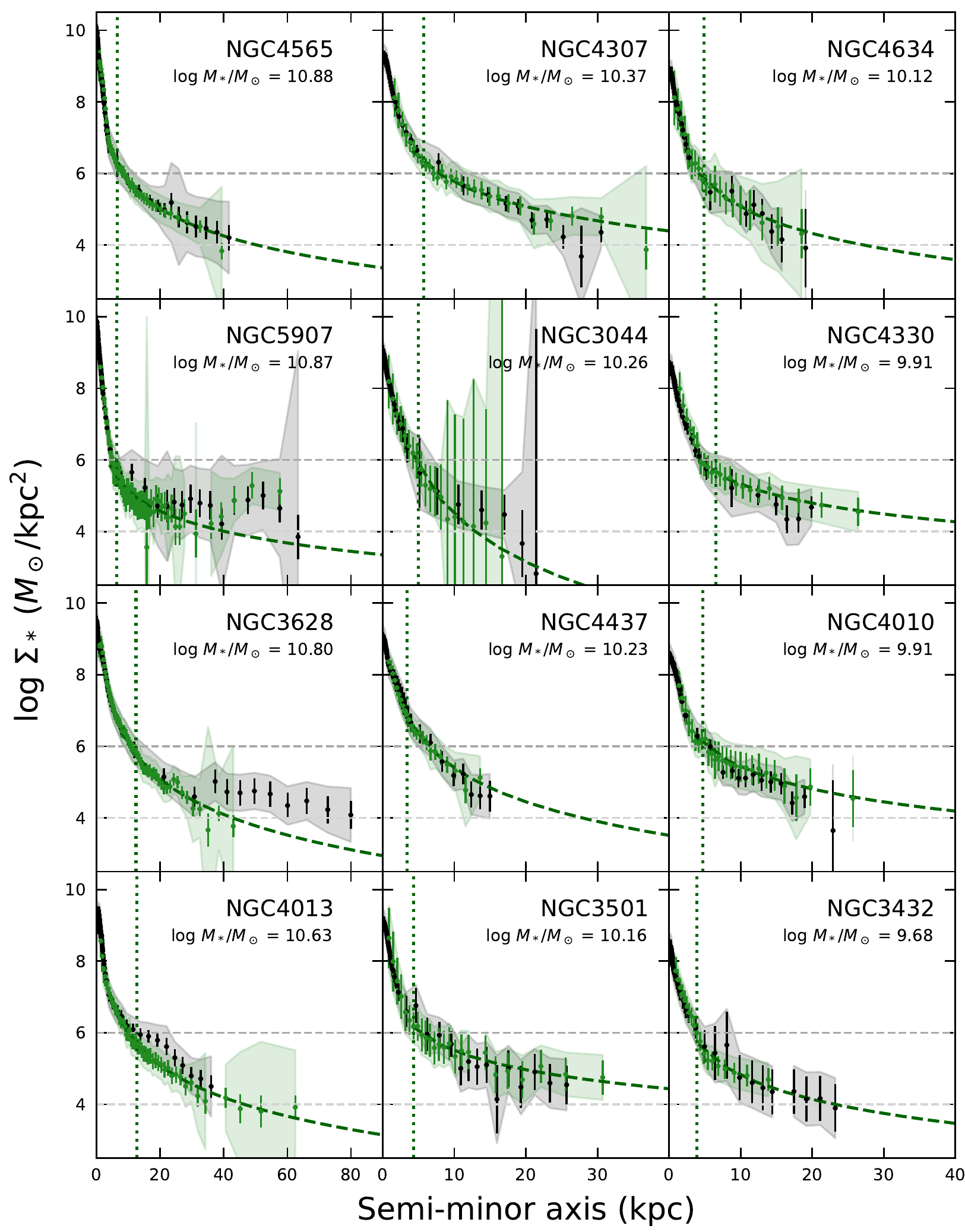}
\caption{{Comparison of the minor axis stellar mass surface density profiles (green) with those obtained from azimuthally averaged profiles (black, originally shown in Figure~\ref{fig:surfacedensity} for all DEGS galaxies). Both sets of profiles are plotted in terms of distance along the minor axis. The power law models fitted to the minor axis profiles are shown in dashed green. The vertical dotted green line indicates the inner limit of the fitting region. The galaxies are shown in order of decreasing stellar mass.}}
\label{fig:minor_surfacedensity}
\end{figure*}

{We repeated our stellar halo analyses using minor axis profiles in order to check that our use of azimuthally averaged profiles did not lead to greater disc contamination in spite of our efforts to model and subtract the disc contribution. We extracted surface brightness profiles in 45$^\circ$ wedges centered on the minor axes of the target galaxies using annular bins of increasing width rather than applying the previously fitted isophotes. The procedure to obtain the minor axis mass density profiles was otherwise analogous to that laid out in Section~\ref{profiles}, except that two minor axis profiles were combined rather than four quadrant profiles.}

{The resulting minor axis stellar mass density profiles are shown in Figure~\ref{fig:minor_surfacedensity} alongside those derived from the azimuthally averaged profiles. There is generally excellent agreement, and the azimuthally averaged profiles are not brighter, more extended, or flatter than the minor axis profiles. Any exceptions are readily explained by stellar halo substructure that lies outside of the minor axis wedges, such as the tidal tail of NGC~3628 and the looping stellar stellar stream surrounding NGC~4013. This favourable comparison demonstrates the validity of our earlier assumption that the outer regions of the azimuthally averaged profiles are dominated by the stellar halo and there is no significant disc contamination along the major axis at large radii. }

\begin{figure}[tb]
\includegraphics[width=0.45\textwidth]{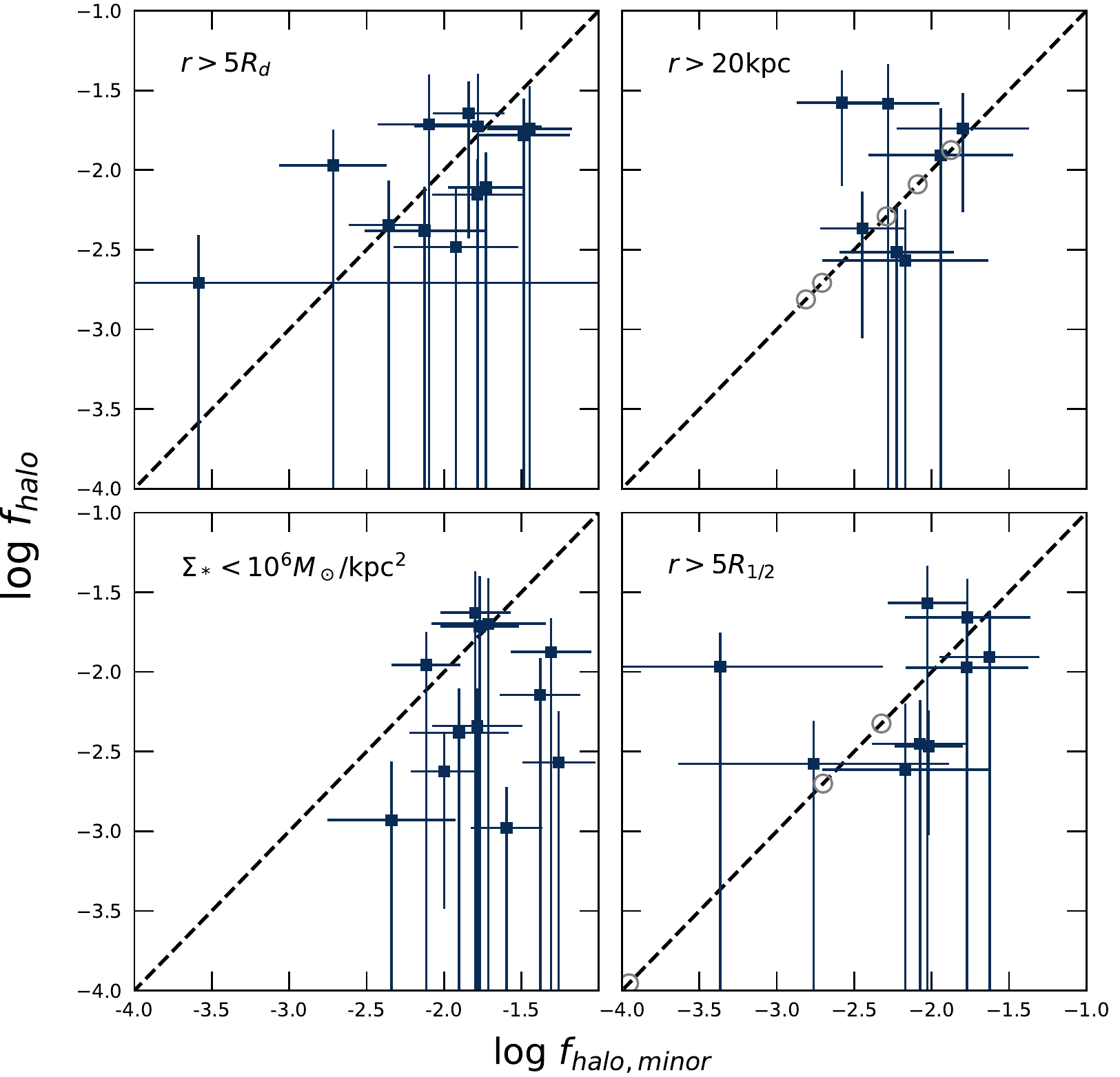}
\caption{{Comparison of stellar halo mass fractions measured from azimuthally averaged versus minor axis profiles for four different definitions of the stellar halo-dominated region. The dashed lines indicate a one-to-one correspondence. Open grey circles indicate missing minor axis measurements, when the onset of the stellar halo-dominated region is beyond the maximum extent of the minor axis stellar mass surface density profile.} }
\label{fig:minorfracs}
\end{figure}

{We re-measured stellar halo masses and fractions from the minor axis profiles according to the four definitions of the onset of the stellar halo-dominated region considered in this work. We assumed that the stellar halos had circular symmetry in projection. Given that minor axis was expected to have minimal disc stars, we did not attempt to model or subtract a disc component. For definitions dependent on a measurement along the major axis (half mass radius and disc scalelength), the according radii were rescaled to effective minor axis lengths using the outer ellipticity of the galaxy's light.}

{The stellar halo mass fractions $f_{halo,minor}$ are tabulated in Table~\ref{tab:minor_results}, and Figure~\ref{fig:minorfracs} compares $f_{halo,minor}$ and $f_{halo}$ for each definition of the stellar halo-dominated region. The two sets of stellar halo mass fractions are often consistent or nearly consistent within uncertainty (barring NGC~3628 and NGC~4013). The $f_{halo,minor}$ measured where $\Sigma_* < 10^6\:  \mathrm{M}_\odot/\mathrm{kpc}^2$ are systematically larger than the azimuthally averaged $f_{halo}$. This definition of the onset of the stellar halo-dominated region typically occurs at the smallest distance and therefore is subject to the most disc contamination. Evidently some disc contamination is also present along the minor axis. Good agreement with greatly reduced scatter is found when the azimuthally averaged measurement \emph{without} disc subtraction ($f_{outer}$) is compared to the minor axis measurement in this case.}

{In addition to these comparative tests, we also fitted the outer minor axis stellar mass surface density profiles with a simple power law such that $\log \Sigma_* = \log \Sigma_{*,0} - \alpha \log(r)$.  The fit region began where the scattered light-corrected minor axis $r$-band profile first became twice as bright as the model profile with no PSF convolution. We estimated the uncertainty of the fitted slope $\alpha$ using \texttt{emcee} \citep{emcee}, adopting the 16th and 84th percentiles of the posterior distribution for $\alpha$ as the 1$\sigma$ limits.  The power law slopes $\alpha$ are found in the last column of Table~\ref{tab:minor_results}. The range of values represented ($2 \lesssim \alpha \gtrsim 4$) fall in the general range often seen for stellar halos \citep[e.g.][]{bullockjohnston05, abadi06, cooper10, font11, monachesi19}}.

\begin{table*}[tb]
\begin{center}
\begin{tabular}{cccccccccc} 
 \hline \hline
 \multirow{2}{*}{Target} & \multicolumn{2}{c}{$r>5R_d$} & \multicolumn{2}{c}{$r>20\mathrm{ kpc}$} & \multicolumn{2}{c}{$\Sigma_* < 10^6\:  \mathrm{M}_\odot/\mathrm{kpc}^2$} & \multicolumn{2}{c}{$r>5R_{1/2}$} & \multirow{2}{*}{$\alpha$} \\
  & {$\log f_{halo}$} & {$\log \mathrm{M}_{halo}$} & {$\log f_{halo}$} & {$\log \mathrm{M}_{halo}$} & {$\log f_{halo}$} & {$\log \mathrm{M}_{halo}$} & {$\log f_{halo}$} & {$\log \mathrm{M}_{halo}$} & \\ \hline \hline
NGC 3044 & $-3.6 \pm 3.2$ & $6.7 \pm 3.2$ & -- &  -- & $-2.3 \pm 0.4$ & $7.9 \pm 0.4$ & -- &  -- & $4.4^{+2.3}_{-2.4}$ \\
NGC 3432 & $-2.1 \pm 0.3$ & $7.6 \pm 0.3$ & -- &  -- & $-1.8 \pm 0.3$ & $7.9 \pm 0.2$ & -- & -- &  $2.2^{+0.8}_{-0.8}$ \\
NGC 3501 & $-1.4 \pm 0.3$ & $8.7 \pm 0.2$ & $-1.9 \pm 0.5$ & $8.2 \pm 0.4$ & $-1.4 \pm 0.3$ & $8.8 \pm 0.2$ & $-1.6 \pm 0.3$ & $8.5 \pm 0.3$ & $1.8^{+0.5}_{-0.5}$ \\
NGC 3628 & $-2.7 \pm 0.3$ & $8.1 \pm 0.3$ & $-2.6 \pm 0.3$ & $8.2 \pm 0.2$ & $-2.1 \pm 0.2$ & $8.7 \pm 0.1$ & $-3.4 \pm 1.1$ & $7.4 \pm 1.0$ & $3.3^{+0.6}_{-0.6}$ \\
NGC 4010 & $-1.5 \pm 0.3$ & $8.4 \pm 0.2$ & -- &  -- & $-1.3 \pm 0.3$ & $8.6 \pm 0.2$ & $-1.8 \pm 0.4$ & $8.1 \pm 0.3$ & $2.1^{+0.7}_{-0.7}$ \\
NGC 4013 & $-1.8 \pm 0.2$ & $8.8 \pm 0.1$ & $-2.3 \pm 0.3$ & $8.3 \pm 0.3$ & $-1.8 \pm 0.2$ & $8.8 \pm 0.1$ & $-2.0 \pm 0.3$ & $8.6 \pm 0.2$ & $3.0^{+0.7}_{-0.8}$ \\
NGC 4307 & $-1.7 \pm 0.2$ & $8.6 \pm 0.1$ & $-2.2 \pm 0.4$ & $8.1 \pm 0.3$ & $-1.6 \pm 0.2$ & $8.8 \pm 0.1$ & $-2.1 \pm 0.3$ & $8.3 \pm 0.2$ & $2.3^{+0.5}_{-0.5}$ \\
NGC 4330 & $-1.9 \pm 0.4$ & $8.0 \pm 0.3$ & $-2.2 \pm 0.5$ & $7.7 \pm 0.5$ & $-1.3 \pm 0.2$ & $8.7 \pm 0.1$ & $-2.2 \pm 0.5$ & $7.7 \pm 0.5$ & $1.7^{+0.7}_{-0.7}$ \\
NGC 4437 & $-1.8 \pm 0.3$ & $8.4 \pm 0.2$ & -- &  -- & $-1.8 \pm 0.3$ & $8.4 \pm 0.2$ & -- &  -- & $3.1^{+0.6}_{-0.6}$ \\
NGC 4565 & $-2.4 \pm 0.3$ & $8.5 \pm 0.2$ & $-2.4 \pm 0.3$ & $8.4 \pm 0.2$ & $-2.0 \pm 0.2$ & $8.9 \pm 0.1$ & $-2.0 \pm 0.2$ & $8.9 \pm 0.1$ & $2.5^{+0.3}_{-0.3}$ \\
NGC 4634 & $-2.1 \pm 0.4$ & $8.0 \pm 0.3$ & -- &  -- & $-1.9 \pm 0.3$ & $8.2 \pm 0.3$ & $-2.8 \pm 0.9$ & $7.4 \pm 0.9$ & $2.4^{+1.5}_{-1.3}$ \\
NGC 5907 & $-1.8 \pm 0.4$ & $9.1 \pm 0.4$ & $-1.8 \pm 0.4$ & $9.1 \pm 0.4$ & $-1.7 \pm 0.4$ & $9.2 \pm 0.3$ & $-1.8 \pm 0.4$ & $9.1 \pm 0.4$ & $1.9^{+0.6}_{-0.6}$ \\
 \hline \hline
\end{tabular}
\end{center}
\caption{{Stellar halo masses and mass fractions measured along the minor axis for DEGS targets.  Missing entries indicate that the minor axis $\Sigma_*$ profile ended before the onset of the stellar halo-dominated region for that definition. The best fit power-law slope $\alpha$ is also included in the final column.} }
\label{tab:minor_results}
\end{table*}

\subsection{Notes on individual DEGS galaxies}

Before moving on to compare our results with previously published work, we pause here to briefly comment on each galaxy in the DEGS sample. This includes mitigating factors and challenges in our analysis (particularly Galactic cirrus), known artifacts of the data reduction pipeline, interfering or potentially misleading background objects, and/or features of note in the galaxy's outskirts or immediate surroundings. The reader interested only in our overall results on stellar halos can safely skip this section.

\subsubsection{NGC 3044}
NGC 3044 is positioned within a hole in the Galactic cirrus. The brightness of the cirrus steadily increases in the galaxy's immediate surroundings, making it necessary to measure the background as close to the galaxy as possible and therefore limiting the area over which the smooth component of the stellar halo can be characterized. This galaxy is flagged with a contrasting colour in Figures~\ref{fig:halofrac}~and~\ref{fig:obs_vs_sim} to emphasize that it must be considered with caution. 

There is a prominent low surface brightness ``blob'' east of the galaxy at a position of 9:54:03.5, +1:38:20. The blob is approximately $2'$ in diameter and is slightly elongated in right ascension. It is bright and large enough to be seen in the false colour composite in Figure~\ref{fig:colour_cutouts}. It does not correspond to any known galaxy or galaxy cluster, to our knowledge. While we cannot rule out the possibility that this feature is an ultra diffuse galaxy, the most likely explanation in this case is cirrus. 100~$\mu$m IRAS/IRIS imaging reveals a larger patch ($\sim5'$ in diameter) of what is likely cirrus directly adjacent to (and partly overlapping) the location of the blob. A connection between the blob and more diffuse emission matching the location of the cirrus patch is visible when viewing Dragonfly images with an extreme stretch. It is plausible that the blob is a dense patch of cirrus connected to a more diffuse filament, and the spatial resolution of the 100~$\mu$m image is insufficient to resolve it while it appears quite distinct in Dragonfly imaging.  

\subsubsection{NGC 3432}

NGC 3432 is the lowest stellar mass galaxy in our sample. It is interacting with the dwarf galaxy UGC~05983. The dwarf's light is very distorted and is smeared $4'$ to the west of its center. NGC 3432 itself is very asymmetric. During our analysis, we masked UGC~05983's core but did not mask its disrupted envelope or trailing debris. Unfortunately the background galaxy  SDSS J105141.30+363653.4 (with a photometrically determined redshift of $z=0.042938$) lies in the path of the trailing debris from UCG 05983 and precludes any search for an extended stream. It is also difficult to search for the stream further towards the northwest due to WISEA J105128.87+363945.5, the central galaxy of cluster MSPM 01624 (spectroscopic redshift of $z=0.042891$). It is difficult to tell whether any low surface brightness emission in the area is intracluster light or belongs to a stream. 

\subsubsection{NGC 3501}

We find no clear structures in the outskirts of NGC 3501. Our search is likely hindered by cirrus and satellite trail artifacts. NGC 3507, a face-on grand design spiral galaxy, is found in this field with a projected separation of $13'$ to the northeast. These two galaxies are almost certainly members of the same group. The distance to NGC 3501 is readily measured via the Tully-Fisher relation since it is inclined relative to the line of sight, but NGC 3507's distance is harder to measure due to its face-on inclination. Its distance is estimated as 19 Mpc by \cite{tully88}. 

\subsubsection{NGC 3628}

NGC 3628 is a member of the Leo Triplet along with M65 and M66. There are a few structures of note in the field that can be attributed to cirrus. An elongated linear feature northwest of M65 and a blob southeast of M66 seen in Dragonfly imaging are both found in 100 $\mu$m IRAS/IRIS imaging. Cirrus structure in the southeast corner of the field also matches well between Dragonfly and IRAS/IRIS. There seems to be a linear feature stretching from south of the east disc limb of NGC 3628 to west side of M65, but it runs parallel to a bright cirrus filament. This is another case where a cirrus map with higher spatial resolution is needed to best determine whether the feature is Galactic or extragalactic in origin. 

The north-south asymmetry about the midplane of NGC 3628 is not likely to be confused with cirrus. The outskirts extend southward from both sides of the disc towards the other massive group members. These (relatively) bright lobes are also accompanied by more extended and fainter light. It may be that some of this faint light is better classified as intragroup light (IGL). Making this distinction and searching for IGL in the rest of the field is beyond the scope of this work as it would require characterization of M65 and M66 in addition to a robust method of cirrus subtraction. 

\subsubsection{NGC 4010}

The stellar halo of NGC 4010 appears asymmetric or structured within $\sim5'$ or $\sim28$~kpc. There is a bright spike-like patch extending southeast of the eastern disc limb and a broader patch opposite it north of the west side of the disc. The latter may be a looping stream, and appears in the profiles as a local increase in $g-r$ (0.3 mag redder than to adjacent points) and $\Sigma_*$ ($\sim0.6$ dex increase) at 237 arcsec (21.7 kpc). It is difficult to assess the shape of the NW feature due to an interfering star. A cirrus filament extending from the west ends near this area of the galaxy and casts some doubt on any features nearby. 

\subsubsection{NGC 4013}

We confirm NGC 4013's stream complex presented in \cite{martinezdelgado09} as well as the box-like shape of its stellar halo. We also detect a large loop on the west side of the galaxy. The loop connects the west and southwest ``wings'' previously noted by \cite{martinezdelgado09} (``D'' and ``E'' in their Figure~3) and extends $\sim8.3'$ in projection ($\sim45$~kpc) from the center of the galaxy at its outer edge. The maximum projected distance between the new west loop and the galaxy midplane is similar to that of the east loop and therefore is compatible with the hypothesis of one connected stream caused by the disruption of a dwarf galaxy on a low inclination orbit. There may also be a $\sim10'$ linear feature extending east from the previously noted ``F'' feature, but this may simply be a coincidental alignment of background galaxies and clusters

Patches of low surface brightness emission north of the galaxy can be attributed to a series of cataloged galaxy clusters with redshifts from 0.29-0.43. Some of these are WHL J115850.5+440536, NSCS J115847+440712, WHL J115823.6+440556, and WHL J115824.1+440532. These are potentially misleading as they have a similar projected separation as the west stream loop and exist in the context of a field with many known tidal features.

\subsubsection{NGC 4307}

The two parallel low surface brightness streaks east of NGC 4307 are image reduction artifacts. On the night of April 28 2019, asteroid 326 Tamara passed through the field coincidentally as we were imaging it. The asteroid was identified as an unwanted bright source and masked automatically in the frames contributing to the final stacked images. The mask was evidently not wide enough, leaving residuals on either side of the asteroid's track. We also note NGC 4307A, a background face-on spiral galaxy immediately south of NGC 4507.

\subsubsection{NGC 4330}

NGC 4330 is a member of the Virgo cluster, likely falling in for the first time \citep{abramson11}. Similar to NGC 3044, NGC 4330 is coincidentally located in a hole in the cirrus. Near the galaxy itself, cirrus contamination is worse northwest of the disc plane (where two bright stars prevent any investigation of the stellar halo anyway) while the southeast is clear. Our profiles and resulting analysis effectively probe the latter region. 

\subsubsection{NGC 4437}

The entire NGC 4437 field has many residual streaks from satellite or airplane trails. These artifacts, combined with cirrus, make it difficult to search for any structures in NGC 4437's stellar halo. We are working to improve our pipeline to reduce the impact of satellite trails.

\subsubsection{NGC 4565}

NGC 4565's disc asymmetry and features at the transition between disc and stellar halo-dominated regions are discussed in \cite{gilhuly20}. 

\subsubsection{NGC 4634}

NGC 4634 is interacting with another spiral galaxy, NGC 4633. Both are members of the Virgo cluster. Structured cirrus is ubiquitous across the field, with an unfortunate filament passing directly across the interacting pair. Without a high resolution map of cirrus and (ideally) a method to subtract it from Dragonfly imaging, it is not realistic to search for tidal features associated with these galaxies. It is certainly an interesting but complicated field. 

The mean 100$\mu$m flux within $5'$ of NGC 4634's position is $1.380\pm0.026$~MJy/sr\footnote{\url{https://irsa.ipac.caltech.edu/applications/DUST/}}; this is the second-highest among our sample. This case shows the limitations of attempting to avoid Galactic cirrus based on this criterion alone. Undoubtedly this is better than making no attempt to avoid cirrus; in other Dragonfly imaging surveys covering wide areas of the sky, the extent and brightness of cirrus is striking at times (see Figures~4~and~5 of \cite{danieli20} for an example). Better criteria for filtering target fields based on the degree of contrast or structure in the cirrus may be worth devising for future surveys. Diffuse or hazy cirrus is adequately handled with extinction corrections and background subtraction, while structured cirrus currently can only be handled by masking and avoidance.

\subsubsection{NGC 5907}

The massive tidal stream that surrounds NGC 5907, first discovered by \cite{martinezdelgado08}, is discussed in \cite{vandokkum19}. Notably, we have discovered a long western arm and detect only one eastern loop where two were previously thought to exist. This single stream appears to be the only genuine (sufficiently bright) tidal structure surrounding the galaxy. Other low surface brightness streaks and patches of note in the field match well with 100$\mu$m emission and are very likely to be cirrus.

\section{Discussion} \label{discussion}

\subsection{Comparison with previous measurements} \label{comparison}

\begin{figure}[tb]
\includegraphics[width=0.45\textwidth]{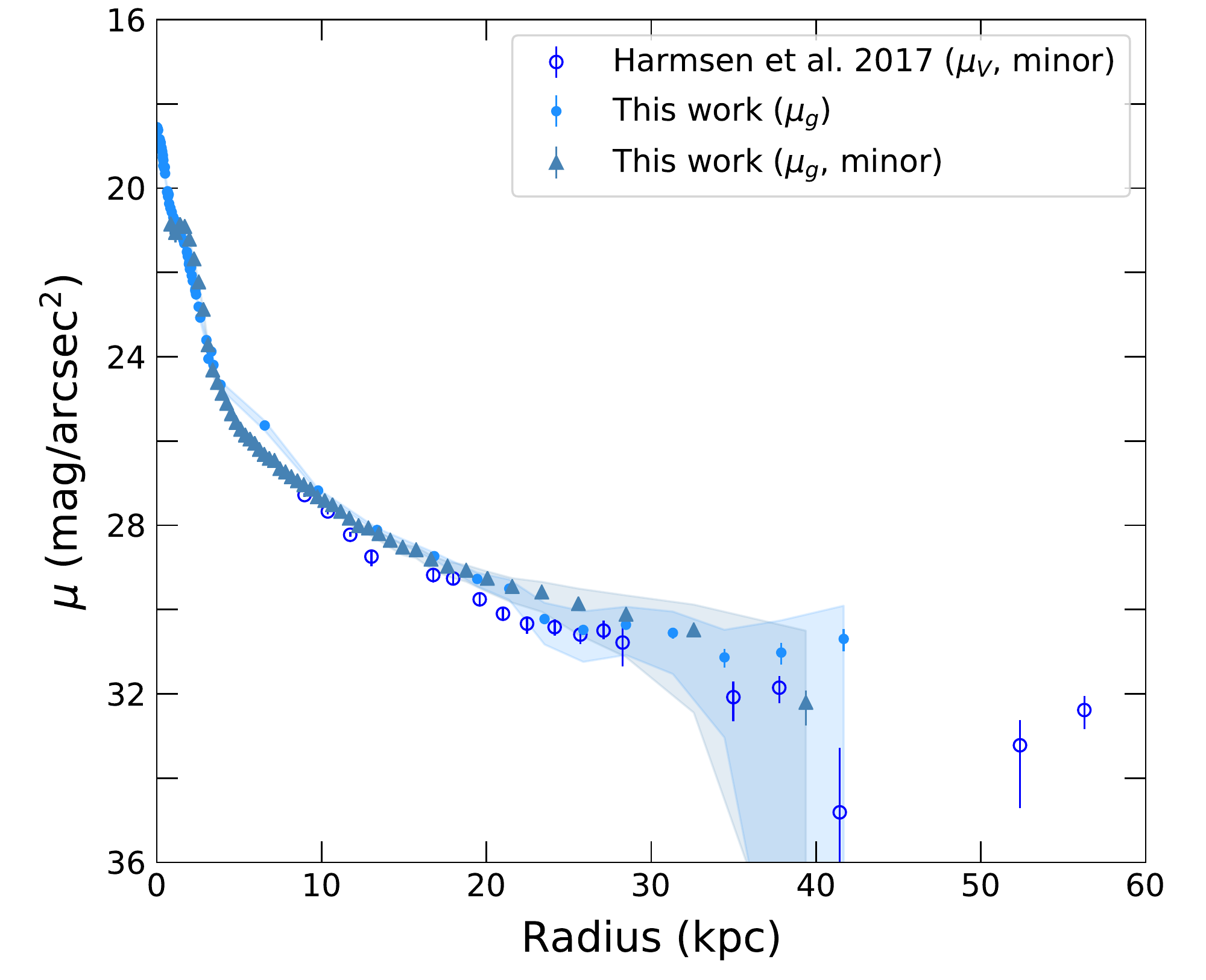}
\caption{{Comparison of the NGC~4565 DEGS $g$-band profiles (azimuthally averaged as dots and minor axis as triangles) with the equivalent $V$-band profile based on the GHOSTS star count profile \protect\citep[shown as open circles]{harmsen17}.}}
\label{fig:NGC4565}
\end{figure}

\textit{NGC 4565}: \cite{harmsen17} report a stellar halo mass of $M_{halo,10-40kpc} = 7.16^{+0.33}_{-0.31} \times 10^8\:  \mathrm{M}_\odot$, corresponding to log $f_{halo} = -2.05\pm0.32$. This is most comparable with our measurements of $f_{outer}$ as they do not attempt to model and subtract the disc contribution in the galaxy's outskirts. Our measurements of log $f_{outer}$ range from {-1.4 to -2.2}, and all are consistent with the GHOSTS value within the statistical error. It is encouraging to find comparable results from two entirely independent analyses. {We also find good agreement between our profiles (see Figure~\ref{fig:NGC4565}).}

\textit{M101}: The underweight or even absent stellar halo of M101 was first highlighted by \cite{vandokkum14}, who found $f_{halo} = 0.3^{+0.6}_{-0.3}\%$. \cite{merritt16} revisit this measurement with a new procedure alongside the rest of the DNGS galaxies and set an upper limit of $f_{halo} < 0.04\pm0.08\%$. \cite{jang20} augment the Dragonfly imaging used in the two previous studies with several HST pointings and provide a more tightly constrained value of $f_{halo} = 0.2^{+0.10}_{-0.08}\%$. 

\begin{figure}[tb]
\includegraphics[width=0.5\textwidth]{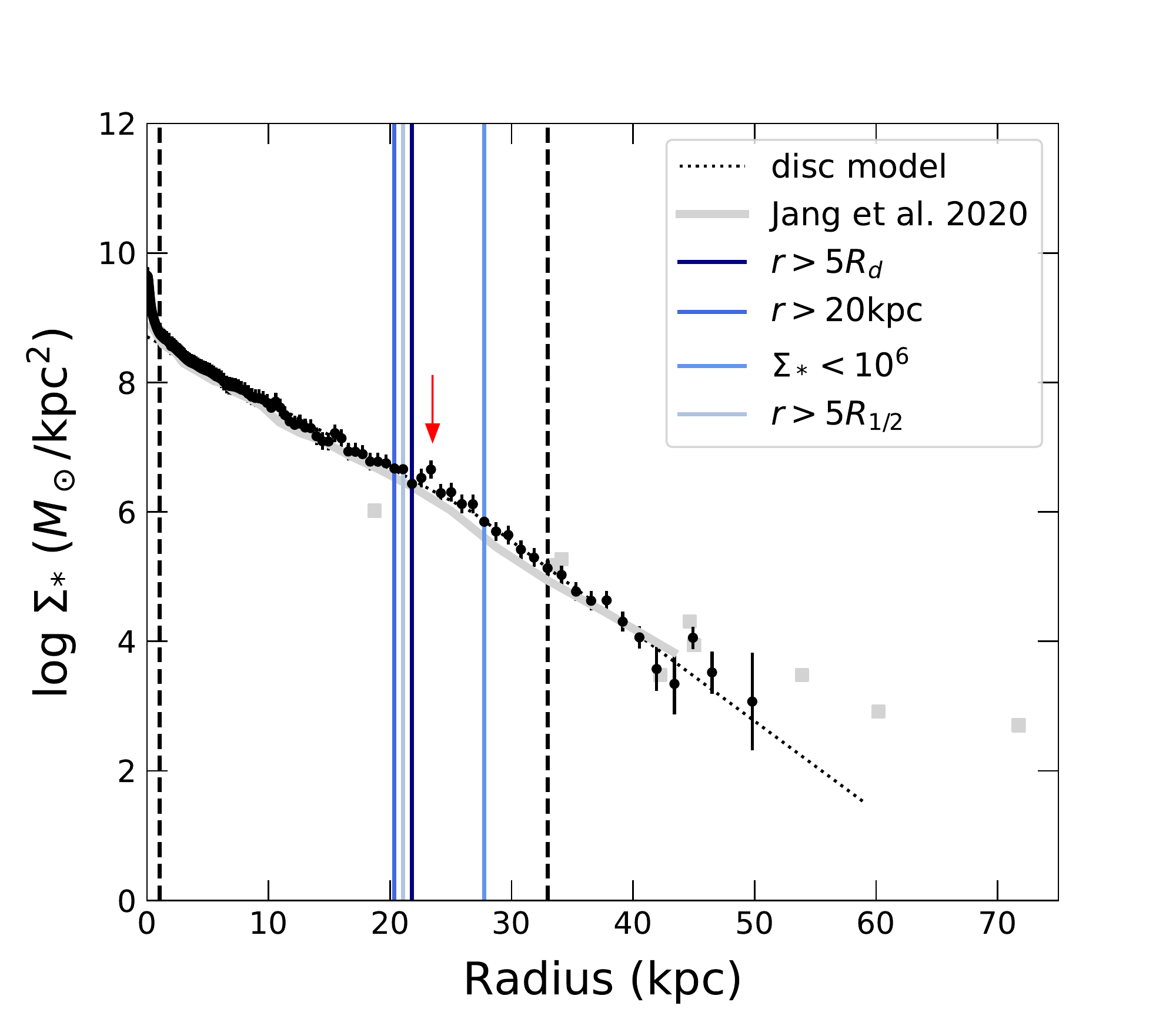}
\caption{Stellar mass surface density profile for M101. The vertical dashed black lines show the region where the broken exponential disc model is fit, and the black dotted line shows this model. The vertical blue lines of various shades show the inner radial limit of the four possible definitions of the stellar halo-dominated region used in this work. We note that these four lines are all within the disc-fitting region for M101. A significant excess corresponding to a spiral arm is marked with a red arrow. Three of four definitions of the stellar halo-dominated region include this excess and therefore have erroneously large stellar halo mass fractions. The fourth definition that does not include this excess results in a stellar halo non-detection. {The integrated light and resolved star profiles from \protect\cite{jang20} for the entire galaxy are shown with a grey line and grey squares, respectively.}}
\label{fig:M101}
\end{figure}

Our values of $f_{halo}$ are high in comparison to these published values; all are in excess of 1\% (except for one non-detection; the generous estimate of the upper limit here is 1.2\%). Upon closer inspection, we find that all four definitions of the onset of the stellar halo-dominated region occur well within the disc-dominated region for this galaxy, and even within the region where the disc model is fitted (refer to Figure~\ref{fig:M101}). We compare the integral of the disc model to the integral of the observed stellar mass surface density profile in order to obtain an estimate of the stellar halo mass; this method breaks down when the disc model does not reasonably represent the disc. In this case, there is a significant excess in $\Sigma_*$ relative to the smooth broken exponential model at 23.5 kpc. This corresponds to a massive spiral arm to the north. We highlight this excess in Figure~\ref{fig:M101}. When this excess mass is attributed to the stellar halo, it leads to a great increase in the stellar halo mass fraction. One of our four definitions of the stellar halo-dominated region begins just outside this spiral arm. The result is a stellar halo non-detection, confirming our suspicions.  This is an excellent example of how rules of thumb can be violated and measurement techniques can break down due to the unique features or properties of individual galaxies. This does not pose a problem when considering statistical samples of galaxies, but we are operating in a small $N$ regime where each galaxy is important. 

For the sake of comparison, we re-measure $f_{halo}$ for M101 beyond 43.5 kpc, where the last few points of the stellar mass surface density profile are higher than the disc model extrapolation. In this outermost region we find a stellar mass of {$10^{6.6^{+0.6}_{-0.6}}\:  \mathrm{M}_\odot$ in excess of the disc model, corresponding to $f_{\mathrm{halo}} = 0.006^{+0.008}_{-0.009}\%$ ($-4.2^{+0.6}_{-0.7}$)}. This is consistent with previous measurements based on Dragonfly imaging alone \citep{vandokkum14, merritt16} but is lower than the \cite{jang20} measurement.

\subsection{Comparison with simulated stellar halos}

\begin{figure*}[tb]
\includegraphics[width=0.9\textwidth]{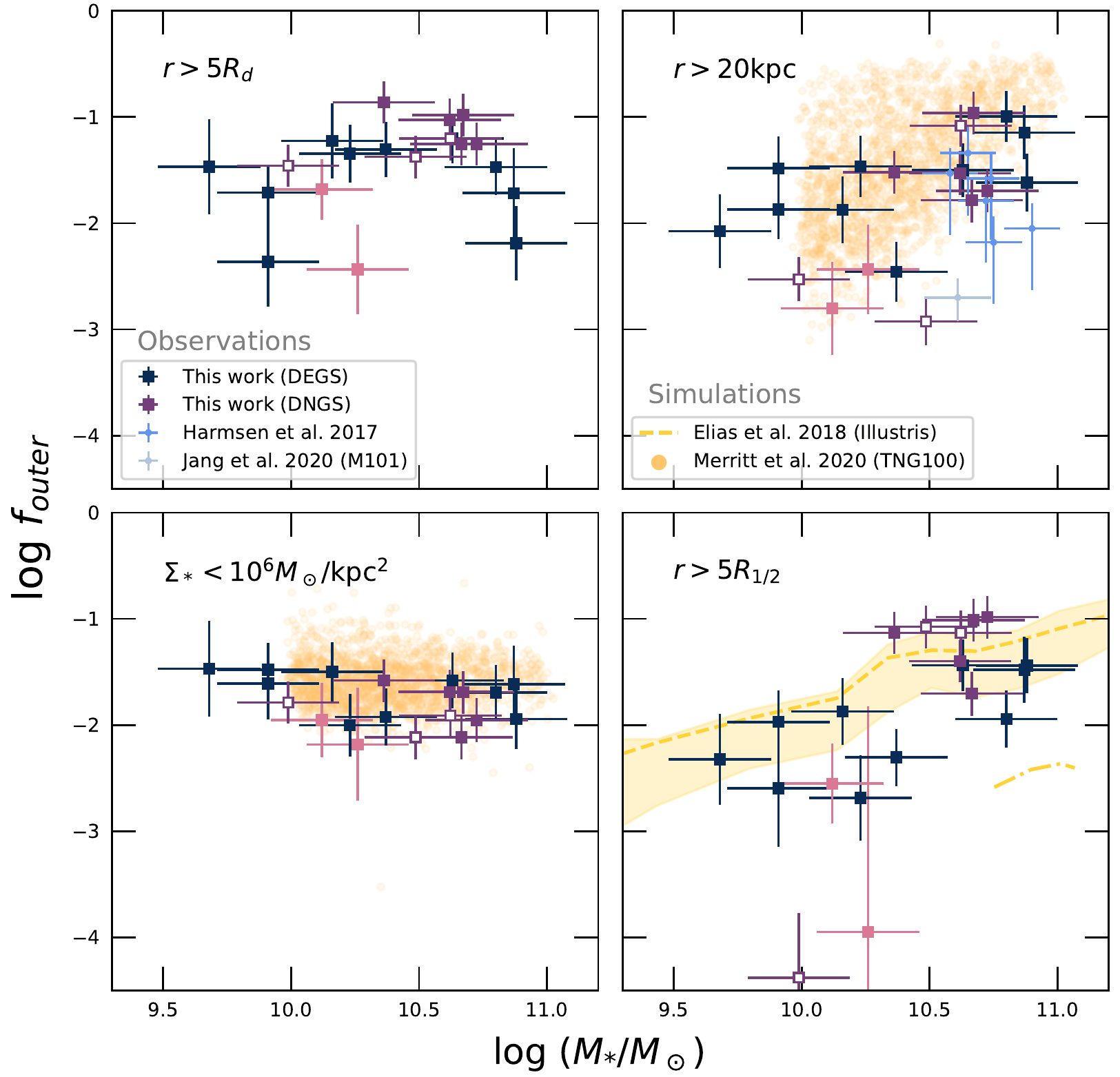}
\caption{Comparison of our observed outer stellar mass fractions (i.e. without subtracting the extrapolated disc contribution) and those of galaxies from the Illustris simulation \protect\citep{elias18} and Illustris TNG100 \protect\citep{merritt20}. The point style and colour convention for observed galaxies is the same as Figure~\ref{fig:halofrac}. The yellow dashed line in the lower right panel corresponds to the median Illustris stellar halo mass fraction, while the light yellow envelope indicates the 25-75 interquantile range. The dash-dotted yellow line towards the right side of the lower right panel indicates the outer range of the lowest $f_{halo}$ galaxies in their primary sample of Milky Way-mass galaxies (see their Figure~5). $f_{halo}$ measurements for individual TNG100 galaxies are plotted in the upper right and lower left panels in transparent orange.}
\label{fig:obs_vs_sim}
\end{figure*}

It is difficult to compare observed stellar halo masses directly with those of simulated galaxies because of the differing amounts of available information. The accreted stellar mass in simulations can be directly obtained by labelling individual star particles at $z=0$ as accreted or in-situ, according to some definition. Whether or not the accreted star particles reside within the host galaxy or somewhere in its stellar halo does not complicate this process. It is even possible to determine the contributions of ejected or heated in-situ stars to the inner stellar halo \citep{zolotov09, purcell10, tissera13}. Of course, such things are not easily done observationally. In integrated light studies such as our own, there is no way to separate accreted stars from stars that formed in-situ in the bulge and disc. 

There are two approaches to bridging this information gap. The first is to empirically correct observed stellar halo masses to be more similar to the accreted stellar masses of simulated galaxies. \cite{harmsen17} adopt this approach, replicating their analysis on mock observations and comparing the resulting stellar halo mass to the known total accreted stellar mass. The second approach is to translate the results from simulations into observer-friendly terms. This can be done to varying degrees, from simply adopting a stellar halo definition while retaining all information, all the way to producing realistic mock observations using simulated instruments from which to measure a stellar halo mass. 

\cite{elias18} and \cite{merritt20} both strive to analyze simulated galaxies in terms that are compatible with observations to facilitate fair comparisons. Between these two works, three of our four definitions of the stellar halo-dominated region are represented. We compare our results with theirs in Figure~\ref{fig:obs_vs_sim}. As neither work subtracts an extrapolated disc contribution from the stellar halo-dominated region, their stellar halo mass fractions are most comparable with our $f_{outer}$. DEGS and DNGS galaxies that were previously stellar halo non-detections are still marked with open symbols. Although $f_{outer}$ is by definition larger than $f_{halo}$ and we report no non-detections for $f_{outer}$, there are still a few galaxies consistent with no stellar mass within these outer regions. 

The Illustris-1 simulation (hereafter ``Illustris'') is a cosmological hydrodynamical simulation of a volume spanning 106 Mpc on each side with a mass resolution of $1.6\times10^6\:  \mathrm{M}_\odot$ for baryonic mass and $6.3\times10^6\:  \mathrm{M}_\odot$ for dark matter \citep{vogelsberger14}. \cite{elias18} adopt a stellar halo mass and mass fraction definition highly compatible with observations for their work with 967 central galaxies of diverse morphologies from Illustris. They do not carry out mock observations or project data in 2D, but motivated by \cite{merritt16} they consider all stellar mass beyond five half mass radii as the stellar halo\footnote{They favour a galaxy-halo separation at two half mass radii for the majority of their analysis, which focuses on Milky Way-like (dark matter) halos, but they also calculate stellar halo mass fractions with a galaxy-halo separation at five half-mass radii over a wider mass range to facilitate comparison with observations.}. The median and 25-75 interquartile range (IQR) of stellar halo mass fractions produced by \cite{elias18} are plotted in the bottom right panel of Figure~\ref{fig:obs_vs_sim}.

 Almost all observed galaxies fall within or below the 25-75 IQR. This is expected, as the sample considered by \cite{elias18} includes galaxies of all morphologies. For Milky Way-mass galaxies, they find that the galaxies with the lowest $f_{halo}$ have greater ongoing star formation and are more rotation-dominated than the galaxies with the highest $f_{halo}$. This tendency almost certainly applies to lower mass galaxies as well.

When the stellar halo is defined as all mass beyond five half-mass radii, \cite{elias18} predict that the median stellar halo mass fraction will decrease with decreasing galaxy stellar mass but with great diversity in stellar halo mass fraction at fixed total stellar mass. Their Milky Way-like sample spans 2 dex in $f_{halo}$ while the median $f_{halo}$ drops by less than 1 dex over 1.5 orders of magnitude in stellar mass. Our modest sample of 20 observed spiral galaxies qualitatively agrees with these predictions. The galaxies with stellar masses $M_* > 10^{10.5}\:  \mathrm{M}_\odot$ all have $f_{outer}$ within the range of values spanned by the Illustris galaxies. The full range of $f_{halo}$ for Illustris galaxies at lower stellar masses is not published but the 25-75 IQR remains close to 0.6 dex wide for galaxies with stellar masses from $10^{9.5} - 10^{11}\:  \mathrm{M}_\odot$. Assuming a similar overall span of $f_{halo}$ for all Illustris galaxies across this mass range, all but two of our galaxies (NGC 3044 and NGC 1042) lie somewhere within the distribution of $f_{halo}$. Both of these galaxies have $\Sigma_*$ profiles with an extent only slightly larger than $5 R_d$, which explains why they have such small $f_{outer}$ under this stellar halo definition.

The Next Generation Illustris ``TNG100'' simulation is an updated version of Illustris \citep{springel18, pillepich18b, nelson18}. With a slightly larger volume (spanning 110.7 Mpc per side) and a slightly finer mass resolution ($1.4\times10^6\:  \mathrm{M}_\odot$ for baryonic mass), the largest changes are improvements to various physical processes contributing to galaxy formation and evolution. The resulting population of simulated galaxies more closely resembles real galaxies \citep{pillepich18a, genel18}. 

\cite{merritt20} studies the outskirts of 1844 disc-dominated galaxies in low density environments from the Illustris TNG100 simulation. In addition to detailed investigations of the assembly histories of various mass-matched and profile-matched samples of simulated galaxies, they calculate stellar halo mass fractions for multiple definitions of the stellar halo-dominated region, two of which match our own. \cite{merritt20} obtain stellar mass surface density profiles by fitting elliptical isophotes to 2D mass ``images'' matching the spatial resolution of DNGS observations at 10 Mpc to facilitate comparison with \cite{merritt16}.  Although DEGS galaxies are on average more distant than DNGS galaxies, the results of \cite{merritt20} nevertheless offer a fairly compatible point of comparison with one of the latest cosmological hydrodynamical simulations. We plot the full distribution of their stellar halo mass fractions measured beyond 20 kpc and for $\Sigma_* < 10^6\:  \mathrm{M}_\odot/\mathrm{kpc}^2$ against our own $f_{outer}$ in the top right and bottom left panels of Figure~\ref{fig:obs_vs_sim}, respectively.  We also include data points from \cite{harmsen17} and \cite{jang20} in the top right panel.

 There is good qualitative agreement between our $f_{outer}$ and those of \cite{merritt20}. Both data sets show a tight distribution of $f_{outer}$ with a weak negative dependence on total stellar mass when considering the stellar halo-dominated region to be where $\Sigma_* < 10^6\:  \mathrm{M}_\odot/\mathrm{kpc}^2$. Meanwhile, both show greater scatter and a stronger positive dependence on total stellar mass when $f_{outer}$ is measured beyond 20 kpc. However, the TNG100 $f_{halo}$ are systematically larger than those of the observed galaxies. \cite{merritt20} note that TNG100 galaxies are systematically more extended than DNGS galaxies, and the outskirts of DNGS galaxies are most consistent with the TNG100 galaxies with the lowest amount of accreted material.

On the surface, we find somewhat better agreement with the $f_{halo}$ measured by \cite{elias18} than with those measured by \cite{merritt20}. The latter draws upon an updated simulation suite, offers a better morphological match in their sample to our own, and analyzes their simulated galaxies in 2D projection with a similar physical resolution and profile-based procedure. And yet, the resulting measurements of $f_{halo}$ are offset higher than our corresponding $f_{outer}$ for both of our common definitions of the stellar halo-dominated region. 

Slight differences in the environment density criteria may be partly responsible for this difference. \cite{merritt20} exclude galaxies with two or more other massive galaxies ($M_* \geq 10^{10}\:  \mathrm{M}_\odot$) within 1 Mpc, while \cite{elias18} exclude galaxies with massive satellites ($M_* \geq 10^{9}\:  \mathrm{M}_\odot$) within 50 kpc or with satellites more than a quarter of the stellar mass of the central galaxy within the virial radius. The exclusion of galaxies with massive satellites likely removes some of the high $f_{halo}$ population present in the Illustris sample. 

The differing stellar halo-dominated region definitions must be considered when attempting to understand the offset (or apparent lack thereof) between the $f_{halo}$ of simulated galaxies and our $f_{outer}$. The two definitions we share with \cite{merritt20} are \emph{independent} of galaxy size, while the definition we share with \cite{elias18} is \emph{dependent} on galaxy size. (Of course, a stellar halo definition independent of galaxy size does not imply that the individual measurements under that definition will be independent of galaxy size. The opposite is likely true; galaxies with smaller sizes are likely to have less stellar mass beyond a fixed physical radius when compared to galaxies with larger sizes.) The population of galaxies produced in TNG100 are substantially more realistic than those of the original Illustris \citep[see Figure~4 in][]{pillepich18a} and have half-mass radii that match observed sizes within the overall statistical uncertainty \citep{genel18, rodriguezgomez19}. However, the unrealistically large size of Illustris galaxies is balanced by the larger radius for the onset of the stellar halo-dominated region when it is defined as $r>5R_{1/2}$. The way that this larger size ultimately affects their stellar halo mass measurements depends on relative overlargeness of the in-situ versus ex-situ stellar mass and the inner stellar mass versus the outer stellar mass. Determining how the large $R_{1/2}$ of Illustris galaxies has impacted the $f_{halo}$ measured by \cite{elias18} is beyond the scope of this paper. All we can say is that while we find good agreement between their $f_{halo}$ and our $f_{outer}$ on a surface level, there are many hidden complicating factors that prevent us from commenting on the similarity of the galaxies and stellar halos behind these measurements. We would expect much worse agreement between the stellar halos of Illustris galaxies and observed galaxies when adopting a stellar halo-dominated region definition that is independent of galaxy size. 

The stellar halo definitions we share with \cite{merritt20}, $r>20$~kpc and $\Sigma_* < 10^6\:  \mathrm{M}_\odot/\mathrm{kpc}^2$, are both independent of galaxy size and so are better at revealing systematic differences between simulated and observed stellar halos. Despite improvements in TNG100 as compared to Illustris, the outskirts of TNG100 galaxies still seem to be overly massive and the reason for this is complicated \citep[see Section~8.4 of][for a discussion of many possible contributing factors]{merritt20}. It is possible to predict the relative impact this will have on different measurements of $f_{outer}$ by considering the simple case of two galaxies of fixed total stellar mass and similar bulge-disc morphology where the characteristic scalelength of the outer component is larger (more extended) for one galaxy. More mass will be found beyond 20 kpc for the galaxy with the more extended outskirts. The same is true for stellar masses measured beyond $\Sigma_* < 10^6\:  \mathrm{M}_\odot/\mathrm{kpc}^2$, but the effect will be reduced because the radius at which this threshold density is met may be larger for the more extended galaxy. Indeed, the difference between the average $f_{outer}$ of DEGS/DNGS galaxies and TNG100 galaxies is much larger for $r>20$~kpc {(0.33 dex)} than for $\Sigma_* < 10^6\:  \mathrm{M}_\odot/\mathrm{kpc}^2$ {(0.20 dex)}. We therefore confirm the conclusion of \cite{merritt20} that there remains a discrepancy between the outskirts of observed spiral galaxies and disc-dominated galaxies from the Illustris TNG100 simulation. Whether this is a case of ``missing'' outskirts in DNGS and DEGS galaxies or overly massive or extended outskirts in TNG100 galaxies remains to be seen, though \cite{merritt20} have made progress in identifying possible origins of this discrepancy if it lies on the side of simulations.

\section{Conclusions} \label{conclusions}
 
In this paper, we have carried out a survey of the stellar halos of twelve edge-on spiral galaxies with stellar masses ranging from $10^{9.68}-10^{10.88}\:  \mathrm{M}_\odot$. We included the DNGS galaxies previously studied by \cite{merritt16} in our analysis for a total of 20 nearby spiral galaxies. {Our main conclusions are as follows:} 
\begin{enumerate}
    \item {The previously observed diversity among the stellar halo mass fractions of Milky Way-mass spirals is also found for galaxies with stellar masses of $\sim10^{10}\:  \mathrm{M}_\odot$ and below.}
    \item {We have begun to probe the underlying trends in stellar halo mass with host galaxy stellar mass, which previously was only possible via stacking \citep{talvandokkum11, dsouza14} or simulations of a cosmologically significant volume. For some definitions of the stellar halo-dominated region ($r>20$~kpc and $r>5R_{1/2}$), a weak positive correlation is found between $f_{halo}$ and total stellar mass.}
    \item {We have demonstrated that our integrated light profiles and measurements of $f_{halo}$ are in line with the results of resolved star observational studies \citep{harmsen17}.}
    \item {Our $f_{outer}$ measurements have good qualitative agreement with predictions from cosmological hydrodynamical simulations (modulo a small offset).}
\end{enumerate}
 
Integrated light studies are an efficient approach to ``weighing'' the stellar halos of nearby galaxies when scattered light, background subtraction,  and the Galactic cirrus foreground can all be managed. We anticipate this method will continue to play a role in expanding the observed sample of stellar halos. The immediate next steps will be to expand investigations to yet lower stellar masses and to include a wider variety of galaxy types, such as lenticular galaxies. Lenticular galaxies are dynamically disc-dominated like spiral galaxies but clearly have experienced some significant differences in their assembly and accretion history that resulted in their comparatively red colours today. It would be interesting to see how they compare with the simulated disc galaxies with particularly massive outskirts.

Dragonfly $g-$ and $r-$band imaging for all the galaxies presented here is now publicly available. In total, DEGS offers approximately 140~deg$^2$ of deep imaging in each band. The fields are generally found at high Galactic latitudes as a consequence of our cirrus avoidance criteria, though some amount of cirrus is often present somewhere in the field. We have only begun to exploit these data with our work on the target galaxies' stellar halos. The vertical structure of the target galaxies' discs, the stellar halos of other galaxies in the field, searches for low surface brightness dwarf or ultra diffuse galaxy candidates, and the intracluster light of distant background clusters are all possible avenues of further exploitation of these data. Even the Galactic cirrus, a major challenge to low surface brightness extragalactic work, may be of interest. Please refer to Appendix~\ref{datarelease} for information on accessing and using these (and other) public Dragonfly data.
 
\begin{acknowledgments}

The authors thank the staff at New Mexico Skies Observatory for their dedication to Project Dragonfly. Support for this work was provided by NSERC, the Dunlap Institute, and MPIA. S.D. is supported by NASA through Hubble Fellowship grant HST-HF2-51454.001-A awarded by the Space Telescope Science Institute, which is operated by the Association of Universities for Research in Astronomy, Incorporated, under NASA contract NAS5-26555.

This work was enabled in part by support provided by the Canadian Advanced Network for Astronomical Research \url{www.canfar.net/en/}) and Compute Canada (\url{www.computecanada.ca}). We acknowledge the usage of the HyperLeda database. This research has made use of the NASA/IPAC Extragalactic Database (NED), which is operated by the Jet Propulsion Laboratory, California Institute of Technology, under contract with the National Aeronautics and Space Administration. This research has made use of the NASA/IPAC Infrared Science Archive, which is funded by the National Aeronautics and Space Administration and operated by the California Institute of Technology. This research has also made use of NASA's Astrophysics Data System Bibliographic Services. 

\textbf{Software}: This work made use of the following \texttt{python} packages:
\texttt{numpy} v.1.19.1 \citep{numpy}, 
\texttt{matplotlib} v.3.2.2 \citep{matplotlib}, 
\texttt{scipy} v.1.5.2 \citep{scipy}, 
\texttt{photutils} v.0.7.2 \citep{photutils}, 
\texttt{astropy} v.4.0.1.post1 \citep{astropy, astropy2}, 
\texttt{emcee} v.3.0.2 \citep{emcee}, and
\texttt{pandas} v.1.1.0 \citep{pandas}.

\textbf{DECaLS}: The Legacy Surveys consist of three individual and complementary projects: the Dark Energy Camera Legacy Survey (DECaLS; Proposal ID \#2014B-0404; PIs: David Schlegel and Arjun Dey), the Beijing-Arizona Sky Survey (BASS; NOAO Prop. ID \#2015A-0801; PIs: Zhou Xu and Xiaohui Fan), and the Mayall z-band Legacy Survey (MzLS; Prop. ID \#2016A-0453; PI: Arjun Dey). DECaLS, BASS and MzLS together include data obtained, respectively, at the Blanco telescope, Cerro Tololo Inter-American Observatory, NSF’s NOIRLab; the Bok telescope, Steward Observatory, University of Arizona; and the Mayall telescope, Kitt Peak National Observatory, NOIRLab. The Legacy Surveys project is honored to be permitted to conduct astronomical research on Iolkam Du’ag (Kitt Peak), a mountain with particular significance to the Tohono O’odham Nation.

NOIRLab is operated by the Association of Universities for Research in Astronomy (AURA) under a cooperative agreement with the National Science Foundation.

This project used data obtained with the Dark Energy Camera (DECam), which was constructed by the Dark Energy Survey (DES) collaboration. Funding for the DES Projects has been provided by the U.S. Department of Energy, the U.S. National Science Foundation, the Ministry of Science and Education of Spain, the Science and Technology Facilities Council of the United Kingdom, the Higher Education Funding Council for England, the National Center for Supercomputing Applications at the University of Illinois at Urbana-Champaign, the Kavli Institute of Cosmological Physics at the University of Chicago, Center for Cosmology and Astro-Particle Physics at the Ohio State University, the Mitchell Institute for Fundamental Physics and Astronomy at Texas A\&M University, Financiadora de Estudos e Projetos, Fundacao Carlos Chagas Filho de Amparo, Financiadora de Estudos e Projetos, Fundacao Carlos Chagas Filho de Amparo a Pesquisa do Estado do Rio de Janeiro, Conselho Nacional de Desenvolvimento Cientifico e Tecnologico and the Ministerio da Ciencia, Tecnologia e Inovacao, the Deutsche Forschungsgemeinschaft and the Collaborating Institutions in the Dark Energy Survey. The Collaborating Institutions are Argonne National Laboratory, the University of California at Santa Cruz, the University of Cambridge, Centro de Investigaciones Energeticas, Medioambientales y Tecnologicas-Madrid, the University of Chicago, University College London, the DES-Brazil Consortium, the University of Edinburgh, the Eidgenossische Technische Hochschule (ETH) Zurich, Fermi National Accelerator Laboratory, the University of Illinois at Urbana-Champaign, the Institut de Ciencies de l’Espai (IEEC/CSIC), the Institut de Fisica d’Altes Energies, Lawrence Berkeley National Laboratory, the Ludwig Maximilians Universitat Munchen and the associated Excellence Cluster Universe, the University of Michigan, NSF’s NOIRLab, the University of Nottingham, the Ohio State University, the University of Pennsylvania, the University of Portsmouth, SLAC National Accelerator Laboratory, Stanford University, the University of Sussex, and Texas A\&M University.

The Legacy Survey team makes use of data products from the Near-Earth Object Wide-field Infrared Survey Explorer (NEOWISE), which is a project of the Jet Propulsion Laboratory/California Institute of Technology. NEOWISE is funded by the National Aeronautics and Space Administration.

The Legacy Surveys imaging of the DESI footprint is supported by the Director, Office of Science, Office of High Energy Physics of the U.S. Department of Energy under Contract No. DE-AC02-05CH1123, by the National Energy Research Scientific Computing Center, a DOE Office of Science User Facility under the same contract; and by the U.S. National Science Foundation, Division of Astronomical Sciences under Contract No. AST-0950945 to NOAO.

\end{acknowledgments}

\bibliographystyle{aasjournal}
\bibliography{refs.bib}

\appendix

\section{Data Release} \label{datarelease}

The Dragonfly Data Access page can be found at \url{https://www.dragonflytelescope.org/data-access.html}. The page contains a list of publicly released fields grouped by their release date. Each field has its own download page with links to the appropriate papers to cite if you make use of the data in your work. Registration is required to facilitate communication of any updated data products, but is open to anyone.

The images for all DNGS targets were previously released in November 2018. Please note that fields are named for the primary target galaxy and some fields contain multiple galaxies of interest. The images for NGC 4565 were released in June 2020. The remaining 11 DEGS targets have now been released as well. 

In the FITS file header, \texttt{NFRAMES} indicates the number of single lens 600 second exposures that passed through all quality checks in the pipeline to be included in the final stacked image. \texttt{REFZP} is the median of the photometric zeropoints of individual frames contributing to the final stacked image. We recommend using \texttt{MEDIANZP} as the image zeropoint instead, which has been calibrated for the final stacked image. 

\section{Mock galaxy tests} \label{mockgalaxies}

{Although our target selection criteria included a proxy for Galactic cirrus avoidance, cirrus is present to varying degrees in all of our fields. In response, we have masked the worst of the cirrus by hand and handle smooth/slowly-varying cirrus as a component of the background. We expect that splitting each galaxy into four quadrants to better measure the local background should also help mitigate cirrus contamination and variability. At the faint surface brightness limits we have reached, one may ask how well the true light distribution can be recovered in the presence of cirrus. We therefore carried out mock galaxy tests in several of our fields spanning the variety of cirrus conditions seen in this survey.}

{A mock galaxy image was created using Imfit \citep{erwin15}, using model parameters similar to NGC~4565 but with an additional S\'ersic component for the stellar halo. This halo is more steeply declining than many of the observed stellar halos in our sample and may be more or less vulnerable to cirrus contamination or confusion. The mock galaxy was injected into several locations in a few fields with a variety of cirrus conditions, as well as one field where there appears to be little variation in the cirrus. These positions were selected manually rather than randomly in order to avoid bright stars, other galaxies, or extremely poor cirrus conditions. The chosen positions have similar local cirrus morphology as the DEGS target galaxies: usually either smooth, flat cirrus or a ``hole'' in the cirrus (an area with flat cirrus surrounded by more bright and often more structured cirrus). Although the cirrus morphology may be similar, the degree of contamination is generally greater than at the position of a typical DEGS target (in terms of cirrus brightness and/or the separation from structured cirrus).}

\begin{figure*}[tb]
     \centering
    \gridline{\fig{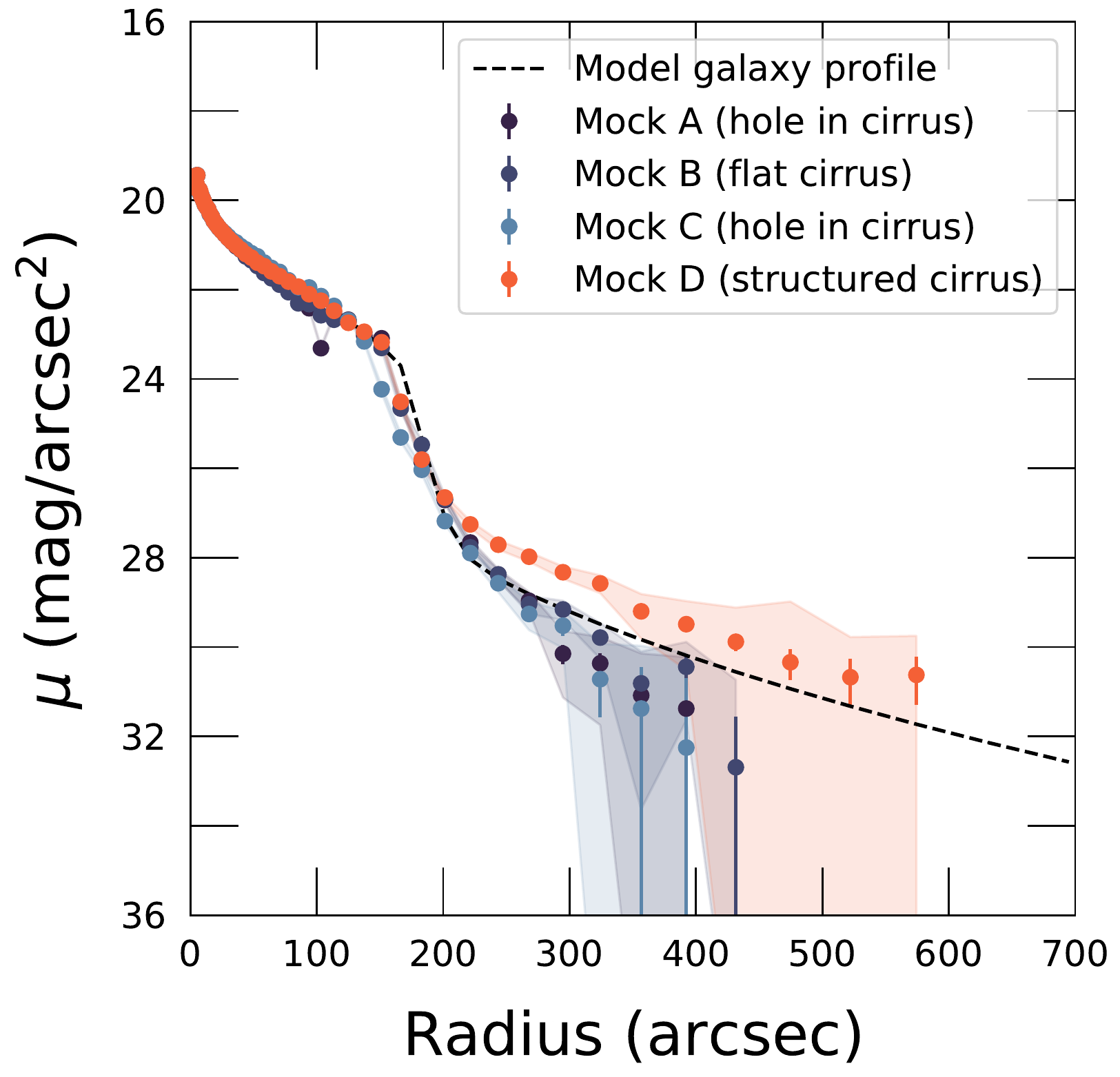}{0.4\textwidth}{(a) NGC 3044}
              \fig{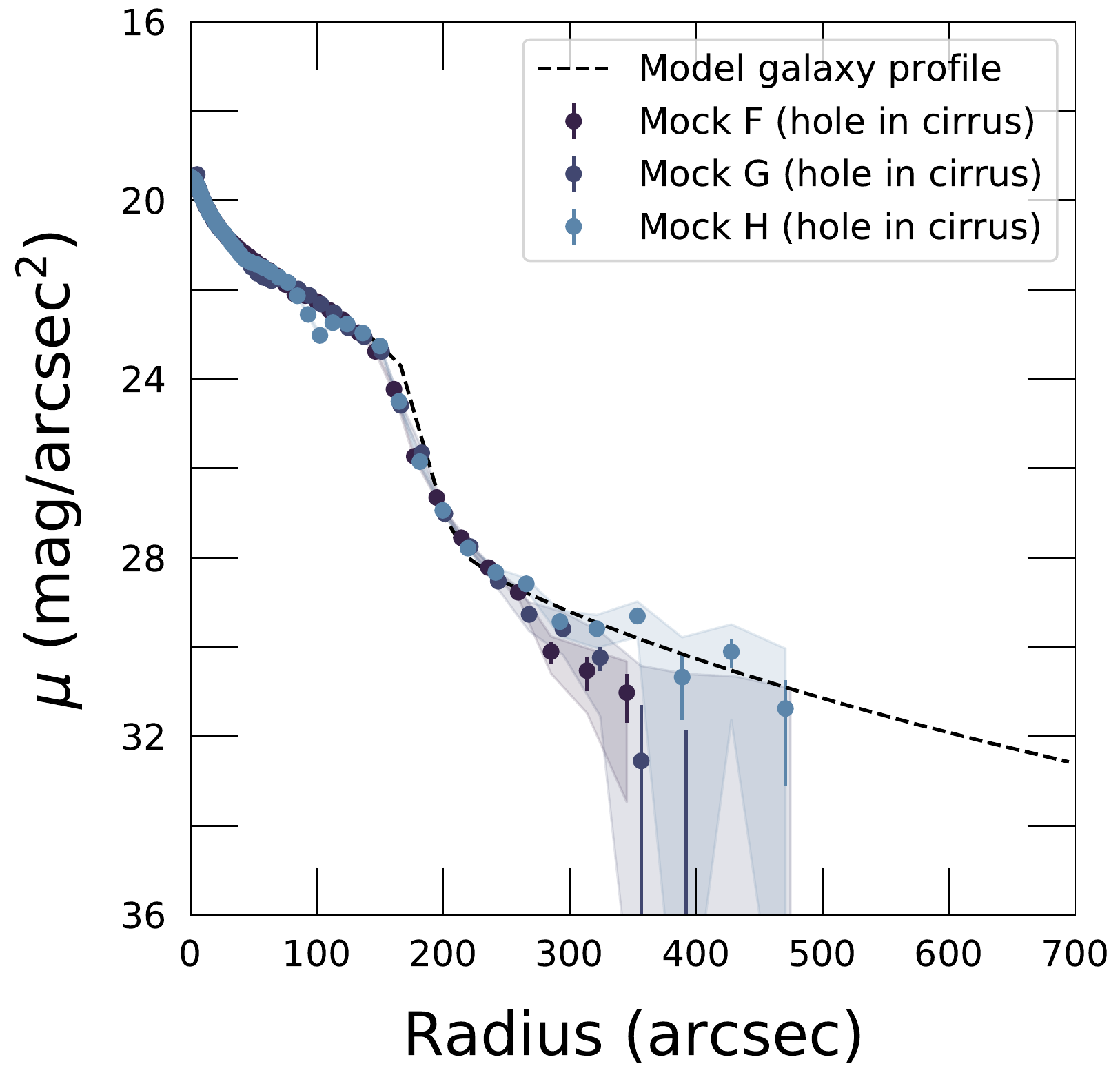}{0.4\textwidth}{(b) NGC 4330}
             }
    \gridline{\fig{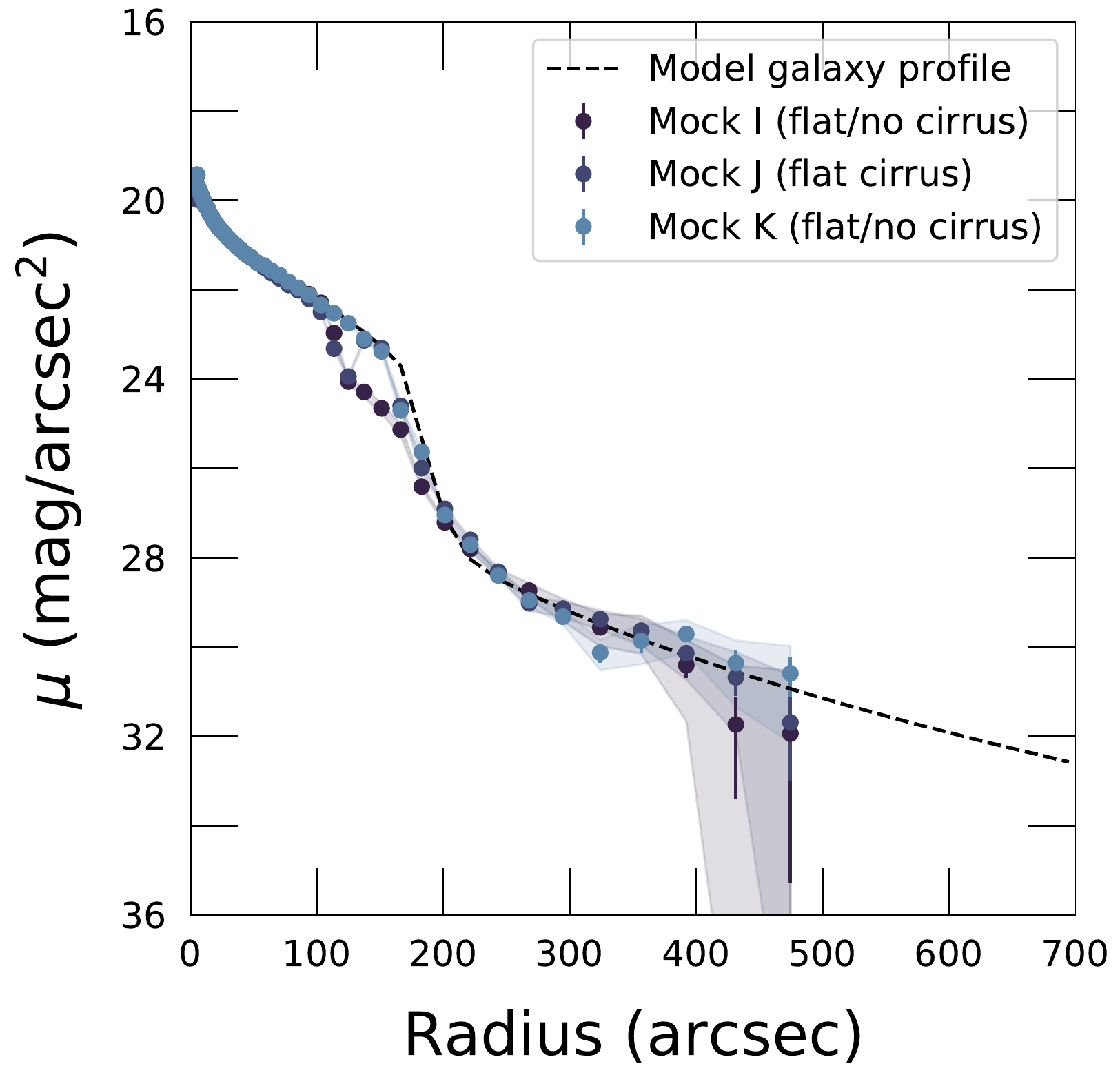}{0.4\textwidth}{(c) NGC 4013 (field with minimal cirrus structure)}
              \fig{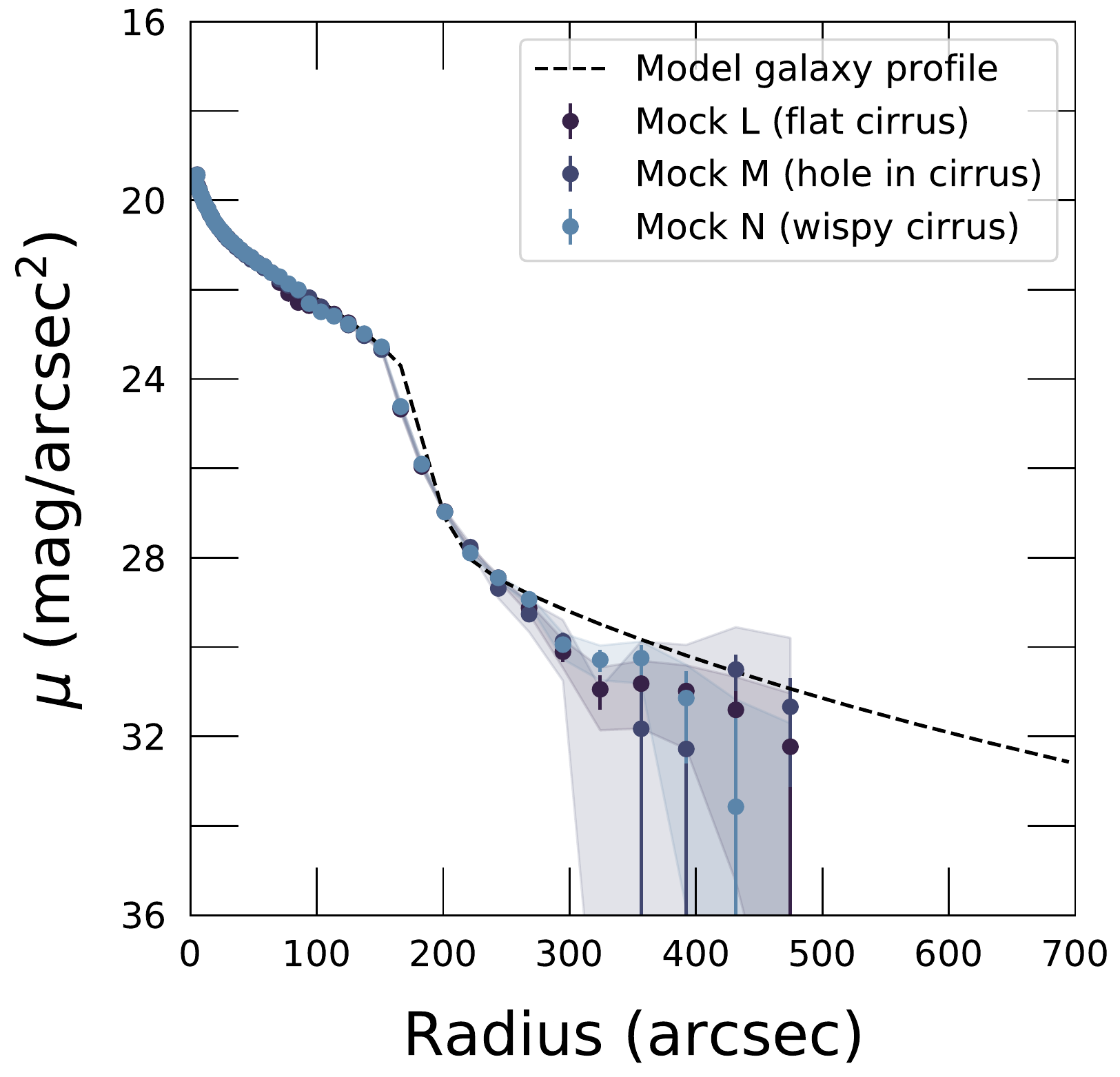}{0.4\textwidth}{(d) NGC 4010}
             }
        \caption{{Recovery of a mock galaxy's surface brightness profile at several different positions with varying cirrus conditions in four DEGS fields. The true profile (extracted directly from the model image) is shown as a dashed black line in each panel. ``Mock E'' is not shown; it was placed too close to the field boundary and suffered from greatly decreased signal-to-noise.}}
        \label{fig:mock_profiles}
\end{figure*}

{Surface brightness profiles for each injected mock galaxy were extracted following the procedure described in Section~\ref{profiles}, including the masking of bright or structured cirrus. The recovered profiles are shown in Figure~\ref{fig:mock_profiles}. In general, the outer profile where the stellar halo component dominates tends to be somewhat oversubtracted in the presence of cirrus. The local background is likely overestimated due to increasingly bright cirrus with increasing separation from the mock galaxy. For the test field where there appears to be little to no structured cirrus (NGC~4013, bottom left panel), the mock galaxy's surface brightness profile was well recovered in all three test locations. This case may be more reflective of the cirrus conditions at the center of DEGS fields than the oversubtracted profiles seen in other test fields, as cirrus interference is typically lowest and smoothest at the position of the target galaxy due to our cirrus avoidance criteria (low $F_{100\mu m}$ within 5~arcmin of the galaxy). }

{We included one worst case scenario (Mock D, orange points in the upper left panel) where the mock galaxy lies partly on a cirrus filament aligned with the major axis. This leads to  overestimation of light in the stellar halo region due to heavy contamination on one side of the midplane and undersubtraction of the background on the other side. This could be similar to the impact of cirrus on NGC~4634 but is not representative of any of the other DEGS galaxies.  }

{The systematic uncertainty in the background is not always great enough for the observed mock profiles to be consistent with the ``true'' mock profile. $\Delta \mathrm{bkgd}_{sys}$ is currently defined as the standard deviation of background values measured within the bins of the background annulus (Equation~C3). The full span of values  within the background annulus ($\mathrm{bkgd}_{max} - \mathrm{bkgd}_{min}$) may be a fairer metric to capture the difficulty in determining the most suitable background value. This is one of the biggest sources of systematic uncertainty in our study, along with that associated with the CMLR used to convert colour and surface brightness profiles into stellar mass density profiles. If we have underestimated $\Delta \mathrm{bkgd}_{sys}$, the uncertainty on our final measurements such as $f_{halo}$ would also be underestimated. } 

{These tests suggest that our profiles are unlikely to be artificially brightened at large radii due to cirrus contamination. If anything, oversubtraction is more likely to occur. However, there are a few important caveats to consider. We only considered one mock galaxy rather than a suite of galaxies with varying discs and halos. Some of our galaxies have larger angular sizes than the mock galaxy, and our ability to handle variations in the background due to cirrus may be worse at larger scales. We only injected mock galaxies into four of twelve DEGS fields, partly due to the lack of locations free from galaxies and bright stars.}

\section{Error calculations} \label{errcalcs}

{In this appendix, we summarize all of the calculations pertaining to the sources of error mentioned throughout Sections}~\ref{profiles}-\ref{halos}. {The corresponding steps in the analysis are briefly described alongside the calculations. }

{We used this identity when converting between linear quantities and their logarithms:}

\begin{equation}
\Delta\log x = \log e \frac{\Delta x}{x} .
\end{equation}

\subsection{Background measurement}
{The background was measured as the median of unmasked pixels within the radial bin (within a given galaxy quadrant) that was identified to have the flattest or sufficiently flat local slope. The random error of the background measurement for each galaxy quadrant was defined by the standard error of the median:}
\begin{equation}
\Delta\textrm{bkgd} = \frac{\sigma(\textrm{pixels})}{\sqrt{\frac{\pi}{2}N}}.
\end{equation}

{The standard deviation of the median background measurements from all radial bins was adopted as the systematic uncertainty of the background measurement:}
\begin{equation}
\Delta\textrm{bkgd}_\textrm{sys} = \sigma(\textrm{bkgd}_\textrm{annulus}).
\end{equation}

\subsection{Surface brightness profiles}

{For each band $\Lambda$ (either $g$ or $r$) we extracted four quadrant profiles $f_{\Lambda,i}$ with random measurement errors $\Delta f_{\Lambda,i}$, internally calculated by \texttt{photutils.isophote.ellipse}. The background was measured separately for each quadrant. The four flux profiles were then averaged, weighted at each radius by number of unmasked pixels, $n_{\Lambda,i}$). The error of combined flux profile is:}
\begin{equation}
\Delta f_\Lambda = \frac{\sqrt{\sum\limits_{i=1}^{4} {n}_{\Lambda,i}^2 (\Delta {f}_{\Lambda,i}^2 + \Delta\textrm{bkgd}_{\Lambda,i}^2 + \Delta\textrm{bkgd}_{\Lambda,i, \textrm{sys}}^2) }}{\sum\limits_{i=1}^4 {n}_{\Lambda,i}}.
\end{equation}

{The flux profiles were then corrected for scattered light by fitting 2D parametric models to the disc, extracting profiles from model images with and without PSF convolution, and then subtracting the difference between the PSF-convolved and unconvolved profiles from the observed flux profiles. Following} \cite{szomoru10, szomoru12}, {we neglected error associated with the goodness of fit of the 2D model and uncertainty in individual parameters because these errors would be difficult to determine while also being small compared to other sources of error (such as the systematic error of the background measurement). We did consider the random measurement errors calculated by} \texttt{photutils.isophote.ellipse} {while extracting profiles from the model images because it was very straightforward to incorporate into our error analysis. The error in $f_{\textrm{model,PSF},\Lambda}$ and $f_{\textrm{model,noPSF},\Lambda}$ was combined in quadrature with the result of Equation~B4:}

\begin{equation}
    \Delta f_{\Lambda,\textrm{corr}} = \sqrt{\Delta f_{\Lambda}^2 + \Delta f_{\textrm{model,PSF},\Lambda}^2 + \Delta f_{\textrm{model,noPSF},\Lambda}^2}.
\end{equation}

{Next, the corrected flux profiles were converted into surface brightness profiles. We neglect uncertainty in the zeropoint. The error of the flux profiles was converted to magnitudes: }

\begin{equation}
    \Delta \mu_\Lambda = 2.5 \log e \frac{\Delta f_{\Lambda,\textrm{corr}}}{f_{\Lambda,\textrm{corr}}}.
\end{equation}

{The error in the $g-r$ colour profile was obtained by combining the individual errors of the $g-$ and $r-$band profiles ($\mu_g$ and $\mu_r$) in quadrature:}

\begin{equation}
    \Delta (g-r) = \sqrt{\Delta \mu_g^2 + \Delta \mu_r^2}.
\end{equation}

{Where the noisy outer colour profile was replaced with a fixed value (the average over the outermost five reliable points), we estimated the uncertainty by combining the standard deviation of the outermost five reliable points with the mean error of those points:}

\begin{equation}
    \Delta (g-r)_\textrm{fixed} = \sqrt{\sigma(g-r)_\textrm{outer}^2 + \textrm{avg}(\Delta (g-r)_\textrm{outer})^2}.
\end{equation}

\subsection{Surface stellar mass density profiles}

{The equation used to calculate $\Sigma_*$ from the colour and surface brightness profiles is worth including here to provide context to our subsequent error calculations:}

\begin{equation}
    \begin{aligned}
        \log \Sigma_* & = \log \Big(\frac{\textrm{M}_*}{\textrm{L}} \times \textrm{L} \times \frac{1}{\Omega^2}\Big) \\
                      & = \log \Big(\frac{\textrm{M}_*}{\textrm{L}}\Big)_g + 0.4(M_{g,\odot} - M_g) + \log\Big(\frac{1}{\Omega^2}\Big) \\
                      & = \log \Big(\frac{\textrm{M}_*}{\textrm{L}}\Big)_g - 0.4(\mu_g - [M-m]) + 0.4(5.11\: \textrm{mag}) + \log\Big(\frac{1}{\Omega^2}\Big) \\
                      & = 2.029(g-r) - 0.984 - 0.4(\mu_g - [M-m]) + 2.044 + \log\Big(\frac{1}{\Omega^2}\Big).
    \end{aligned}
\end{equation}

{The solar absolute magnitude ($M_{g,\odot}$) was obtained from} \cite{willmer18}. {In addition to the propagated uncertainty of $(g-r)$ and $\mu_g$, we included the error of the CMLR itself, $\Delta \textrm{CMLR} = 0.15$ and $\Delta \textrm{CMLR}_\textrm{sys} = 0.3$} \citep{roediger15}. {We combined these sources of error in quadrature:}

\begin{equation}
    \Delta \log \Sigma = \sqrt{\Big(0.4\cdot\Delta \mu_{g,\textrm{corr}}\Big)^2 + \Big( 2.029 \cdot\Delta (g-r)\Big)^2 + \Delta \textrm{CMLR}^2 + \Delta \textrm{CMLR}_\textrm{sys}^2}.
\end{equation}

\subsection{Outer masses, disc masses, and stellar halo mass fractions}

{The $\Sigma_*$ profiles were integrated from radii corresponding to various empirical definitions of the stellar halo-dominated region to the outer extent of the profile to determine the stellar mass in these regions. The area reported by} \texttt{photutils.isophote.ellipse}  {for each isophote, $a_r$, was rounded to nearest pixel so we adopted $\Delta a_r = 1\:\textrm{pixel} \: \forall \: r$. The error in integrating the $\Sigma_*$ profile from a radius of $r_i$ to $r_f$ was calculated as: }

\begin{equation}
    \Delta \log \textrm{M}_{*,\textrm{outer}} = \frac{\sqrt{\sum\limits_{r=r_i}^{r_f}(a_r\Sigma_{*,r})^2\Big( (\frac{\Delta a_r}{a_r})^2 + (\frac{1}{\log e} \Delta \log \Sigma_{*,r})^2 \Big)}}{\sum\limits_{r=r_i}^{r_f}a_r \Sigma_{*,r}}
\end{equation}

{A broken exponential disc was fitted to the bright inner regions of the $\Sigma_*$ profile in order to estimate the stellar mass contributed by the disc in these outer regions. The uncertainty in the disc mass contribution (found by integrating $\Sigma_{*,\textrm{disc}}$ from $r_i$ to $r_f$) was estimated by using} \textrm{emcee} {to generate a large number of independent samples from the posterior probability density function of the disc model parameters. Each sample (or set of possible disc model parameters) was used to calculate the stellar mass of the disc in the outer region, $\textrm{M}_{*,\textrm{disc}}$. The $16-84$ interquartile range of the ensemble of $\textrm{M}_{*,\textrm{disc}}$ was used to estimate the $1\sigma$ upper and lower limits for the $\textrm{M}_{*,\textrm{disc}}$ calculated using the maximum likelihood disc model parameters. }

{The stellar halo masses, $\textrm{M}_{*,\textrm{halo}}$, were obtained by subtracting $\textrm{M}_{*,\textrm{disc}}$ from $\textrm{M}_{*,\textrm{outer}}$. Upper and lower error bounds for $\textrm{M}_{*,\textrm{halo}}$ and $f_{\textrm{halo}}$ were calculated according to the equations below. {The \SfourG{} total stellar masses, $\textrm{M}_*$, were assumed to have a combined uncertainty of 0.2 dex \citep{querejeta15}.} We note that as a larger value of $\textrm{M}_{*,\textrm{disc}}$ would result in a lower value of $\textrm{M}_{*,\textrm{halo}}$, $\Delta \textrm{M}_{*,\textrm{disc,upper}}$ was used to calculate $\Delta \textrm{M}_{*,\textrm{halo,lower}}$ while $\Delta \textrm{M}_{*,\textrm{disc,lower}}$ was used to calculate $\Delta \textrm{M}_{*,\textrm{halo,upper}}$. }

\begin{equation}
    \Delta \log \textrm{M}_{*,\textrm{halo}} = \frac{\sqrt{\Big(\textrm{M}_{*,\textrm{outer}}\cdot \Delta \log \textrm{M}_{*,\textrm{outer}}\Big)^2 + \Big(\textrm{M}_{*,\textrm{disc}}\cdot \Delta \log \textrm{M}_{*,\textrm{disc}}\Big)^2}}{\Big(\textrm{M}_{*,\textrm{outer}} - \textrm{M}_{*,\textrm{disc}}\Big)}
\end{equation}

\begin{equation}
    \Delta \log f_\textrm{halo} = \sqrt{\Delta \log \textrm{M}_{*,\textrm{halo}}^2 + \Delta \log \textrm{M}_*^2}
\end{equation}

\medskip

\section{Additional figures} \label{fig_appendix}

The following pages contain figures gathering all of the final profiles and images for each DEGS galaxy. {The galaxies are sorted by increasing NGC number.} All panels on the left side of a figure share a common horizontal scale. The extent of the top right and middle right panels is shown with a white box in the lower right panel. \textit{Top left:} $g$- and $r$-band surface brightness profiles. Error bars on points depict random error. The shaded envelopes depict upper and lower limits considering random error and systematic uncertainty in the sky background measurement. \textit{Middle left:} $g-r$ colour profile. As with the surface brightness profiles above, error bars on points depict random error and the shaded envelope depicts upper and lower limits considering random error and systematic uncertainty. The grey vertical lines indicate the region where the average outer colour is determined. Beyond the second grey vertical line, the average outer colour is favoured over the raw observed colours due to large error and/or gaps in either the $g$- or $r$-band profile. The grey points show this average outer colour. \textit{Bottom left:} Stellar mass surface density profile, in units of $\mathrm{M}_\odot/\mathrm{kpc}^2$. As with the profiles above, error bars on points depict random error and the shaded envelope depicts upper and lower limits considering random error and systematic uncertainty. The vertical grey line is included to mark the radius beyond which the average outer colour is used over the raw observed colour. \textit{Top right:} A central cutout of the $r$-band source-subtracted image {(as produced using the} \texttt{mrf} {package; see Section~}\ref{analysis} {for further details}), centered on the target galaxy. Masked areas are shown in white. {The $r-$band surface brightness profile shown in the top left panel is extracted from this image along elliptical isophotes fitted using Photutils} \citep{photutils}. {The ellipse fitting procedure is described in Section~}\ref{profiles}. {A subset of the best fit elliptical isophotes are shown in black.} The outermost {ellipses} are labelled with their approximate semi-major axis length. \textit{Middle right:} A false colour composite matching the footprint of the central cutout in the top right panel. {We generated the false colour composite using the} \cite{lupton04} {method as implemented in} \texttt{astropy.visualization.make\_lupton\_rgb()}{, with $Q=7$ and} \texttt{stretch}$=10$. {The three image colour channels are as follows: $R=r$, $G=0.5(r+g)$, $B=1.2g$.} \textit{Bottom right:} A false colour composite of the {full Dragonfly field}, centered on the target galaxy. The white box shows the extent of the central cutouts shown above. A scale bar indicating one degree is included.

\begin{figure}[hbp]
\begin{centering}
\includegraphics[width=0.95\textwidth]{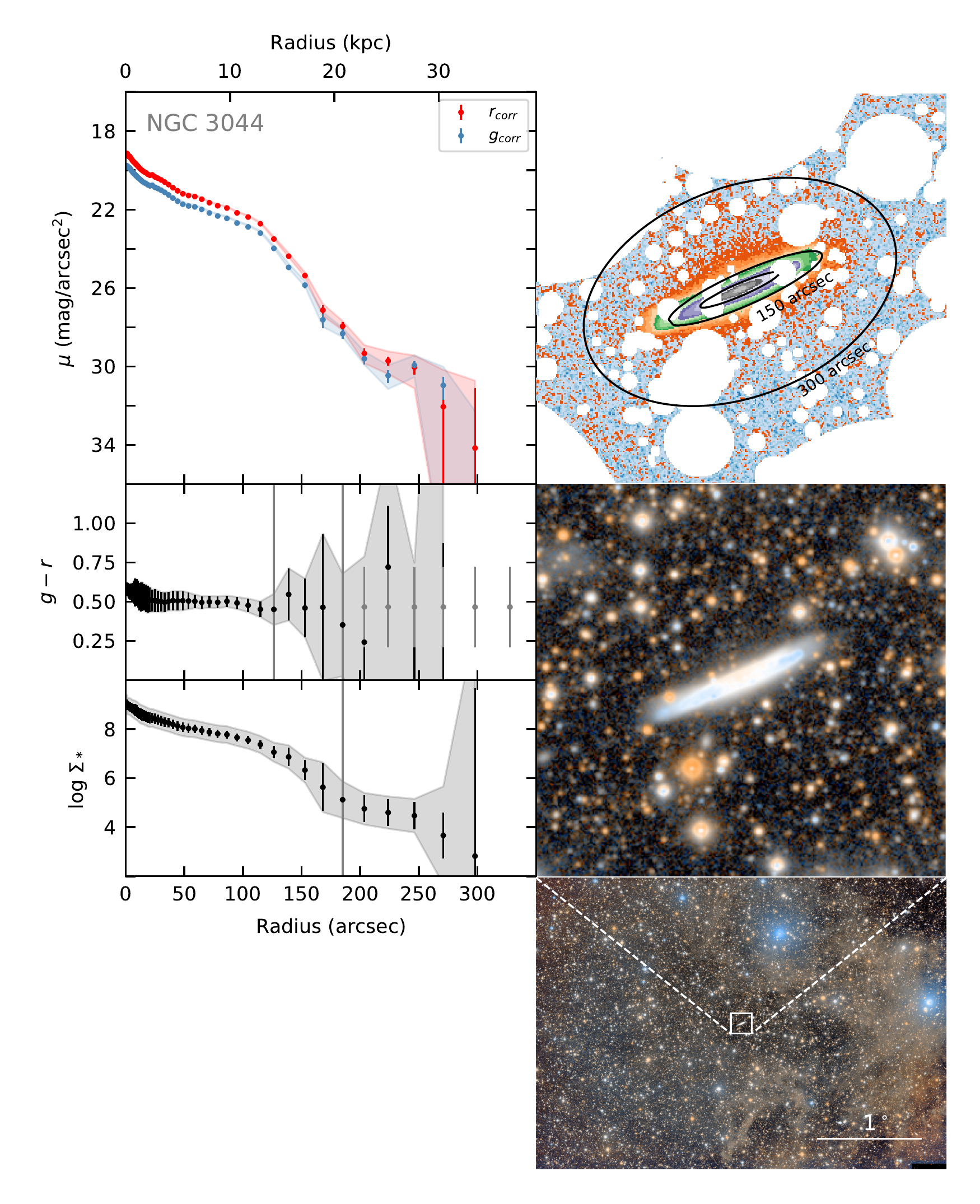}
\caption{A detailed summary figure for DEGS galaxy NGC 3044. Please refer to the beginning of Appendix~\ref{fig_appendix} for a description of this figure.}
\label{fig:NGC3044}
\end{centering}
\end{figure}

\begin{figure}[tbp]
\begin{centering}
\includegraphics[width=0.95\textwidth]{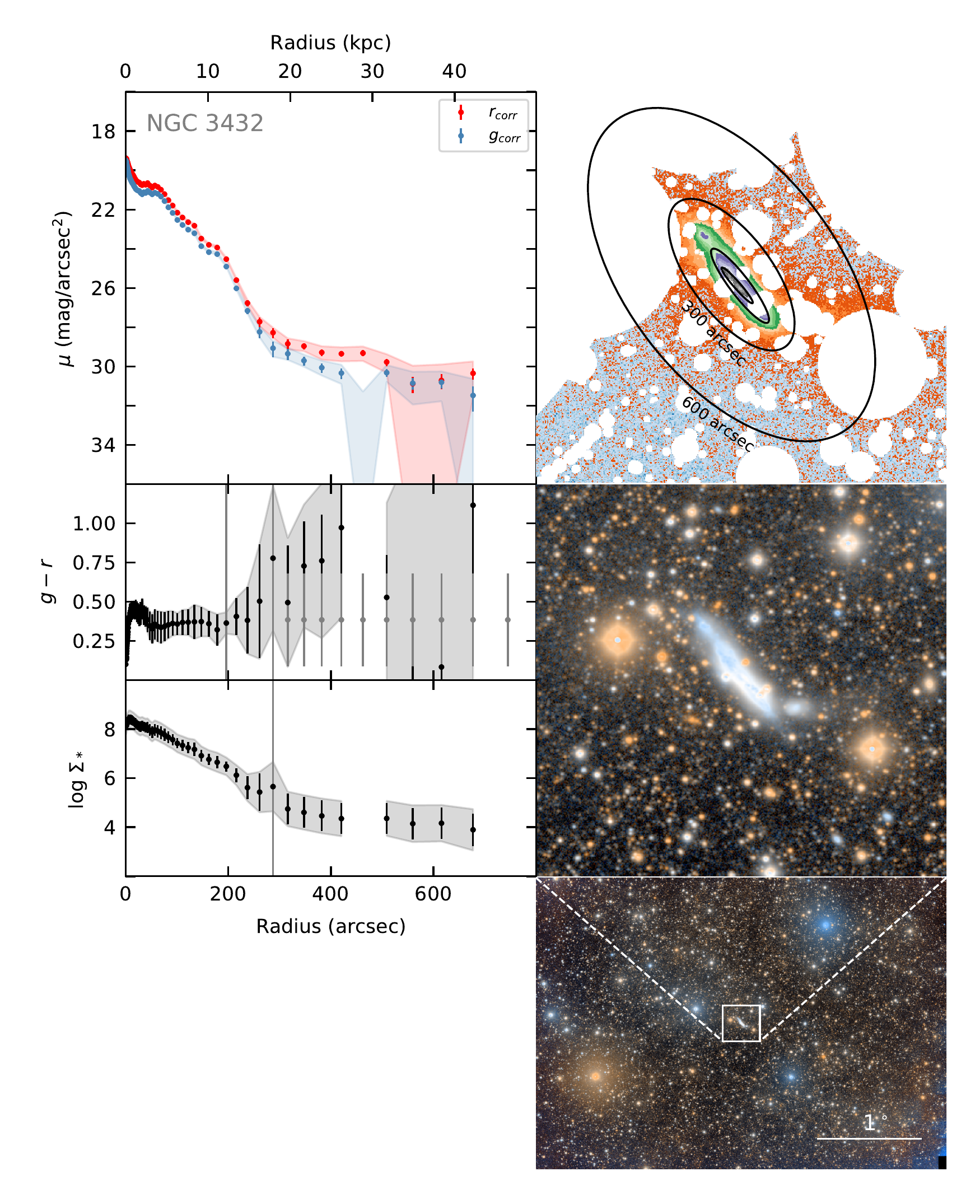}
\caption{Same as Figure~\ref{fig:NGC3044} for NGC 3432.}
\end{centering}
\end{figure}

\begin{figure}[tbp]
\begin{centering}
\includegraphics[width=0.95\textwidth]{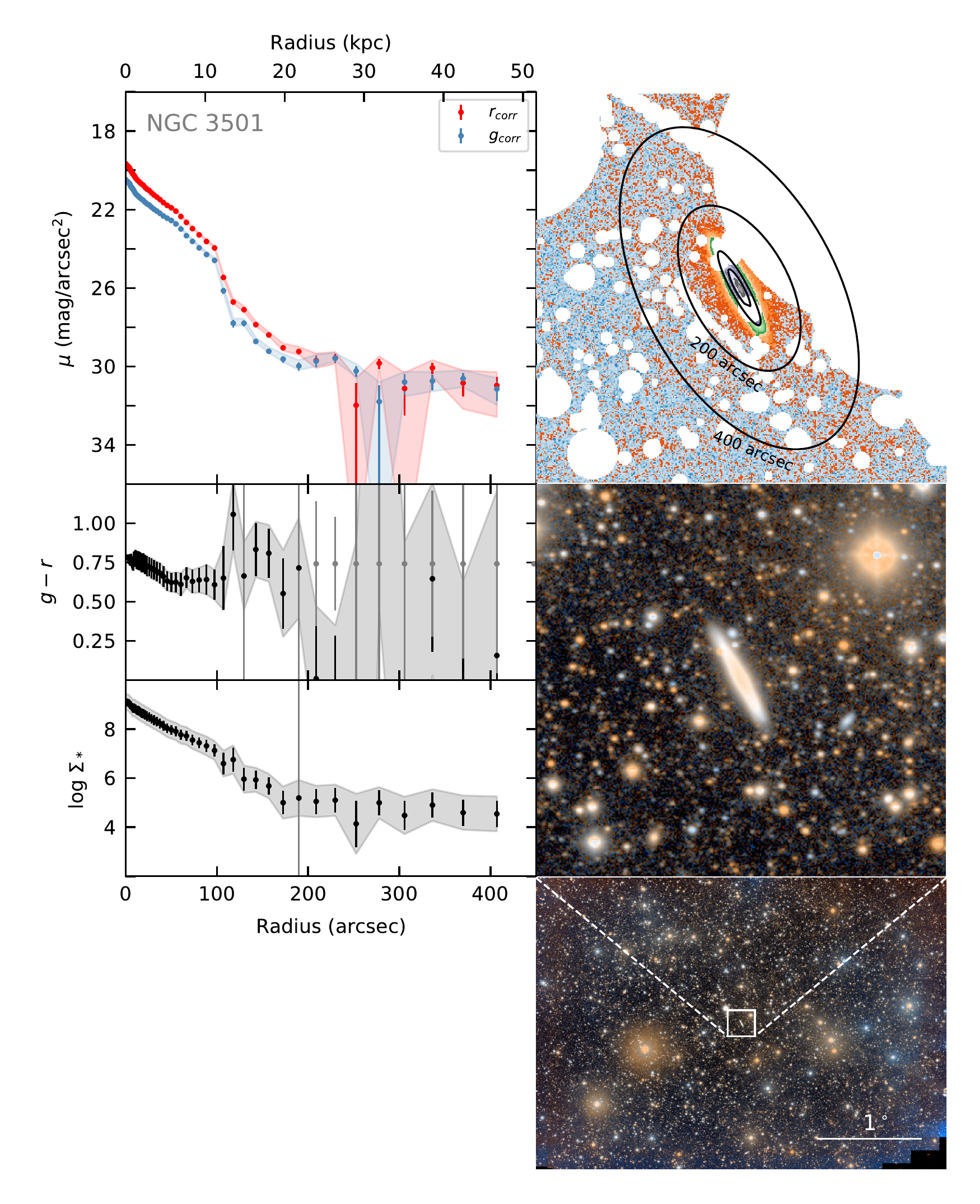}
\caption{Same as Figure~\ref{fig:NGC3044} for NGC 3501.}
\end{centering}
\end{figure}

\begin{figure}[tbp]
\begin{centering}
\includegraphics[width=0.95\textwidth]{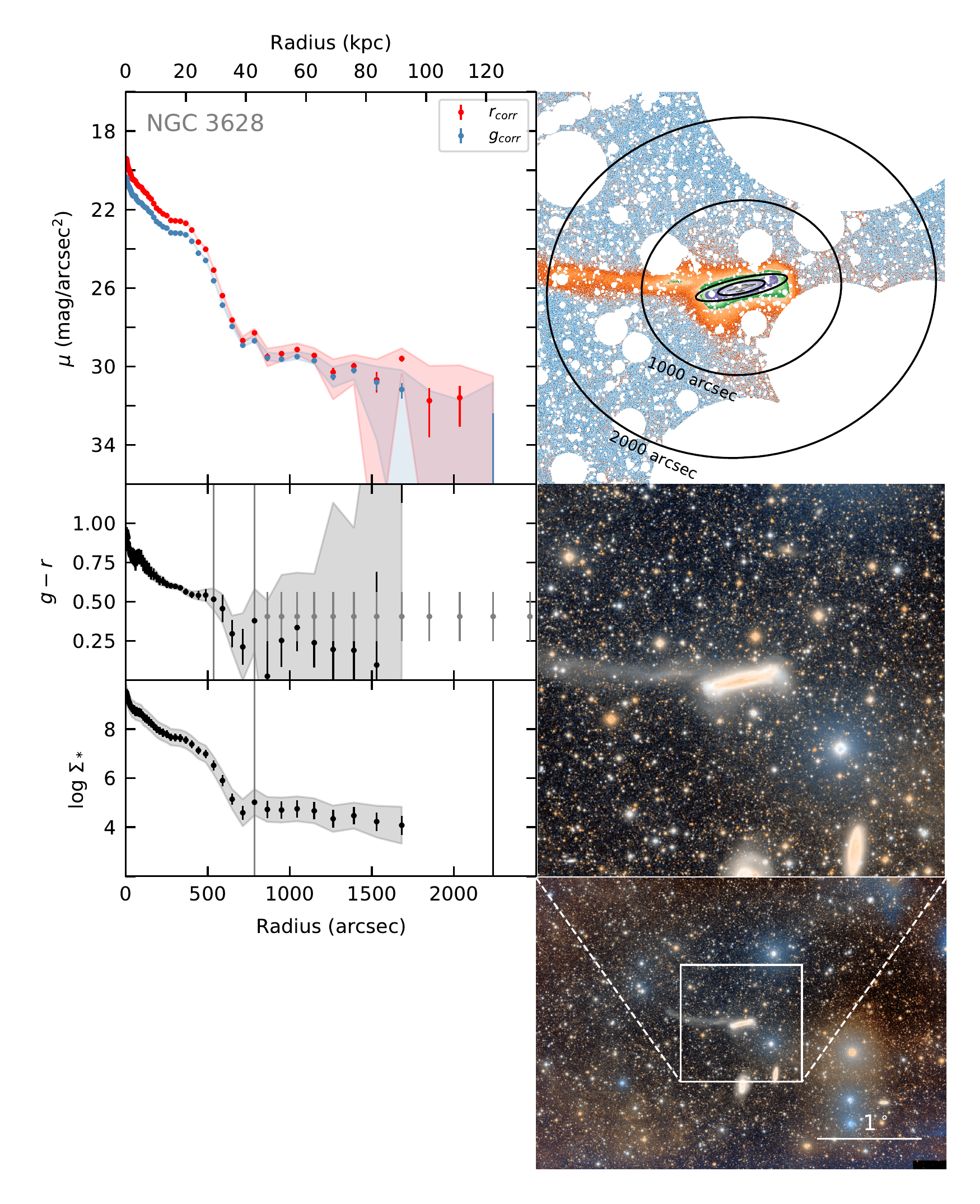}
\caption{Same as Figure~\ref{fig:NGC3044} for NGC 3628.}
\end{centering}
\end{figure}

\begin{figure}[tbp]
\begin{centering}
\includegraphics[width=0.95\textwidth]{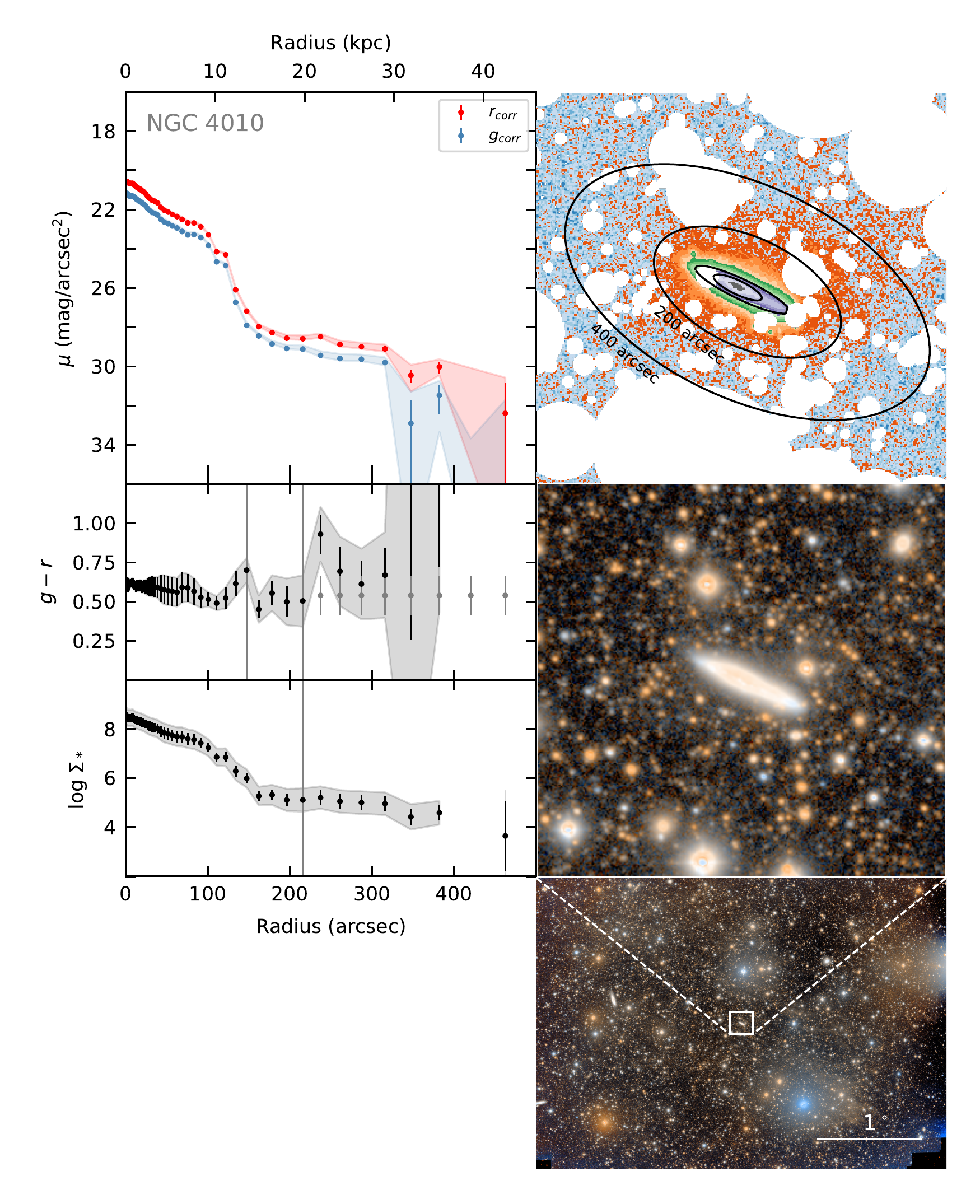}
\caption{Same as Figure~\ref{fig:NGC3044} for NGC 4010.}
\end{centering}
\end{figure}

\begin{figure}[tbp]
\begin{centering}
\includegraphics[width=0.95\textwidth]{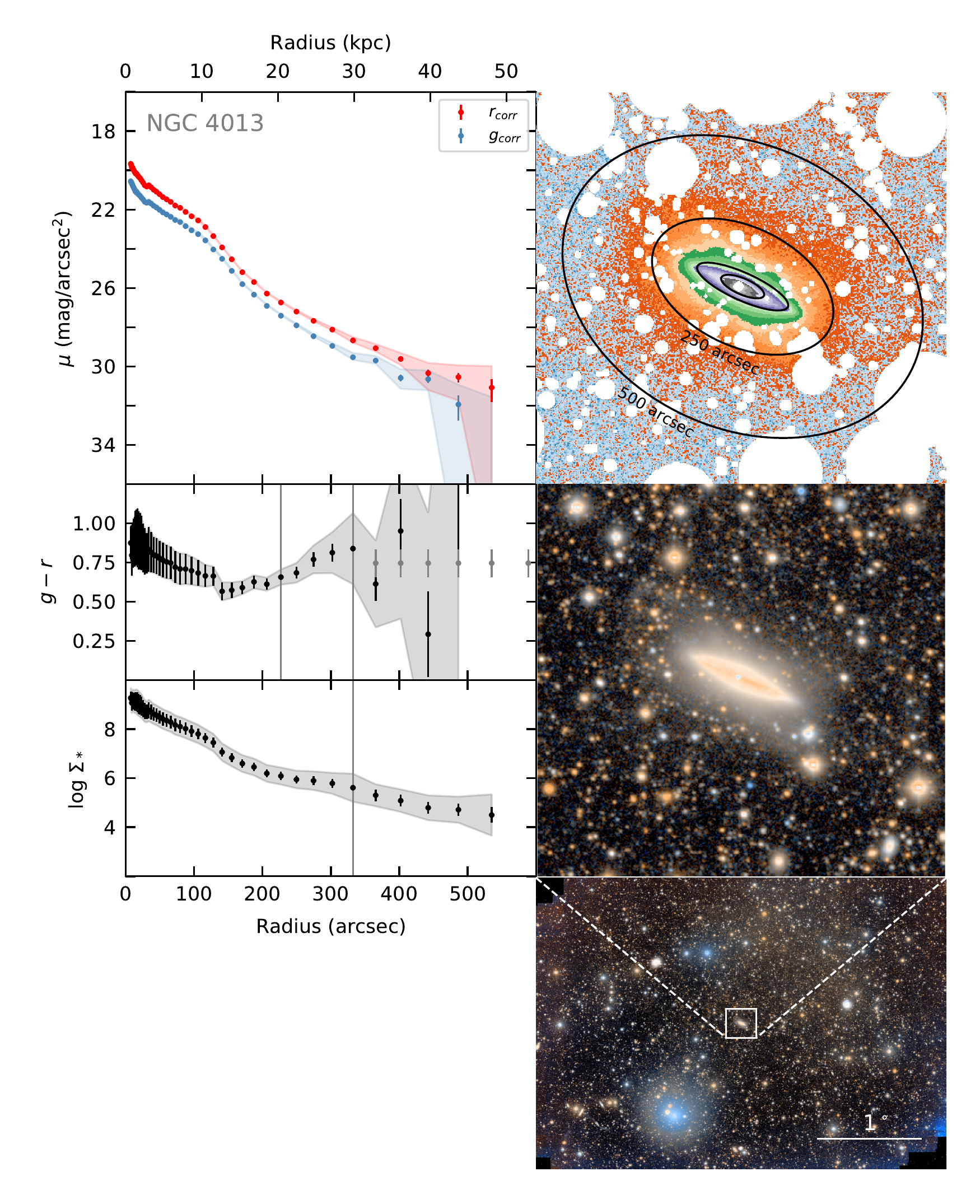}
\caption{Same as Figure~\ref{fig:NGC3044} for NGC 4013.}
\end{centering}
\end{figure}

\begin{figure}[tbp]
\begin{centering}
\includegraphics[width=0.95\textwidth]{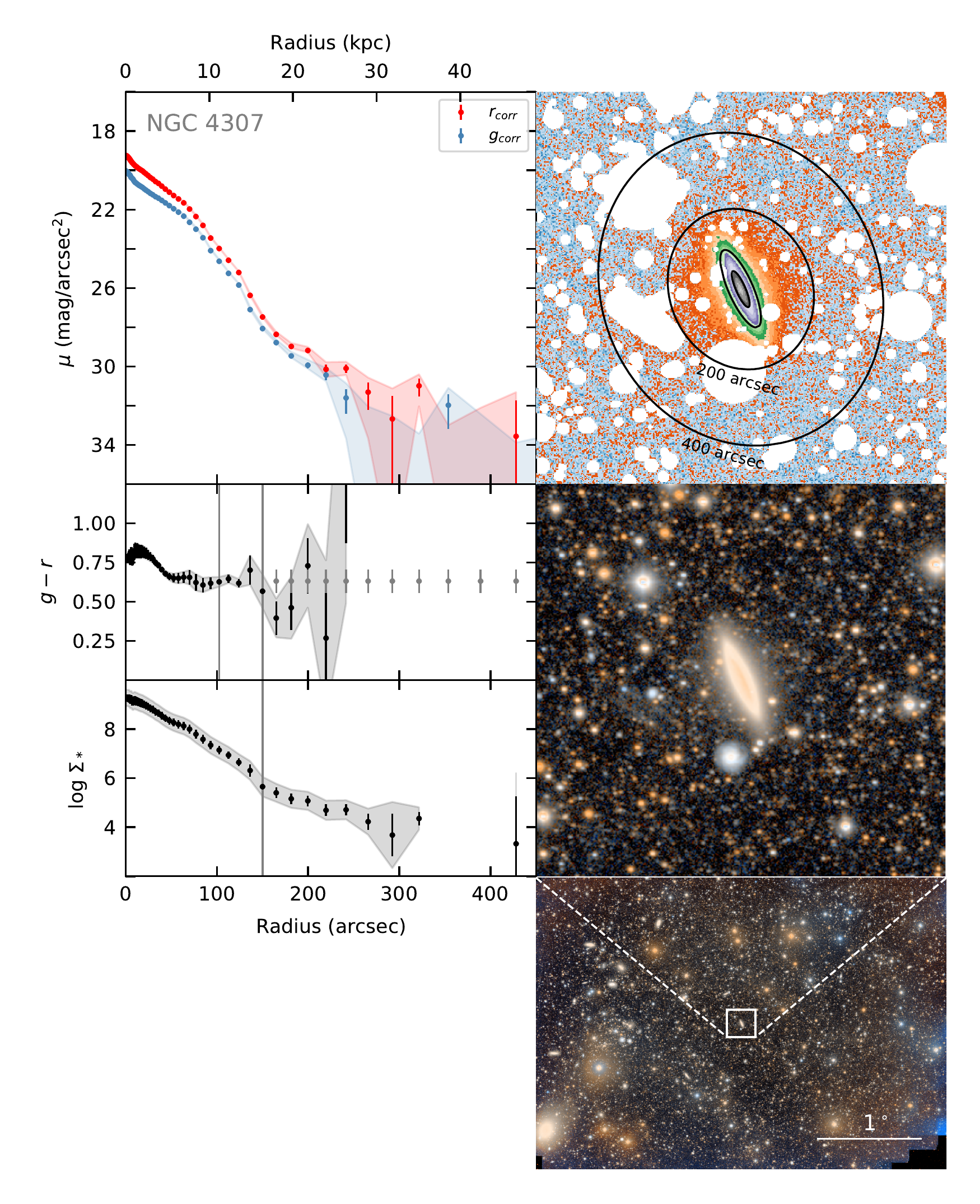}
\caption{Same as Figure~\ref{fig:NGC3044} for NGC 4307.}
\end{centering}
\end{figure}

\begin{figure}[tbp]
\begin{centering}
\includegraphics[width=0.95\textwidth]{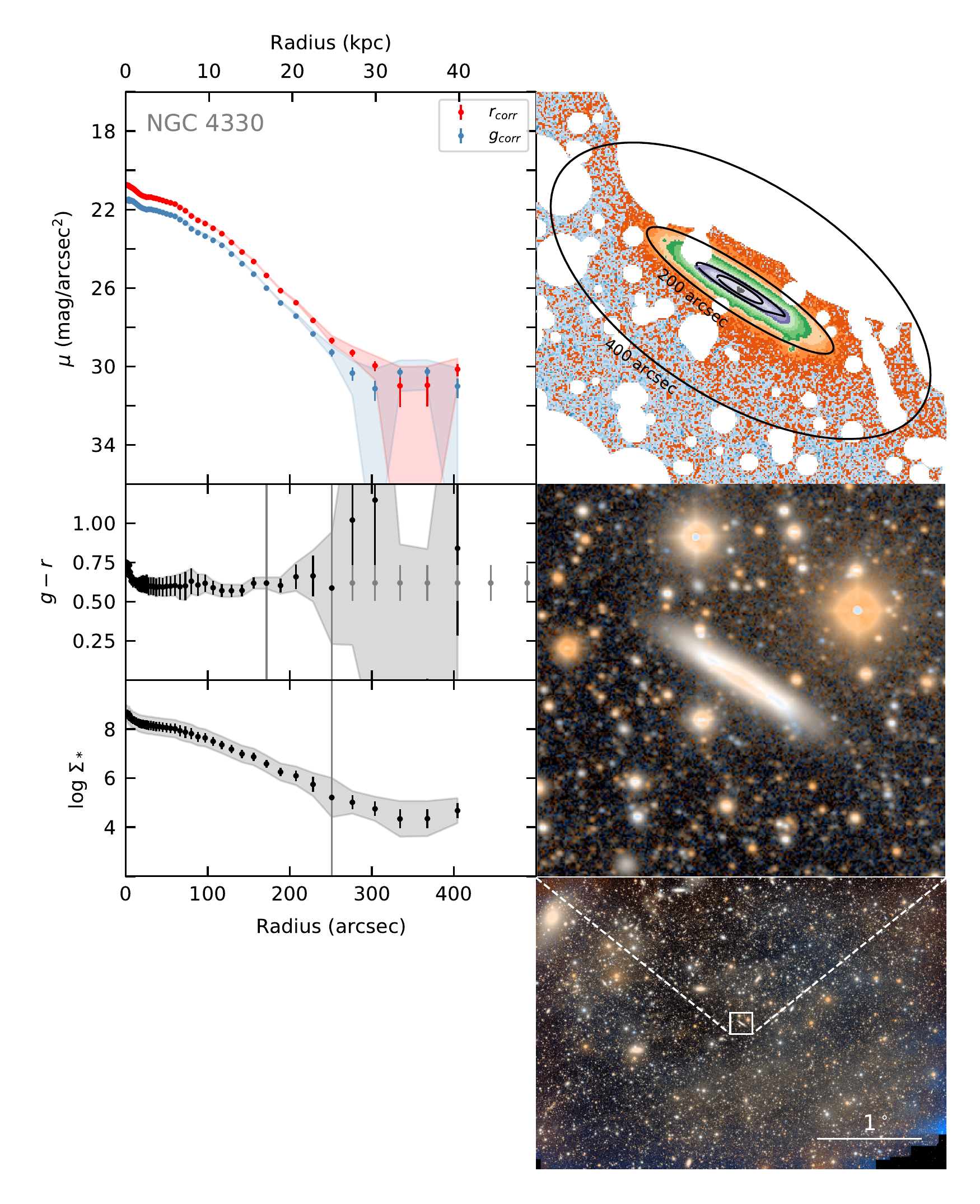}
\caption{Same as Figure~\ref{fig:NGC3044} for NGC 4330.}
\end{centering}
\end{figure}

\begin{figure}[tbp]
\begin{centering}
\includegraphics[width=0.95\textwidth]{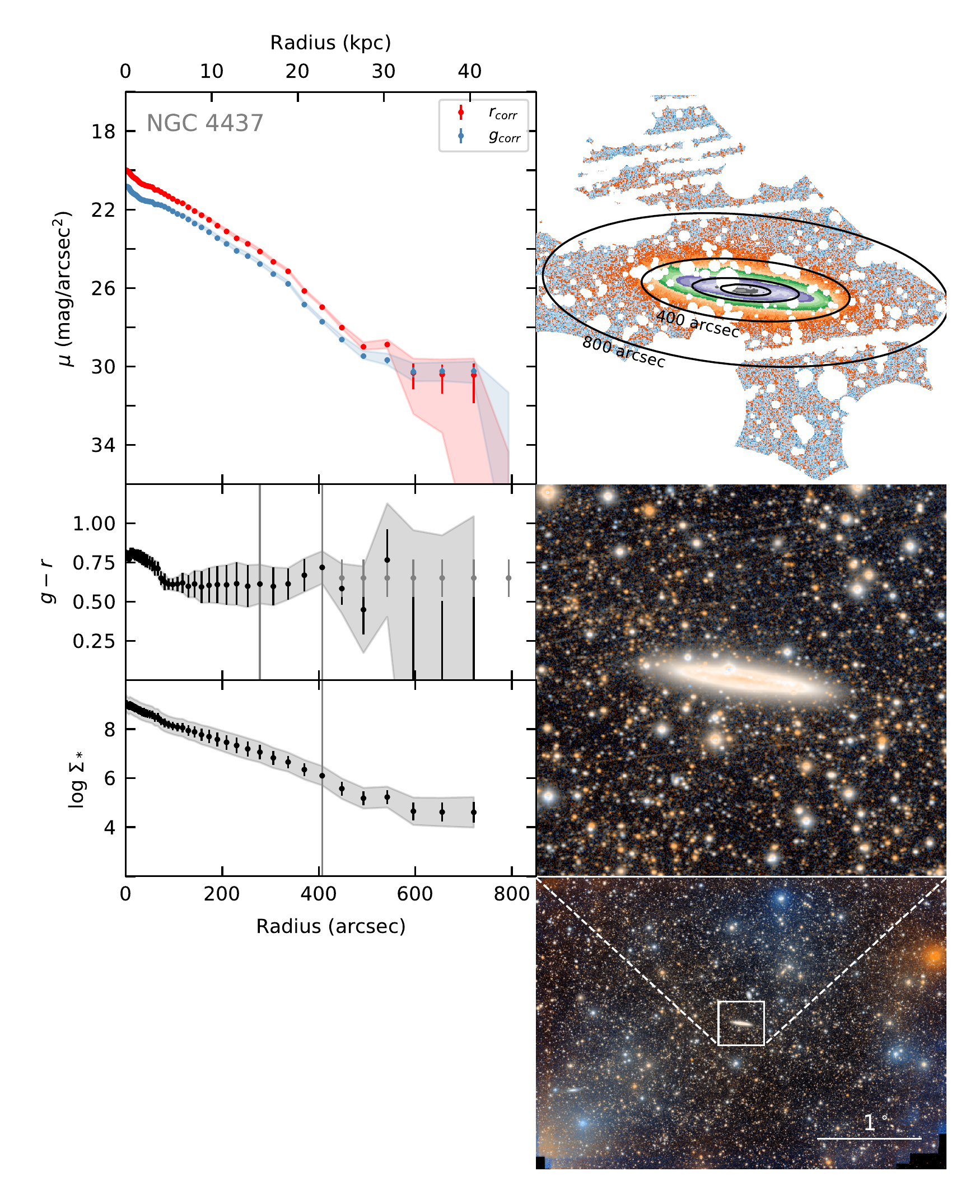}
\caption{Same as Figure~\ref{fig:NGC3044} for NGC 4437.}
\end{centering}
\end{figure}

\begin{figure}[tbp]
\begin{centering}
\includegraphics[width=0.95\textwidth]{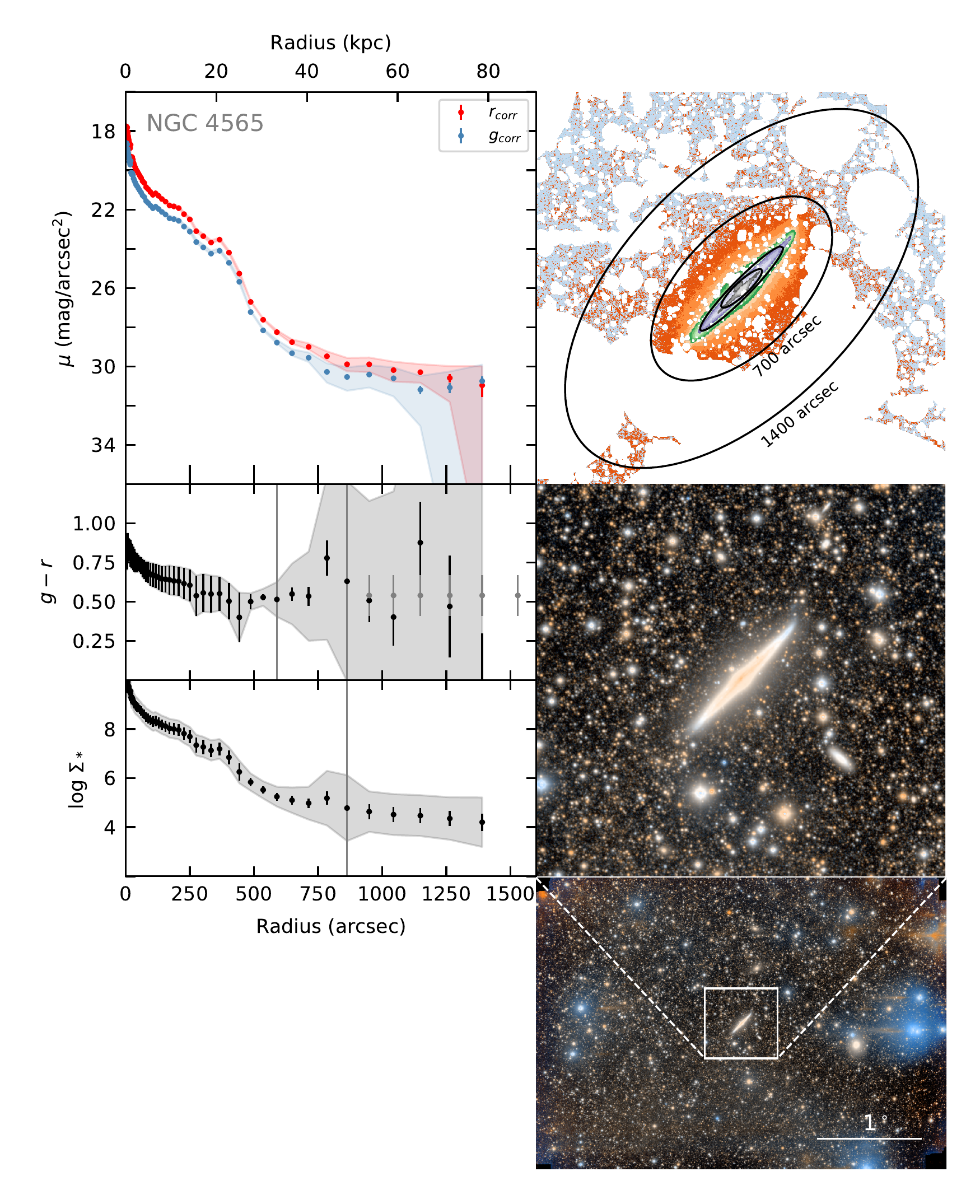}
\caption{Same as Figure~\ref{fig:NGC3044} for NGC 4565.}
\end{centering}
\end{figure}

\begin{figure}[tbp]
\begin{centering}
\includegraphics[width=0.95\textwidth]{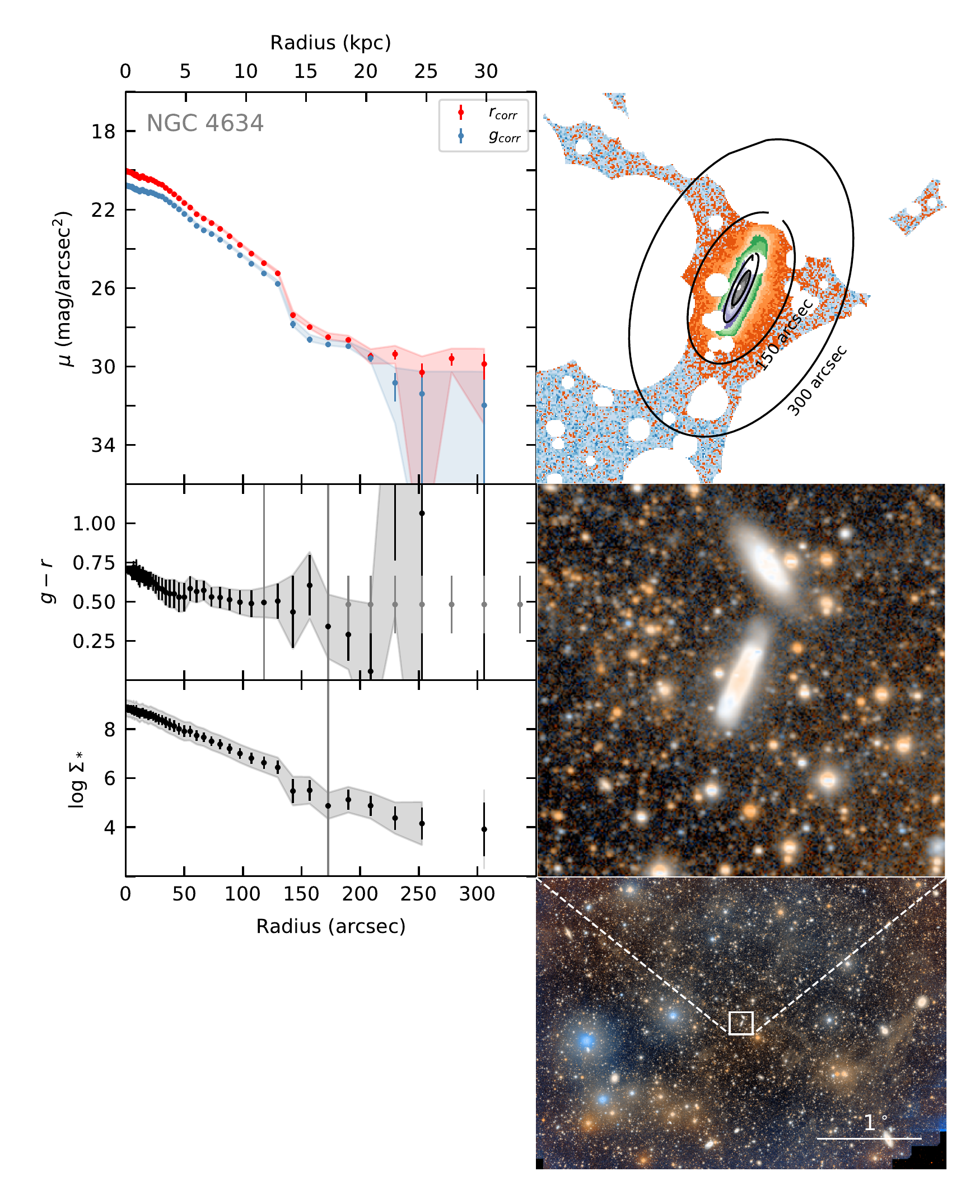}
\caption{Same as Figure~\ref{fig:NGC3044} for NGC 4634.}
\end{centering}
\end{figure}

\begin{figure}[tbp]
\begin{centering}
\includegraphics[width=0.95\textwidth]{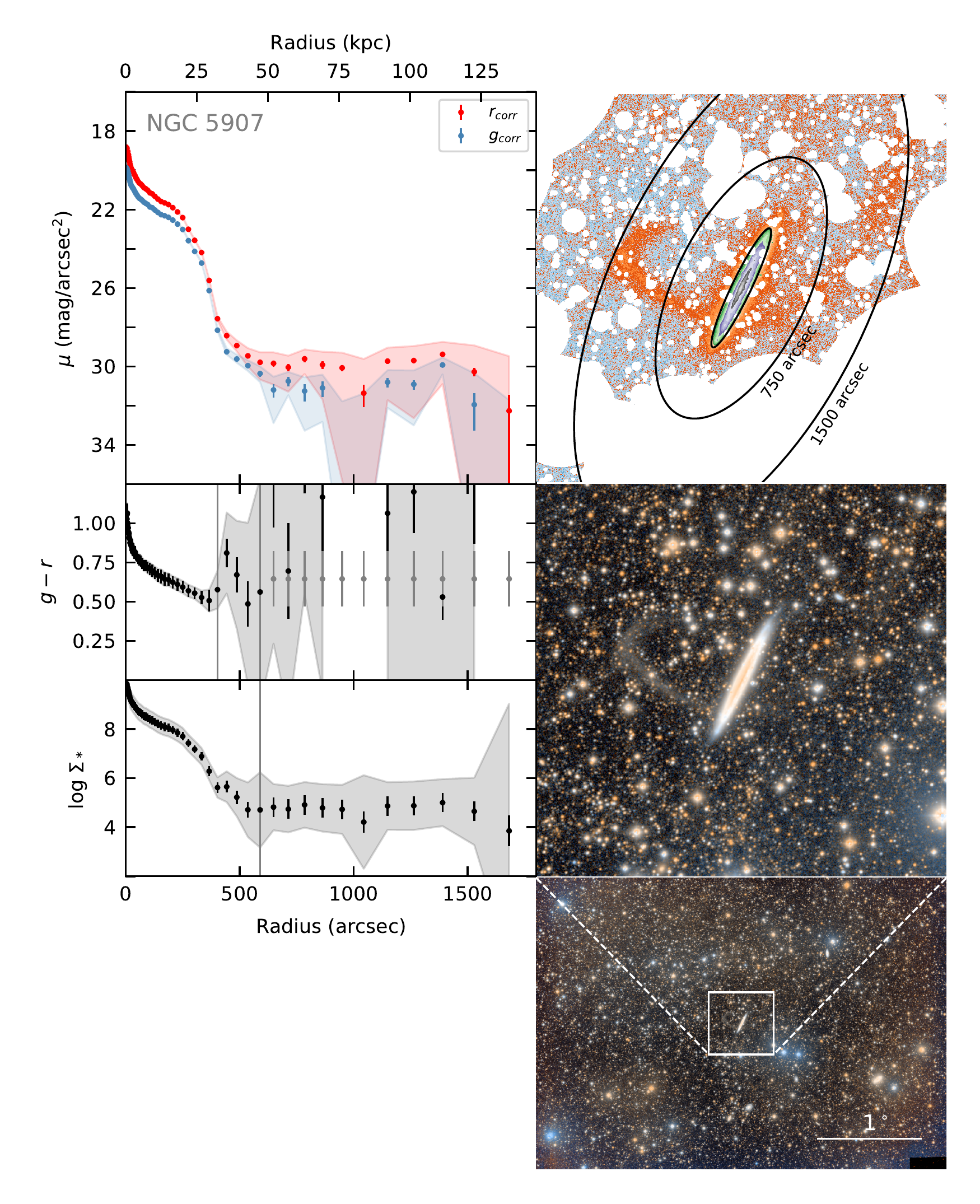}
\caption{Same as Figure~\ref{fig:NGC3044} for NGC 5907.}
\end{centering}
\end{figure}

\end{document}